\documentclass[12pt]{article}

\usepackage[margin=2cm]{geometry}
\usepackage{cite}
\usepackage[hidelinks]{hyperref}
\usepackage{amsmath,amssymb,amsfonts,amsthm}
\usepackage{algorithmic}
\usepackage{graphicx}
\usepackage{textcomp}
\usepackage{xcolor}
\usepackage{authblk}
\usepackage{multirow}
\usepackage{tabularx}
\usepackage{diagbox}
\usepackage{makecell}
\usepackage{caption}
\usepackage{mathtools}
\usepackage{float}
\usepackage{tocloft}
\usepackage{setspace}
\usepackage{lineno}
\usepackage[numbers,sort&compress]{natbib}
\usepackage{etoc}
\providecommand{\JournalTitle}[1]{#1}

\begin{document}


\pagenumbering{arabic}
\title{\textbf{Evolution of cooperation in multiplex interaction networks}}

\author[1]{Zijie Chen}
\author[2, 3, $\ast$]{Xingru Chen}
\author[4, 5, 6, $\ast$]{Feng Fu}
\affil[1]{School of Mathematical Sciences, Beijing University of Posts and Telecommunications, Beijing 100876, China}
\affil[2]{School of Artificial Intelligence, Beihang University, Beijing 100191, China}
\affil[3]{Key Laboratory of Mathematics, Informatics and Behavioral Semantics (Ministry of Education), Beihang University, Beijing 100191, China}
\affil[4]{Department of Mathematics, Dartmouth College, Hanover, NH 03755, USA}
\affil[5]{Department of Biomedical Data Science, Geisel School of Medicine at Dartmouth, Lebanon, NH 03756, USA}
\affil[6]{Department of Applied Mathematics, School of Engineering \& Applied Science, Yale University, New Haven, CT 06520, USA}
\affil[$\ast$]{Corresponding Author: \texttt{xingrucz@gmail.com, feng.fu@dartmouth.edu}}

\renewcommand{\Affilfont}{\small}
\date{}

\maketitle

\begin{abstract}
Across biological and social systems, preferential interactions among similar individuals can promote cooperation. To account for real-world interactions unfolding across multiple, simultaneous dimensions, we develop a framework for the evolution of cooperation in multiplex networks. Each layer is defined by a distinct phenotypic trait, with homophily favoring encounters between similar partners. We derive analytical conditions for natural selection to favor cooperation when phenotypes are independent or exhibit epistasis, allowing for different modes of mutation coupling. Despite payoffs being aggregated across layers, the cooperation conditions separate by layer under weak selection. Each condition depends on local payoffs and a structure coefficient determined by the effective number of phenotypes and the mutation rates. Epistasis restricts the phenotypes available on a layer and can thus shift its strategy distribution relative to that of its isolated counterpart. We show that phenotypic diversity fosters cooperation by partitioning populations into assortative niches. Furthermore, in finite populations, intensifying the prisoner's dilemma shifts the dependence of cooperation on strategy mutation from a monotonic decrease, through a U-shaped response, to a monotonic increase. Beyond weak selection, simulations show that accumulated payoffs across layers can alter mean cooperation relative to isolated systems even when phenotypes are independent. Our study unravels how homophily based on multiple phenotypes can shape the evolution of cooperation in heterogeneous populations.

\end{abstract}

\section{Introduction}

Cooperation is a cornerstone of biological and social organization, from animal societies to collective responses to global challenges such as climate change~\cite{wang2009emergence, vasconcelos2014climate, barfuss2020caring}, biodiversity loss~\cite{heal2020economic}, and pandemics~\cite{jia2020population, chen2022highly, glaubitz2024social}. Why individuals cooperate despite immediate incentives to defect remains a central challenge in evolutionary dynamics~\cite{axelrod1981evolution, nowak2006five, dreber2008winners, allen2017evolutionary, hilbe2018evolution, su2022evolutionPNAS, traulsen2023future, daniel2023evolution}. A powerful mechanism for sustaining cooperation is homophily, the tendency to interact with similar others. Widely observed in nature and society~\cite{lusseau2004identifying, bizzozzero2019tool, mcpherson2001birds, moody2004structure,  fowler2011correlated, apicella2012social, shen2025neural}, homophily promotes assortment and is generally thought to favor the emergence of cooperation~\cite{cohen2021non}.

Previous studies have explored the evolutionary origins of homophily~\cite{fu2012evolution, chen2025evolution}, its role in promoting in-group favoritism~\cite{fu2012evolutionof}, and its coevolution with phenotypic diversity~\cite{wu2017coevolutionary}. Others have applied the principle of homophily to understand social link formation via content similarity~\cite{aiello2012friendship}, competitive information diffusion~\cite{liu2020homogeneity}, and patterns of scientific collaboration~\cite{santos2024homophily}. However, most existing models base interaction preferences on similarity in a single trait, leaving the interplay among multiple phenotypic dimensions largely unexplored.

In practice, similarity is rarely one-dimensional. Individuals possess a suite of morphological, behavioral, and social phenotypes that jointly dictate interaction patterns. These traits are often linked by genetic, functional, or social constraints~\cite{otto1997deleterious, armbruster2014integrated, mech2019wolves, cheney2019baboon, simons2022agonism}. Even before several traits are considered together, a single phenotype may reflect more than one mode of transmission, including genetic and non-genetic inheritance~\cite{feldman1985gene}. As more traits enter social assortment, individuals become embedded in multiple interaction layers whose dynamics may be coupled~\cite{venkateswaran2019evolutionary}. Related work on multiplex networks also shows how behavior can ``spill over'' across distinct layers to reshape cooperation beyond isolated domains~\cite{wang2014degree, khoo2018spillover, su2022evolutionNHB}. Consequently, homophily operates not along a single axis, but across multiple, potentially interacting axes.

One might expect that such phenotypic complexity would complicate evolutionary dynamics. This challenge remains even when phenotypes on different layers are independent. Results for an isolated layer do not automatically extend to a multiplex system, because fitness depends on payoffs summed across layers. Constraints between traits can further intertwine the layers, since the phenotype expressed on one layer may narrow the range of phenotypes accessible on another. Nevertheless, ecology offers a different perspective: rich dynamics need not preclude simple organizing variables, provided the relevant scales are identified~\cite{levin1992problem, hastings2018transient}. The central question is therefore whether increasing phenotypic complexity inevitably entangles the conditions for cooperation, or whether these conditions can remain structurally separable despite interdependence among traits.

We propose a multi-layer framework where distinct phenotypic dimensions form different interaction layers. Homophily dictates interactions within each layer, whereas strategy updating is population-wide, as in earlier models incorporating phenotypic similarity~\cite{antal2009evolution, fu2012evolution}. While research on multi-layer networks has already shown how interlayer structure, migration, and payoff coupling influence cooperation~\cite{battiston2017determinants, zhang2015cooperation, zhu2025evolution}, these models typically bind interaction and imitation to the same network structure. Recent studies have begun to partition these processes, revealing that asymmetries between social contacts and role models can alter evolutionary outcomes~\cite{su2019spatial, inaba2023evolution}. Here, our framework employs a multi-layer structure specifically to encode phenotype-dependent interactions, with evolution driven by the aggregate payoffs across all layers. In a social interpretation, individuals interact with similar partners but can learn from successful role models throughout the population. This separation isolates the role of phenotypic assortment in shaping evolutionary outcomes.

We first derive the conditions for the evolution of cooperation when phenotypes are independent and then extend the analysis to unidirectional and bidirectional epistatic dependencies. Under weak selection, the analytical results are expressed in terms of layer-specific structure coefficients. This formulation is also known as the $\sigma$-rule, introduced by Tarnita et al.~\cite{tarnita2009strategy} and further developed by McAvoy and Wakeley~\cite{mcavoy2022evaluating}. Although the payoff conditions are disentangled by layer, the structure coefficients retain the effects of constraints between traits. These constraints determine which phenotypes are accessible and can shift strategy distributions relative to those in the corresponding isolated systems. For the social dilemmas considered, increasing phenotypic diversity within each dimension facilitates cooperation by relaxing payoff requirements. Our results suggest that phenotypic complexity does not simply translate into evolutionary complexity. Rather, it reveals an interplay between diversity and structure: complexity decomposes across dimensions but promotes cooperation at the level of individual traits.

\section{Model}

\begin{figure}[htpb!]
	\centering
	\includegraphics[width=\textwidth]{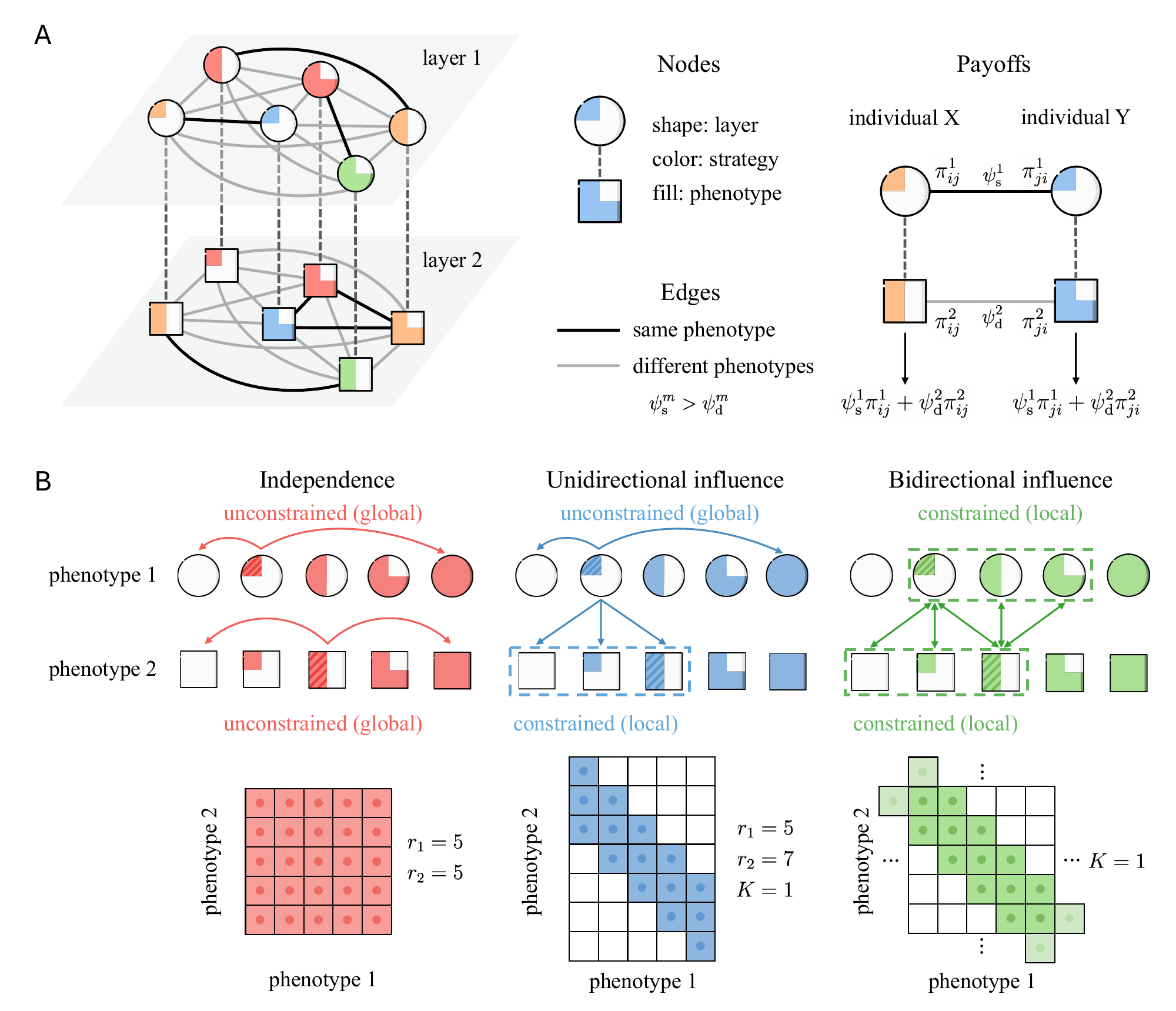}
	\caption{\small\small \textbf{Model schematic and epistatic constraints.} (A) Phenotypic assortment and payoff accumulation across two layers. Circles and squares represent individuals on layers 1 and 2, respectively. Vertical dashed lines connect the same individual across layers. Node color identifies the strategy. The filled fraction indicates the phenotype on each layer. Black and gray edges represent interactions between individuals with matching and different phenotypes, respectively. In the example, individuals X and Y adopt strategies $i$ and $j$, respectively. Payoffs from the two layers are weighted by the corresponding interaction probabilities and then summed (the main analysis sets $\psi_s^m = 1$ and $\psi_d^m = 0$). (B) Phenotypic dependencies determine which combinations are admissible at initialization and during mutation. Under independence, phenotypes on the two layers impose no constraints on each other. Under unidirectional influence, the first-layer phenotype restricts the second-layer phenotype to the corresponding state and its $2K$ nearest neighbors. The first layer remains unconstrained by the second. Under bidirectional influence, each phenotype restricts the other to $2K + 1$ candidate states. Hatched symbols indicate the current phenotypes, and dashed boxes indicate the permitted mutation ranges. Shaded cells in the lower grids mark admissible phenotype combinations. For the unidirectional case, we use $r_2 \geq r_1 + 2K$, ensuring that each neighborhood contains $2K + 1$ phenotypes on the second layer. For the bidirectional case, both phenotype spaces are unbounded, preserving this neighborhood size on both layers. The examples use $r_1 = r_2 = 5$ for independence; $r_1 = 5$, $r_2 = 7$, and $K = 1$ for unidirectional influence; and $K = 1$ for bidirectional influence.
}
\label{fig:Fig_1}
\end{figure}

We consider phenotype-based evolutionary game dynamics with multiple strategies in a well-mixed population of size $N$. Each individual carries two phenotypic traits that define interaction preferences across separate layers (Fig.~\ref{fig:Fig_1}A). An illustration using blood type and hair color is provided in Fig.~S1 of the SI Appendix.

On layer $m$ ($m = 1, 2$), there are $r_m$ possible phenotypes, indexed by $k = 1, 2, \cdots, r_m$. To model homophily~\cite{fu2012evolution}, we assume that individuals preferentially interact with others possessing similar phenotypes on the same layer. This interaction bias on layer $m$ is represented by an adjacency matrix $\Psi^m = \{\psi^m_{kl}\} \in \mathbb{R}^{r_m \times r_m}$, where $\psi^m_{kl}$ denotes the relative probability that an individual with phenotype $k$ interacts with an individual with phenotype $l$ on that layer. By construction, $\psi_{kk}^m > \psi_{kl}^m$ whenever $k \ne l$. 

In the simplest scenario, we take
\begin{align}
	\psi_{kl}^m = \left\{
	\begin{matrix}
		\psi_{\mathrm{s}}^m, \qquad k = l \\
		\psi_{\mathrm{d}}^m, \qquad k \ne l
	\end{matrix}\right.
\label{eq:Psi}
\end{align}
and further set $\psi_{\mathrm{s}}^m = 1$ and $\psi_{\mathrm{d}}^m = 0$, so that interactions on each layer occur exclusively between individuals sharing the same phenotype. Our main analysis focuses on this minimal form of phenotype-based connectivity, which captures perfect homophily and allows for analytical tractability. In the SI Appendix, we also examine imperfect homophily, with interaction probabilities $\psi_{kl}^m$ decreasing with phenotypic distance.

For the interaction dynamics on the $m$-th layer ($m = 1, 2$), we consider a two-player, two-action game with the payoff matrix $A_m = [R_m, S_m, T_m, P_m]$. These payoffs may vary across layers, as the costs and benefits of cooperation can depend on the form of social interaction~\cite{smukalla2008flo1, gore2009snowdrift}. Let $p$ and $q$ denote the probabilities of cooperation associated with the first and second layers, respectively. Accordingly, the strategy of an individual can be represented by a point $(p, q) \in [0, 1] \times [0, 1]$ in a two-dimensional strategy space. We further assume that the strategy space contains $n$ distinct strategies. The $i$-th strategy is specified by $(p_i, q_i)$, where $i = 1, 2, \dots, n$. When an individual taking strategy $i$ interacts with an individual taking strategy $j$ on the $m$-th layer, the expected payoff to the individual of strategy $i$ is denoted by $\pi_{ij}^m$.

For a focal individual, the payoff against another individual is the sum of the expected payoffs from the two layers. As illustrated in Fig.~\ref{fig:Fig_1}A, individuals X and Y adopt strategies $i$ and $j$, respectively. They share a phenotype on layer 1 but differ on layer 2, giving X an expected payoff of $\psi_{\mathrm{s}}^1\pi_{ij}^1 + \psi_{\mathrm{d}}^2\pi_{ij}^2$. Here, we use an unweighted sum across layers for simplicity.

Based on individual payoffs, the population evolves according to a Moran process with mutation~\cite{moran1958random}. At each update event, individuals reproduce in proportion to their fitness~\cite{smith1988evolution, hofbauer1998evolutionary}, defined as $f_i = \exp(\beta\pi_i)$. Here, $\pi_i$ is the expected payoff of an individual with strategy $i$, obtained by aggregating the expected contributions from all layers, and $\beta$ is the selection strength. The offspring replaces an individual chosen uniformly at random from the population. Large values of $\beta$ correspond to strong selection, under which fitness differences induced by payoffs dominate the dynamics, whereas small values of $\beta$ correspond to weak selection, where such differences have a perturbative effect. Throughout this work, our analytical results concern the limit of weak selection. We also perform agent-based simulations to investigate how stronger selection affects cooperation.

To prevent the population from reaching fixation, we incorporate mutation into the evolutionary process. Upon reproduction, an offspring inherits the strategy of the parent with probability $1 - u$, or adopts a strategy drawn uniformly at random (including that of the parent) with probability $u$. For phenotypic mutations, we distinguish two cases: (i) non-concurrent mutations, in which the phenotypic mutation events occur independently across layers, and (ii) concurrent mutations, in which the mutation events occur simultaneously. In the non-concurrent case, on each layer, the offspring inherits the phenotype of the parent with probability $1 - v$ or adopts a random phenotype (including that of the parent) with probability $v$. In the concurrent case, the offspring inherits the parental phenotype sequence with probability $1 - v$ or adopts a phenotype sequence drawn at random from all admissible combinations (again including the parental sequence) with probability $v$. Phenotypic inheritance and mutation are subject to the interlayer constraints described below. Finally, we define the rescaled mutation rates $\mu = Nu$ and $\nu = Nv$.

We further consider relationships between phenotypes across the two layers. We classify these relationships into three categories: independence, unidirectional influence (non-reciprocal epistasis), and bidirectional influence (reciprocal epistasis). The corresponding illustrations are provided in Fig.~\ref{fig:Fig_1}B. The left column of Fig.~\ref{fig:Fig_1}B depicts phenotypic independence, where the phenotype on the first layer neither constrains nor determines that on the second layer, and vice versa~\cite{amundadottir2009genome, avent2000rh, sulem2007genetic}. Thus, phenotypes on the two layers are initialized independently and undergo mutation without any restrictions.

The middle column of Fig.~\ref{fig:Fig_1}B depicts unidirectional influence, in which the phenotype on the first layer affects that on the second layer, but not conversely~\cite{locke2015genetic}. This influence imposes an epistatic constraint on both initialization and mutation. The first-layer phenotype may represent a background biological state that constrains the second-layer phenotype. Inherited defects in DNA repair provide one example: germline variants can alter the spectrum of somatic mutations~\cite{robinson2022inherited}. In our model, the phenotype on the second layer is assumed to lie within the $K$-neighborhood of the phenotype $k$ on the first layer, where $K$ is a nonnegative integer. Accordingly, any phenotypic mutation on the second layer is restricted to the $2K + 1$ candidates consisting of phenotypes $\{k - K, \cdots, k, \cdots, k + K\}$. The admissible range on the constrained layer remains centered on the current driving-layer phenotype. If a mutation occurs only on the driving layer, the constrained-layer phenotype shifts by the same amount to preserve this relation (see Fig. S9 for more details).

The right column of Fig.~\ref{fig:Fig_1}B shows bidirectional influence, where phenotypes on the two layers mutually constrain each other~\cite{hu1999heritabilities}. Likewise, the epistatic constraint applies to both initialization and mutation. The phenotype on each layer must lie within the $K$-neighborhood of the phenotype expressed on the other layer. Consequently, a phenotypic mutation on either layer is restricted to $2K + 1$ candidates determined by the phenotype on the opposite layer. An example with $K = 1$ for both unidirectional and bidirectional influence is shown in Fig.~\ref{fig:Fig_1}B. Additional empirical examples from plants, animals, and humans are provided in the SI Appendix.

\section{Results}

Under weak selection, the condition for the emergence of cooperation in our model can be expressed using a structure coefficient~\cite{tarnita2009strategy}, denoted by $\sigma_m$. Remarkably, despite the distinct mechanisms considered in the above scenarios, all our analytical results can be unified into the following form:
\begin{align}
	\sigma_m R_m + S_m > T_m + \sigma_m P_m, \qquad m=1,2.
\label{eq:sigma_rule}
\end{align}
Here, $\sigma_m$ is a coefficient that encapsulates the combined effects of population structure, mutation rates, and phenotypic dependencies on layer $m$. We next present the explicit expressions for $\sigma_m$ in each case.

\subsection*{Non-concurrent mutation}

When mutation events occur separately across layers, the structure coefficient on layer $m$ is given by
\begin{equation}
\sigma_m = \dfrac{(\mu + \nu + 1) \left[ H_m(\mu + 2\nu + 3)+\nu(\mu + \nu + 2) \right]}{(\mu + \nu + 3)\left[ H_m(\mu + 1)+\nu(\mu + \nu + 2) \right]}. 
\label{eq:sigma_non-concurrent}
\end{equation}
Here, $H_m$ represents the number of candidate phenotypes accessible on layer $m$, which depends on the type of inter-layer dependency:
\begin{equation}
H_m = 
\begin{cases}
r_m, & \text{independence} \\ 
r_m, & \text{unidirectional influence} \\
&\text{(layer $m$ is the driving layer)} \\
2K + 1, & \text{unidirectional influence} \\
&\text{(layer $m$ is the constrained layer)} \\
2K + 1. & \text{bidirectional influence}
\end{cases}
\label{eq:H_m_non-concurrent}
\end{equation}
In particular, for $H_m = 1$ or $\nu = 0$, the coefficient reduces to $\sigma_m = 1$. In these degenerate cases, the additional effect of phenotypic assortment on layer $m$ is removed, so that the evolutionary condition reduces to the baseline form $R_m + S_m > T_m + P_m$.

We then turn to the nondegenerate cases in which multiple phenotypes are admissible on each layer. In the first two cases (independence and unidirectional influence), the number of phenotypes $r_m$ on each layer may be either finite or infinite. By contrast, under bidirectional influence, we have assumed that the number of phenotypes on both layers is infinite. In the limit of $H_m \to \infty$, the coefficient simplifies to
\begin{equation}
\sigma_m = \dfrac{(\mu + \nu + 1)(\mu + 2\nu + 3)}{(\mu + \nu + 3)(\mu + 1)}.
\end{equation}

Epistasis changes the diversity entering the structure coefficient, even though the payoff conditions remain separate by layer. For independent phenotypes, this diversity is the total number of phenotypes, $r_m$. On a constrained layer, it is instead the number permitted by the driving layer, $2K + 1$. Thus, in the unidirectional case, increasing $r_2$ at fixed $K$ and mutation rates leaves $\sigma_2$ unchanged. Under bidirectional influence, both structure coefficients likewise depend on $K$ and the mutation rates, rather than on the total sizes of the phenotype spaces.

\begin{figure}[htpb!]
	\centering
	\includegraphics[width=\linewidth]{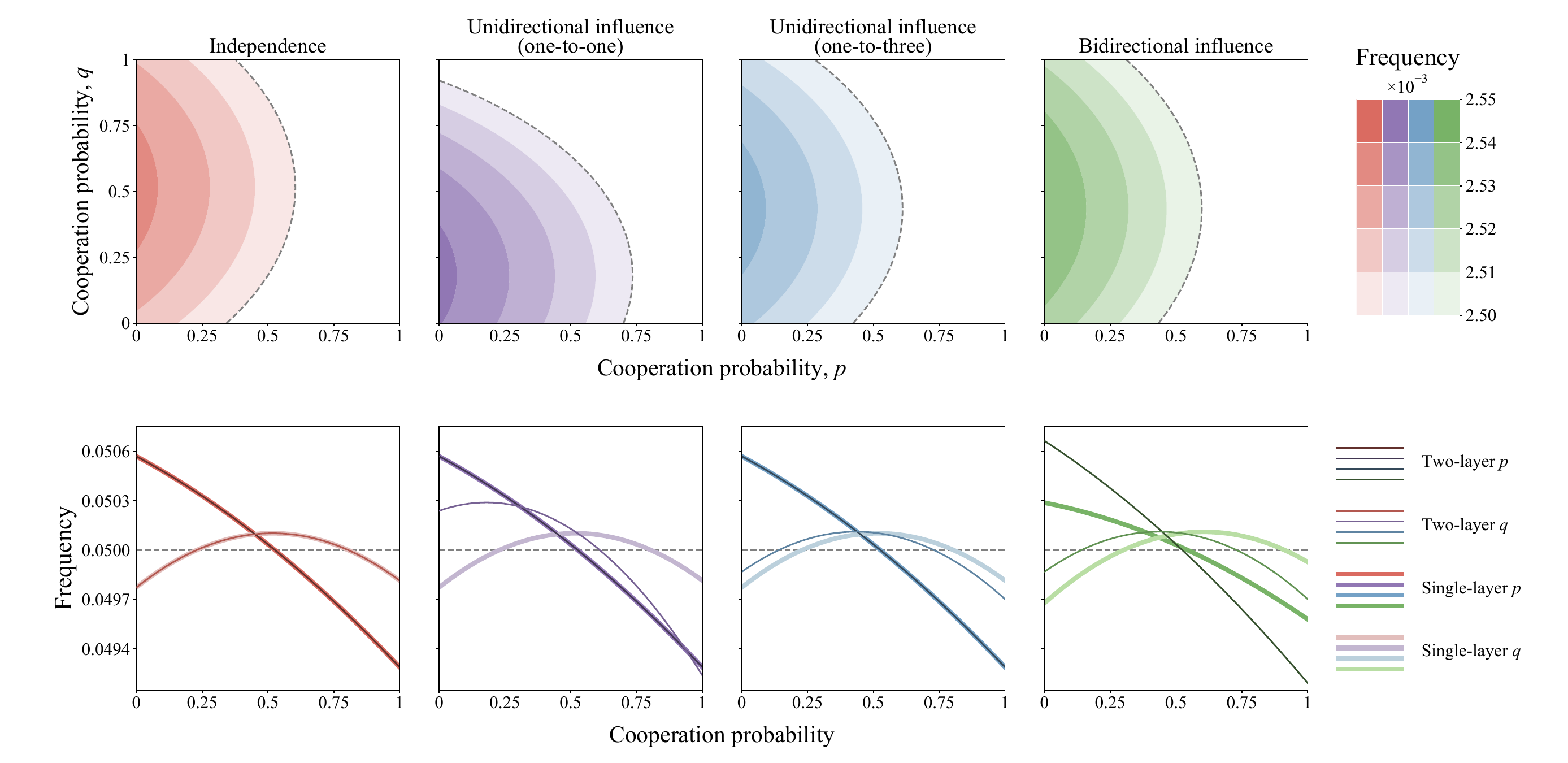}
	\caption{\small\small \textbf{Joint and marginal distributions of cooperation probabilities under different phenotypic dependencies.} From left to right, the columns show phenotypic independence, unidirectional influence with $K = 0$ (one-to-one) and $K = 1$ (one-to-three), and bidirectional influence with $K = 1$. All distributions are calculated to first order in selection strength at mutation-selection equilibrium under non-concurrent phenotypic mutation. The top row shows the stationary mean frequency of strategy pairs $(p, q)$, where $p$ and $q$ are the cooperation probabilities on layers 1 and 2, respectively. In each panel, the gray dashed contours mark the neutral frequency. Darker shading indicates higher frequencies; values below neutrality are white. The bottom row compares each two-layer marginal (thin line) with the distribution for the corresponding isolated layer (thick line). Darker curves denote $p$, whereas lighter curves denote $q$. The gray dashed line marks the neutral marginal frequency $1/n_m$, where $n_m$ is the number of possible cooperation probabilities on layer $m$. For the first three columns, $r_1 = 4$ and $r_2 = 6$ in both the two-layer model and the isolated system. For bidirectional influence, phenotype spaces are unbounded, and the isolated-layer distributions are evaluated as $r_m \to \infty$. Common parameters: $N = 50$, $n_1 = n_2 = 20$, giving $n = 400$, $u = 0.04$, $v = 0.02$, $\beta = 0.001$, and $[R_1, S_1, T_1, P_1] = [3, 0, 5, 1]$ and $[R_2, S_2, T_2, P_2] = [3, 1, 5, 0]$.
}
\label{fig:Fig_2}
\end{figure}

It can be observed from Fig.~\ref{fig:Fig_2} how phenotypic constraints shape the joint strategy distribution and its marginals. With independent phenotypes, each marginal coincides with its isolated counterpart to first order in selection strength. Under unidirectional influence, only the marginal on the constrained layer changes, with a larger departure at $K = 0$ than at $K = 1$ in the examples shown. Under bidirectional influence, both marginals differ from their isolated counterparts.

\subsection*{Concurrent mutation}

When the two phenotypes mutate concurrently, the structure coefficient $\sigma_m$ preserves the exact same form as derived for the non-concurrent case in~\eqref{eq:sigma_non-concurrent}. The distinction between the two mutation cases is absorbed into the structural parameter $H_m$. Under the concurrent regime, this term becomes
\begin{equation}
H_m = H/(1 + \delta_m).
\label{eq:H_m_concurrent}
\end{equation} 
Here, $H$ is the number of possible composite phenotypes, and $\delta_m$ is the fraction of ordered pairs of composite phenotypes that agree on layer $m$ but differ on the other layer. Their precise definitions and derivations are provided in the SI Appendix.

For games with $R_m > P_m$, larger values of the structure coefficient $\sigma$ correspond to more permissive conditions for the evolution of cooperation in~\eqref{eq:sigma_rule}. From~\eqref{eq:sigma_non-concurrent}, we further show that $\sigma_m$ increases with $H_m$ (see SI Appendix for a proof). Notably, $H_m$ increases with $r_m$ or $K$, which quantify the richness of phenotypes $k_m$ or composite phenotypes $(k_1, k_2)$ in the population and thus capture phenotypic diversity. Collectively, these results indicate a positive association between phenotypic diversity and the conditions for cooperation: increased diversity relaxes the payoff requirements under which cooperation can emerge and persist.

\begin{figure}[htpb!]
	\centering
	\includegraphics[width=\textwidth]{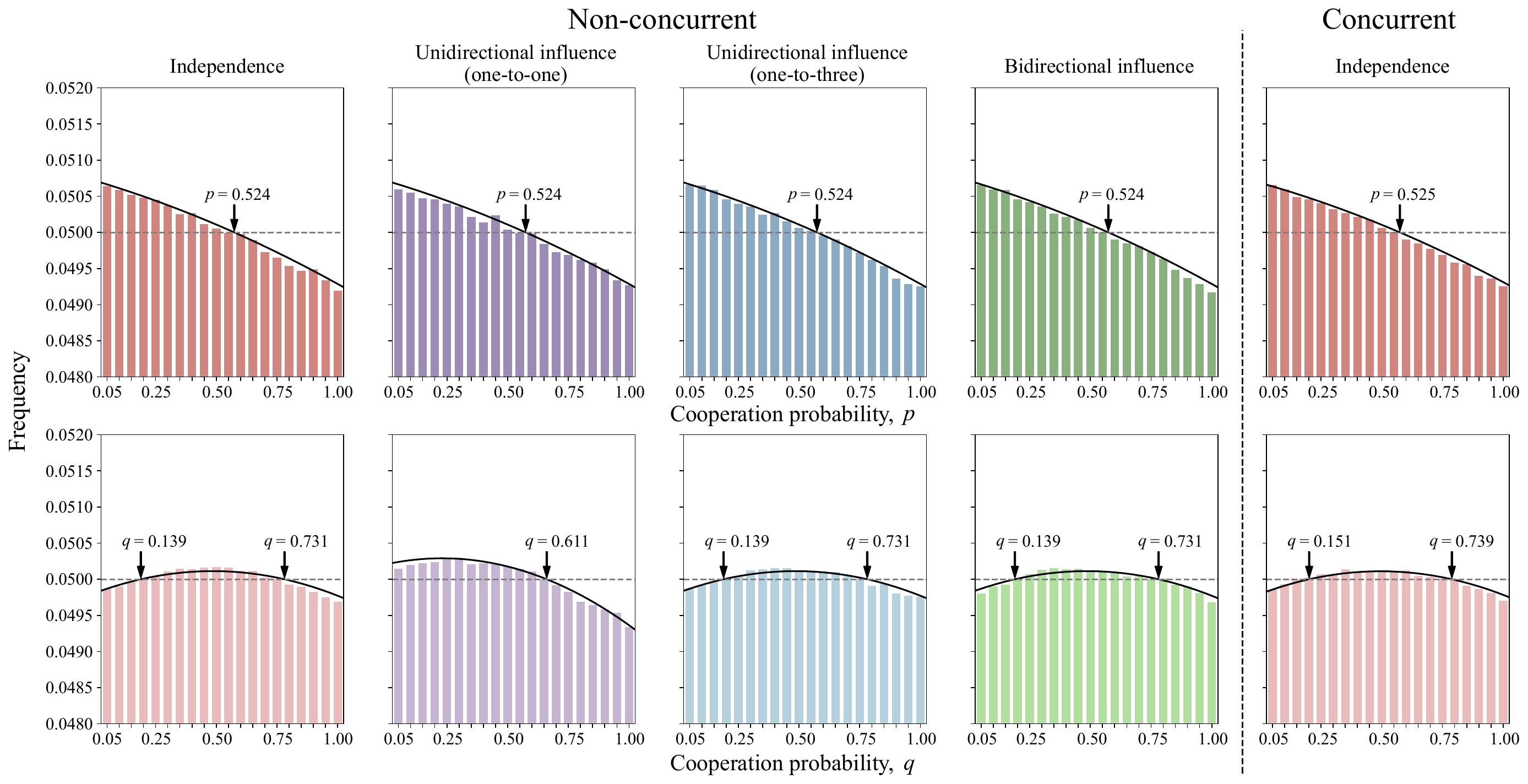}
	\caption{\small\small \textbf{Cooperation probabilities under non-concurrent mutation with different phenotypic dependencies, and under concurrent mutation with independent phenotypes.} Each column corresponds to a specific phenotypic dependency and mutation regime, as indicated at the top. The two rows show the stationary distributions of cooperation probabilities on layer 1 and layer 2, as denoted by $p$ (first row) or $q$ (second row), respectively. In each panel, the solid curve depicts the theoretical prediction of the stationary distribution, while the dashed line indicates the neutral expectation, that is, the uniform distribution with proportion $1/n_m$. Their intersection defines the theoretical critical point at which the predicted abundance equals the neutral baseline. The histograms display results from agent-based simulations. For unidirectional influence, ``one-to-one'' means that each phenotype on the first layer maps to a single phenotype on the second layer ($K = 0$), whereas ``one-to-three'' means that each phenotype maps to a corresponding phenotype as well as two neighboring phenotypes ($K = 1$). For bidirectional influence, $K = 1$. The last column shows the concurrent mutation case with independent phenotypes. Parameters: $N = 50$, $n = 20 \times 20 = 400$, $u = 0.04$, $v = 0.02$, $\beta = 0.001$, $r_1 = 3$ and $r_2 = 3, 3, 5, 3$ from left to right (excluding the bidirectional influence case), and $[R_1, S_1, T_1, P_1] =  [3, 0, 5, 1]$ and $[R_2, S_2, T_2, P_2] = [3, 1, 5, 0]$. Results are averaged over $T = 5 \times 10^8$ time steps.
}
\label{fig:Fig_3}
\end{figure}

\begin{figure}[htpb!]
	\centering
	\includegraphics[width=\textwidth]{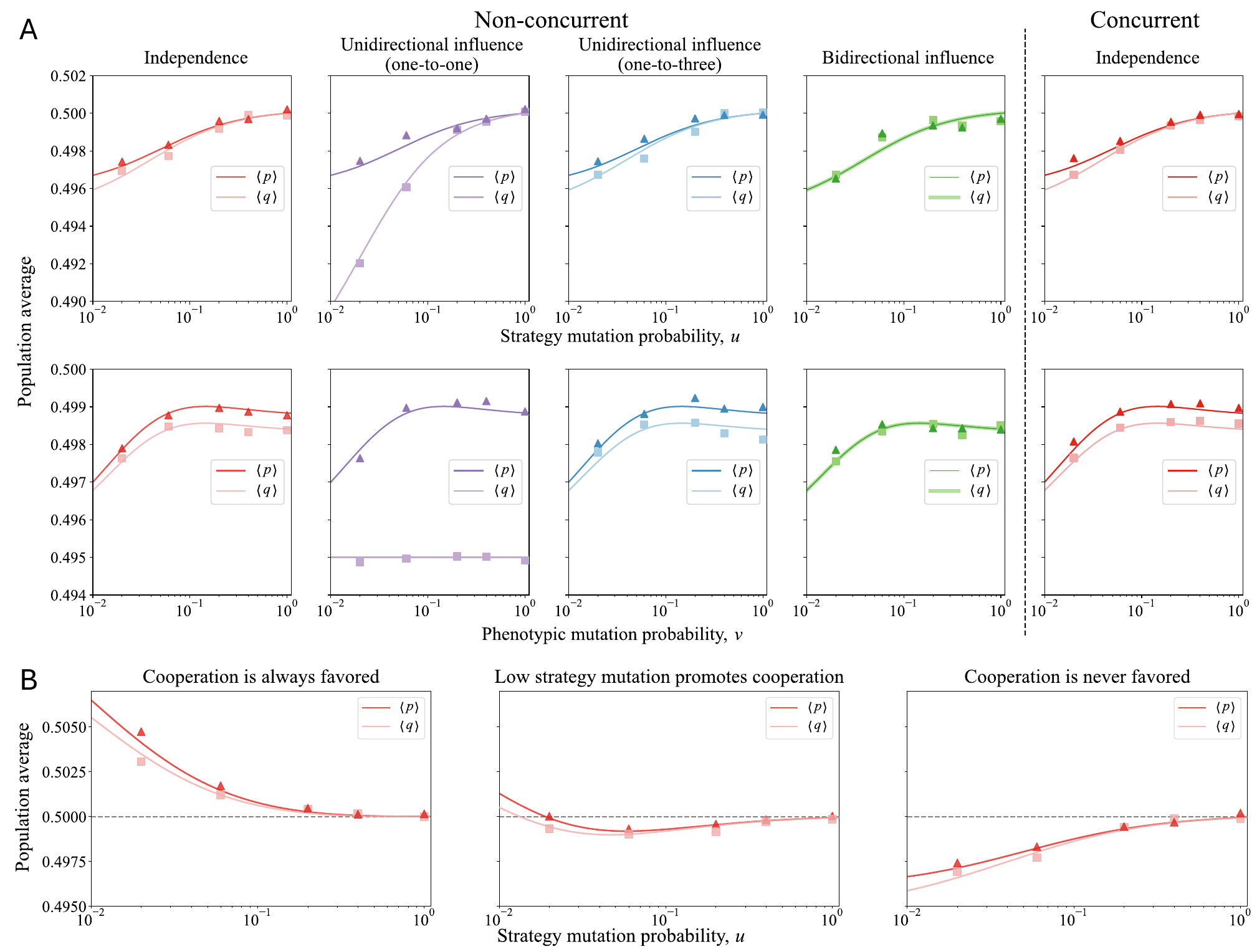}
	\caption{\small\small \textbf{Effects of mutation on cooperation in finite populations.} (A) Population averages $\langle p \rangle$ and $\langle q \rangle$ as functions of the strategy mutation probability $u$ (first row) and the phenotypic mutation probability $v$ (second row). Each column corresponds to a phenotypic dependency and mutation regime, as indicated at the top. For unidirectional influence, ``one-to-one'' and ``one-to-three'' correspond to $K = 0$ and $K = 1$, respectively. For bidirectional influence, $K = 1$.
(B) Population averages $\langle p \rangle$ and $\langle q \rangle$ as functions of $u$ under non-concurrent mutation and phenotypic independence. From left to right, mean cooperation decreases monotonically, first decreases and then increases, or increases monotonically. In each panel, solid curves depict theoretical predictions, with darker curves denoting $\langle p \rangle$ and lighter curves denoting $\langle q \rangle$. Triangles and squares display the corresponding agent-based simulation results, respectively. Dashed lines in (B) indicate the neutral value of $1/2$. Both layers use the same payoff matrix $[R, S, T, P]$ within each plot.
Parameters for (A): $N = 50$, $r_1 = 4$ and $r_2 = 3, 4, 6, 3$ from left to right (excluding the bidirectional influence case), and $[R, S, T, P] = [3, 0, 5, 1]$. The first row uses $v = 0.02$, and the second row uses $u = 0.04$. Parameters for (B), from left to right: $N = 100, 100, 50$; $r_1 = 15, 15, 4$, $r_2 = 10, 10, 3$, $v = 0.3, 0.025, 0.02$, and $[R, S, T, P] = [15, 0, 16, 1]$, $[3, 0, 4, 1]$, and $[3, 0, 5, 1]$. Common parameters: $n = 20 \times 20 = 400$, and $\beta = 0.001$. Simulation points use values $\{0.02, 0.06, 0.2, 0.4, 1\}$ for the varying mutation probability. Results are averaged over $T = 10^9$ time steps.
}
\label{fig:Fig_4}
\end{figure}

To validate the theoretical predictions, we perform agent-based simulations under five representative scenarios: non-concurrent mutation with independent phenotypes, unidirectional influence with $K = 0$, unidirectional influence with $K = 1$, bidirectional influence with $K = 1$, and concurrent mutation with independent phenotypes. The stationary distributions of cooperation probabilities $p$ and $q$ are shown in Fig.~\ref{fig:Fig_3}, while their population averages $\langle p \rangle$ and $\langle q \rangle$ as functions of the mutation probabilities $u$ and $v$ are shown in Fig.~\ref{fig:Fig_4}A. Across all scenarios, simulation results are in close agreement with analytical predictions, confirming the validity of our framework under both non-concurrent and concurrent mutation regimes. 

In finite populations, the severity of a social dilemma shapes the response of cooperation to strategy mutation. For independent phenotypes under non-concurrent mutation, our analysis identifies three distinct responses to increasing $u$ in the prisoner's dilemma: a monotonic decrease for mild dilemmas, a decrease followed by an increase for dilemmas of intermediate severity, and a monotonic increase for harsh dilemmas (Fig.~\ref{fig:Fig_4}B). All three approach the neutral value of $1/2$ as strategy mutation becomes frequent. We also characterize the dependence of cooperation on strategy and phenotypic mutation in the limit of large populations (SI Appendix).

Building on the above results, we further examine synchronous updating dynamics, as described by the Wright-Fisher process~\cite{fisher1930genetical, wright1931evolution}, in contrast to the asynchronous updating considered so far. For independent phenotypes, the Wright-Fisher process yields evolutionary outcomes that are consistent with our previous results after an appropriate rescaling of mutation rates. A detailed derivation is given in the SI Appendix. This consistency indicates that asynchronous Moran updating and synchronous Wright-Fisher updating exhibit dynamical equivalence when viewed through the lens of coalescent theory.

For the donation game, the payoff matrix is $[R_m, S_m, T_m, P_m] = [b - c, -c, b, 0]$. Under non-concurrent mutation and phenotypic independence, we obtain a critical benefit-to-cost ratio expressed as
\begin{equation}
\begin{aligned}
	\left( \dfrac{b}{c} \right)^{\ast} &= \dfrac{1}{r_m - 1}(\mu + \nu + 2)  \\
	&+ \dfrac{r_m}{r_m - 1} \dfrac{\mu^2 + 2\mu(\nu + 2) +\nu^2 + 3\nu + 3}{\nu(\mu + \nu + 2)}.
\label{eq:b_c_critical_non_independence}
\end{aligned}
\end{equation}
The above result can be readily compared with that obtained for set-structured populations. According to~\cite{tarnita2009evolutionary}, in a population of $N$ individuals distributed over $M$ sets, where each individual belongs to exactly $K$ sets ($K \le M$), the critical benefit-to-cost ratio is
\begin{equation}
\begin{aligned}
	\left( \dfrac{b}{c} \right)^{\ast} &=\dfrac{K}{M - K}(\mu + \nu + 2) \\
	&+\dfrac{M}{M - K} \dfrac{\mu^2 + 2\mu(\nu + 2) +\nu^2 + 3\nu + 3}{\nu(\mu + \nu + 2)}.
\label{eq:b_c_critical_tarnita}
\end{aligned}
\end{equation}
The close formal agreement between~\eqref{eq:b_c_critical_non_independence} and~\eqref{eq:b_c_critical_tarnita} implies a correspondence between the cooperation thresholds of the two models. In fact, the population is partitioned into $r_m$ sets, and each individual belongs to exactly one set.

We also collapse the interaction layers into a single layer using either strict matching, which requires both phenotypes to match, or permissive matching, which requires at least one match. For independent phenotypes under weak selection, the corresponding structure coefficients satisfy
\begin{align}
	\sigma_{\mathrm{permissive}} < \max_{m = 1, 2}\{\sigma_m\} \leq \sigma_{\mathrm{strict}}.
\end{align}
For a common payoff matrix with $R > P$, the strict rule is at least as favorable to cooperation as either multiplex layer, whereas the permissive rule is less favorable than the layer with the larger phenotype space.

\begin{figure}[htpb!]
	\centering
	\includegraphics[width=0.4\linewidth]{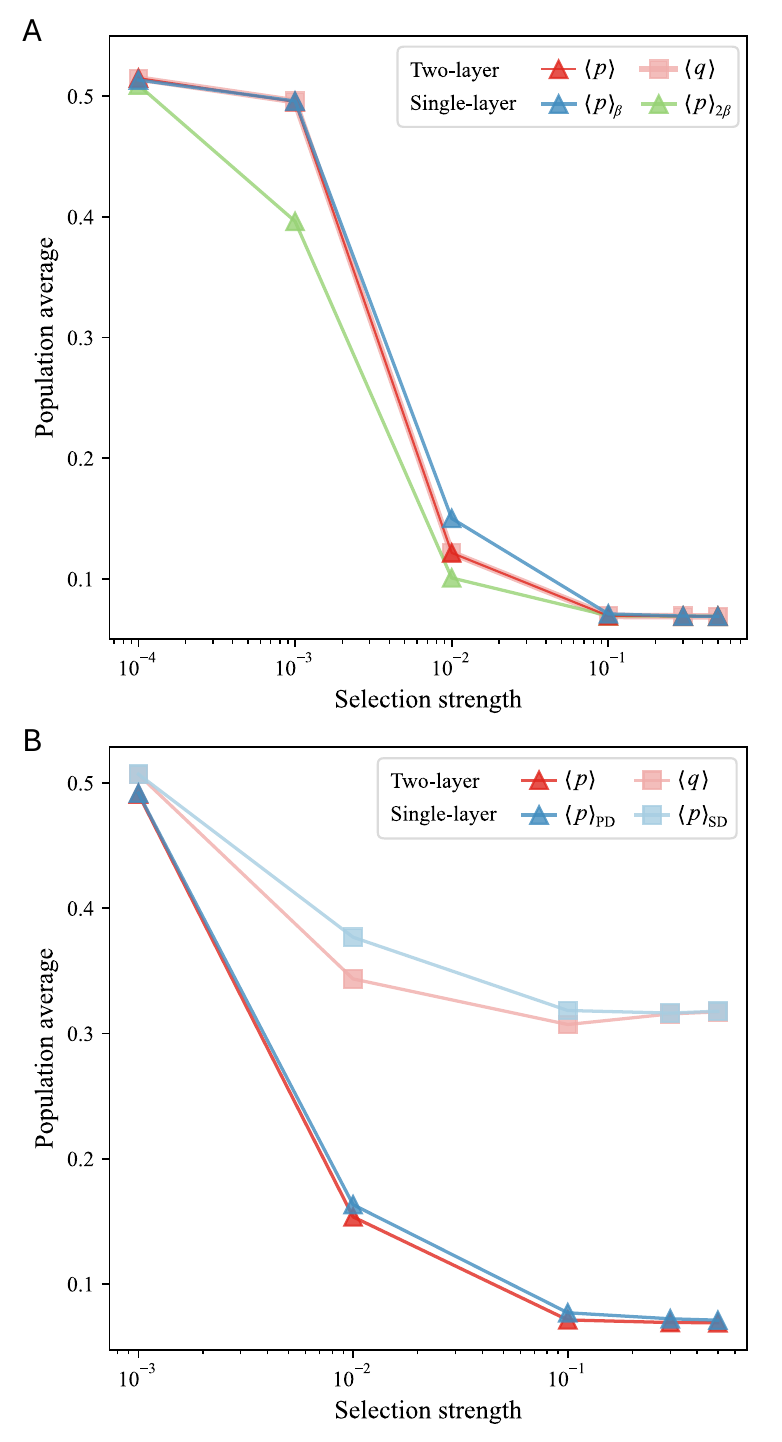}
	\caption{\small\small \textbf{Cooperation levels (population average) in two-layer and isolated single-layer systems across selection strengths under non-concurrent phenotypic mutations and independence.} Each isolated system shares the same number of phenotypes, payoff matrix, and mutation rates as its corresponding multiplex layer. The multiplex fitness is $f_i = \exp[\beta(\pi_i^1 + \pi_i^2)]$. In both panels, red triangles and light red squares denote $\langle p \rangle$ and $\langle q \rangle$ on the first and second layers of the two-layer system, respectively. (A) Symmetric-game benchmark. Both layers of the multiplex system use the same prisoner's dilemma payoff matrix and contain the same number of phenotypes, which are assigned independently across layers. Blue triangles denote the isolated single-layer result $\langle p \rangle_{\beta}$, obtained with $f_i = \exp(\beta\pi_i)$. Green triangles denote $\langle p \rangle_{2\beta}$, obtained with $f_i = \exp(2\beta\pi_i)$. Parameters: $\beta \in \{0.0001, 0.001, 0.01, 0.1, 0.3, 0.5\}$. For the two-layer system, $r_1 = r_2 = 3$ and $[R_1, S_1, T_1, P_1] = [R_2, S_2, T_2, P_2] = [3, 0, 5, 1]$. (B) Heterogeneous-game benchmark. The first layer uses the prisoner’s dilemma (PD), whereas the second uses the snowdrift game (SD). Blue triangles and light blue squares denote $\langle p \rangle_{\mathrm{PD}}$ and $\langle p \rangle_{\mathrm{SD}}$ in the corresponding isolated PD and SD systems, respectively. Parameters: $\beta \in \{0.001, 0.01, 0.1, 0.3, 0.5\}$. For the two-layer system, $r_1 = 3$, $r_2 = 4$, $[R_1, S_1, T_1, P_1] = [3, 0, 5, 1]$, and $[R_2, S_2, T_2, P_2] = [3, 1, 5, 0]$. Common parameters: $N = 50$, $n = n_1 \times n_2 = 400$, $u = 0.04$, and $v = 0.02$. Results are averaged over $1\times10^{10}$ generations.
}
\label{fig:Fig_5}
\end{figure}

Beyond weak selection, we perform agent-based simulations to compare the multiplex system with isolated layers under non-concurrent mutation and phenotypic independence. When both layers have identical prisoner's dilemma payoffs and the same number of phenotypes, the mean cooperation at intermediate selection strengths is lower than in an isolated layer at $\beta$ but higher than in one at $2\beta$ (Fig.~\ref{fig:Fig_5}A). Hence, replicating the game on another layer differs from simply doubling selection strength in the isolated system. When a prisoner's dilemma is paired with a snowdrift game, the experiments show lower cooperation on both layers than in their corresponding isolated systems, with the most pronounced differences at intermediate selection strengths (Fig.~\ref{fig:Fig_5}B). These comparisons indicate that selection based on payoffs accumulated across layers can alter cooperation even when phenotypes are independent.

\section{Discussion}

We have developed a general framework for the evolution of cooperation in multiplex networks with phenotypic heterogeneity and inter-layer dependence. Our central finding is that, despite the presence of multiple phenotypes and potential epistatic constraints, the condition for cooperation on each layer collapses into a simple rule involving its payoff matrix and a structure coefficient $\sigma_m$ under weak selection. This rule has the same form across layers, while $\sigma_m$ captures the effects of phenotypic structure, mutation regimes, and inter-layer relations. The conditions for cooperation on different layers are \emph{formally disentangled}. Although the total fitness of an individual integrates payoffs from all layers, whether selection favors cooperation on a given layer depends solely on that layer's payoffs and its corresponding $\sigma_m$~\cite{mcavoy2022evaluating}.

Epistasis changes cooperation through the diversity of phenotypes available on each layer. In Fig.~\ref{fig:Fig_2}, the independent and unidirectional cases have the same phenotype numbers and payoff matrices, yet constraining the second phenotype shifts its strategy distribution. Under bidirectional influence, both marginal distributions differ from their isolated counterparts. Indeed, the relevant diversity is determined by the relations between traits, not by phenotype counts alone. Beyond weak selection, an additional interaction layer can reduce cooperation even when phenotypes are independent (Fig.~\ref{fig:Fig_5}). This reduction occurs both when the two layers host identical prisoner's dilemmas and when a prisoner's dilemma is paired with a snowdrift game. Hence, cooperation within a given layer can change even when its own payoff structure and phenotype space remain fixed.

Under weak selection, the cooperation rule retains its form under both concurrent and non-concurrent phenotypic mutation. As given in~\eqref{eq:H_m_non-concurrent} and~\eqref{eq:H_m_concurrent}, the distinction between these two mutation regimes enters through their respective diversity terms $H_m$. Furthermore, the formal equivalence between asynchronous Moran updates and synchronous Wright-Fisher updating, contingent on an appropriate rescaling of mutation rates, indicates that the threshold for cooperation remains invariant to the microscopic temporal structure of life cycles. When specialized to donation games, our framework recovers the classic results of single-layer and set-structured populations as limiting cases. In this light, the number of distinct phenotypes mirrors the number of sets, providing a unified interpretation of assortment.

This perspective also connects our framework to the classical role of relatedness in the evolution of cooperation. Hamilton's rule expresses this idea by weighting social benefits by genetic relatedness~\cite{hamilton1964genetical}. Later, population-genetic treatments placed kin-structured and group-structured interactions in a common framework~\cite{uyenoyama1980theories}, while inclusive-fitness models of viscous populations showed how limited dispersal can increase relatedness yet also intensify local competition~\cite{taylor1992altruism}. In our model, phenotypic homophily plays a complementary role, generating assortment by organizing who meets whom across layers without requiring common ancestry to be specified explicitly.

A key insight from this work is that phenotypic diversity determines how readily cooperation is favored. The structure coefficient $\sigma_m$ grows with the number of admissible phenotypes, quantified by $H_m$, which in turn increases with either the phenotypic richness $r_m$ or the inter-layer tolerance $K$. This finding is consistent with earlier theoretical work~\cite{wu2017coevolutionary, santos2012role} and experimental observations in microbial populations~\cite{brockhurst2006character}. By partitioning the population into distinct groups, phenotypic diversity reduces the effective number of neighbors and thereby facilitates assortment. This interpretation is also reflected in the comparison with flattened representations. For a given phenotypic composition, strict matching restricts interaction neighborhoods, whereas permissive matching expands them. Each multiplex layer lies between these two extremes in terms of mean degree. Analogous to the advantage of low degree in social networks~\cite{ohtsuki2006simple}, the phenotypic segregation in our model allows cooperators to cluster and flourish through mutual reciprocity. 

Beyond phenotypic niches, the effect of strategy mutation depends on the severity of the social dilemma. We find that strategy mutation draws cooperation toward its neutral level, with varying consequences across different prisoner's dilemmas (Fig.~\ref{fig:Fig_4}B). In a mild environment, it weakens the cooperation bias generated by selection. As the environment becomes harsher, it serves as a source of variety that prevents the population from being trapped in a state dominated by defection. This leads to a transition through a non-monotonic, U-shaped profile toward a monotonic increase, where higher mutation rates actually facilitate the emergence of cooperation.

By yielding exact conditions for cooperation expressed through the structure coefficient, the discrete interaction mode provides a tractable starting point from which broader, continuous social kernels can be explored. Phenotypic heterogeneity is thus not merely background variation, but can enter directly into population-level dynamics~\cite{wachter2015population}. A natural extension would consider a broader range of relationships between phenotypic distance and interaction probability. An interaction probability that decreases with phenotypic distance represents homophily (see Fig.~S22 in the SI Appendix), which favors proximity and can reinforce culturally patterned cooperation~\cite{boyd2009culture}. In contrast, a probability that increases with distance represents heterophily, which favors complementarity. This latter force catalyzes the division of labor in social insects~\cite{ribbands1952division, wilson1971social} and the increasing complexity of role specialization in human societies~\cite{moody2004structure, gintis2005moral}. Homophily and heterophily could also coexist, with interaction probabilities lowest at intermediate phenotypic distances. The next question is whether the weak-selection separability found here survives when cooperation crosses phenotypic divides: diversity may help not only by sorting like with like, but also by bringing unlike types together~\cite{wuchty2007increasing}.

{\small\small
\section*{Materials and Methods}
\label{mat_a_met}
\subsection*{A. Expected payoff of a strategy}

By combining phenotypes and strategies, the population can be partitioned into finer subgroups. Let $x_i^{k_1k_2}$ denote the proportion of individuals adopting strategy $(p_i, q_i)$ and possessing phenotype $k_1$ on the first layer and phenotype $k_2$ on the second layer, respectively. The total proportion of individuals using strategy $i$ is then given by
\begin{align}
	x_i = \sum_{k_1 = 1}^{r_1} \sum_{k_2 = 1}^{r_2} x_i^{k_1k_2}.
\label{eq:x_i}
\end{align}
When an individual taking strategy $i$ interacts with an individual taking strategy $j$ on the $m$-th layer, the expected payoff to the individual of strategy $i$ is denoted by $\pi_{ij}^m$. For the two layers, we have
\begin{align}
\begin{split}
	\pi_{ij}^{1} &= R_1p_ip_j + S_1p_i(1 - p_j) + T_1(1 - p_i)p_j + P_1(1 - p_i)(1 - p_j), \\
	\pi_{ij}^{2} &= R_2q_iq_j + S_2q_i(1 - q_j) + T_2(1 - q_i)q_j + P_2(1 - q_i)(1 - q_j).
\end{split}
\label{eq:payoff}
\end{align}

Based on the adjacency matrix $\Psi^m$ for phenotypes defined in \eqref{eq:Psi}, we derive the effective adjacency matrix for strategies on layer $m$, denoted as $\Phi^m = \{\varphi_{ij}^m\} \in \mathbb{R}^{n \times n}$. The entry $\varphi_{ij}^m$ represents the expected total number of interactions between strategy-$i$ focal individuals and strategy-$j$ partners on layer $m$. Specifically,
\begin{align}
	\varphi_{ij}^m = N^2 \sum_{k_1=1}^{r_1} \sum_{l_1=1}^{r_1} \sum_{k_2=1}^{r_2} \sum_{l_2=1}^{r_2} \psi_{k_ml_m}^m x_i^{k_1k_2} x_j^{l_1l_2} .
\label{eq:interaction_probability}
\end{align}
The expected payoff of an individual with strategy $i$ is then given by
\begin{align}
	\pi_i = \dfrac{1}{Nx_i} \sum_{m=1}^2 \sum_{j=1}^n \varphi_{ij}^m \pi_{ij}^m.
\label{eq:payoff_i}
\end{align}
This normalization averages over all individuals using strategy $i$. We can then obtain the corresponding fitness $f_i = \exp(\beta \pi_i)$ and follow the update process in the \textbf{Model} section. The SI Appendix provides further details of this payoff calculation and discusses alternative payoff formulations, including normalization by the number of interactions.

\subsection*{B. Condition for the evolution of cooperation}

We aim to identify the conditions under which cooperation is favored by natural selection. Rather than working with a finite set of discrete strategies $(p_i, q_i)$, we further consider a continuous strategy space $(p, q) \in [0, 1] \times [0, 1]$. Using coalescent theory~\cite{wakeley2009coalescent} together with the perturbation approach for evolutionary games~\cite{antal2009evolution, antal2009mutation}, we derive first-order approximations to the stationary distributions of $p$ and $q$. In practice, we deal with rescaled deviations $\mathcal{D}^1(p)$ and $\mathcal{D}^2(q)$, which are obtained from the corresponding stationary probability densities by subtracting the neutral baseline and rescaling by a positive constant factor (see SI Appendix for details). By construction, both functions integrate to zero. 

Although fitness is determined by the total payoff aggregated across layers, the evolutionary conditions for cooperation on each layer remain disentangled under weak selection. The contribution of each layer enters additively and can be evaluated separately to first order in the selection strength $\beta$. Specifically, cooperation is favored on layer 1 when the average probability of cooperation satisfies $\langle p  \rangle > 1/2$, and on layer 2 when $\langle q  \rangle > 1/2$. Equivalently, these conditions can be expressed as
\begin{align}
\begin{split}
	\int_0^1 p\mathcal{D}^1(p) dp > 0,\\
	\int_0^1 q\mathcal{D}^2(q) dq > 0.
\end{split}
\label{eq:favor_condition}
\end{align}
For discrete strategies, this approach also allows us to study the long-run abundance of any strategy pair $(p_i, q_i)$.
}

\subsection*{Data accessibility}

The source code required to reproduce the results and the data used to generate the figures are available in the \href{https://github.com/xingrucz/Cooperation-Multiplex}{GitHub repository}, along with detailed usage instructions.
Supplementary material is available online.


\newpage

\begin{center}

{\LARGE Supplementary Information for \\ \textbf{Evolution of cooperation in multiplex interaction networks} \par}

\vspace{1em}

{\large
Zijie Chen$^{1}$,
Xingru Chen$^{2,3,*}$,
Feng Fu$^{4,5,6,*}$
}

\vspace{1em}

{\small
$^{1}$ School of Mathematical Sciences, Beijing University of Posts and Telecommunications, 

Beijing 100876, China

$^{2}$ School of Artificial Intelligence, Beihang University, Beijing 100191, China

$^{3}$ Key Laboratory of Mathematics, Informatics and Behavioral Semantics (Ministry of Education), Beihang University, Beijing 100191, China

$^{4}$ Department of Mathematics, Dartmouth College, Hanover, NH 03755, USA

$^{5}$ Department of Biomedical Data Science, Geisel School of Medicine at Dartmouth, Lebanon, 

NH 03756, USA

$^{6}$ Department of Applied Mathematics, School of Engineering \& Applied Science, Yale University, 

New Haven, CT 06520, USA

$^{*}$ Corresponding Author: \texttt{xingrucz@gmail.com, feng.fu@dartmouth.edu}
}

\end{center}

\pagenumbering{arabic}
\setcounter{page}{1}

\renewcommand{\thefigure}{S\arabic{figure}}
\setcounter{figure}{0}
\renewcommand{\thetable}{S\arabic{table}}
\setcounter{table}{0}
\renewcommand{\theequation}{S\arabic{equation}}
\setcounter{equation}{0}

\setcounter{section}{0}
\setcounter{subsection}{0}
\setcounter{subsubsection}{0}
\setcounter{NAT@ctr}{0}

\section{Model}
\label{section:model}

\subsection{Population structure}

We consider a population of $N$ individuals, where each individual is characterized by two traits, such as blood type and hair color. They are represented on two distinct network layers: the first layer corresponds to the first trait (e.g., blood type), and the second layer corresponds to the second trait (e.g., hair color). Thus, each individual simultaneously possesses one trait on each layer. For example, an individual may have blood group O on the first layer and brown hair on the second layer.  

We further assume that each trait can take multiple discrete phenotypes. In the case of blood type, the trait admits several phenotypes such as A, B, AB, and O. On the $m$-th layer, where $m = 1, 2$, the number of possible phenotypes is denoted by $r_m$, and each phenotype is indexed by $k = 1, 2, \ldots, r_m$. To capture the tendency toward homophily, we assume that individuals are more likely to interact with others holding similar phenotypes on the same layer~\cite{fu2012evolution}. Formally, we represent the connection bias on layer $m$ by an adjacency matrix $\Psi^m = \{\psi^m_{kl}\} \in M_{r_m \times r_m}(\mathbb{R})$, where $\psi^m_{kl}$ describes the relative probability of interaction between individuals with phenotypes $k$ and $l$. By construction, $\psi_{kk}^m > \psi_{kl}^m$ whenever $k \ne l$. In the simplest scenario, we set
\begin{align}
	\psi_{kl}^m =
	\begin{cases}
		\psi_{\mathrm{s}}^m, & k = l, \\
		\psi_{\mathrm{d}}^m, & k \neq l .
	\end{cases}
\label{eq:Psi}
\end{align}
Here, the subscripts $\mathrm{s}$ and $\mathrm{d}$ stand for ``same'' and ``different'', respectively. In the main analysis, we specify $\psi_{\mathrm{s}}^m=1$ and $\psi_{\mathrm{d}}^m=0$, so that individuals are connected on each layer if and only if they share the same phenotype on that layer. We relax this assumption in Section~\ref{section:imperfect_homophily} to consider imperfect homophily and distance-dependent interactions. 

\subsection{Game dynamics}

\subsubsection{Payoff}

For the interaction dynamics on the $m$-th layer of the network, we focus on a two-player game with a payoff matrix 
\begin{equation}
A_m = 
\begin{bmatrix}
R_m & S_m \\
T_m & P_m
\end{bmatrix},
\end{equation} 
where $m=1, 2$. Let the probabilities of cooperation for individuals on the two layers be denoted by $p$ and $q$, respectively. We represent the strategy of an individual as a point $(p,q) \in [0,1] \times [0,1]$ in a two-dimensional strategy space. Moreover, we assume that there are $n$ distinct strategies, with the $i$-th strategy represented by $(p_i, q_i)$, where $i = 1, 2, \dots, n$.

By combining strategies and phenotypes, the population can be further classified into finer subgroups. Let $x_i^{k_1k_2}$ denote the proportion of individuals adopting strategy $(p_i, q_i)$ and exhibiting phenotypes $k_1$ and $k_2$ on the first and second layer, respectively. The total proportion of individuals using strategy $i$ is then
\begin{align}
	x_i = \sum_{k_1 = 1}^{r_1} \sum_{k_2 = 1}^{r_2} x_i^{k_1k_2}.
\label{eq:x_i}
\end{align}
When an individual using strategy $i$ interacts with another individual using strategy $j$, their expected payoffs on layers 1 and 2 can be written as
\begin{equation}
\begin{dcases}
	\pi_{ij}^1 = R_1p_ip_j + S_1p_i(1 - p_j) + T_1(1 - p_i)p_j + P_1(1 - p_i)(1 - p_j), \\
	\pi_{ij}^2 = R_2q_iq_j + S_2q_i(1 - q_j) + T_2(1 - q_i)q_j + P_2(1 - q_i)(1 - q_j).
\end{dcases}
\label{eq:payoff}
\end{equation}

Based on the adjacency matrix $\Psi^m$ for phenotypes in \eqref{eq:Psi} and the distribution of strategies in \eqref{eq:x_i}, we can derive the adjacency matrix for strategies on layer $m$, denoted as $\Phi^m = \{\varphi_{ij}^m\} \in M_{n \times n}(\mathbb{R})$. Here, $\varphi_{ij}^m$ represents the total number of meetings between individuals with strategies $i$ and $j$ on layer $m$ per unit of time:
\begin{align}
	\varphi_{ij}^m = N^2 \sum_{k_1=1}^{r_1} \sum_{l_1=1}^{r_1} \sum_{k_2=1}^{r_2} \sum_{l_2=1}^{r_2} x_i^{k_1k_2} x_j^{l_1l_2} \psi_{k_ml_m}^m.
\label{eq:interaction_probability}
\end{align}
Accordingly, the expected payoff of an individual with strategy $i$ can be expressed as
\begin{align}
	\pi_i = \dfrac{1}{Nx_i} \sum_{m=1}^2 \sum_{j=1}^n \varphi_{ij}^m \pi_{ij}^m.
\label{eq:payoff_i}
\end{align}
This formulation averages accumulated payoffs over individuals sharing strategy $i$. In Section~\ref{section:normalized}, we examine alternative payoff calculations, including normalizing each individual's payoff on each layer by their number of interactions on that layer.

\begin{figure}[htpb!]
	\centering
	\includegraphics[width=0.9\columnwidth]{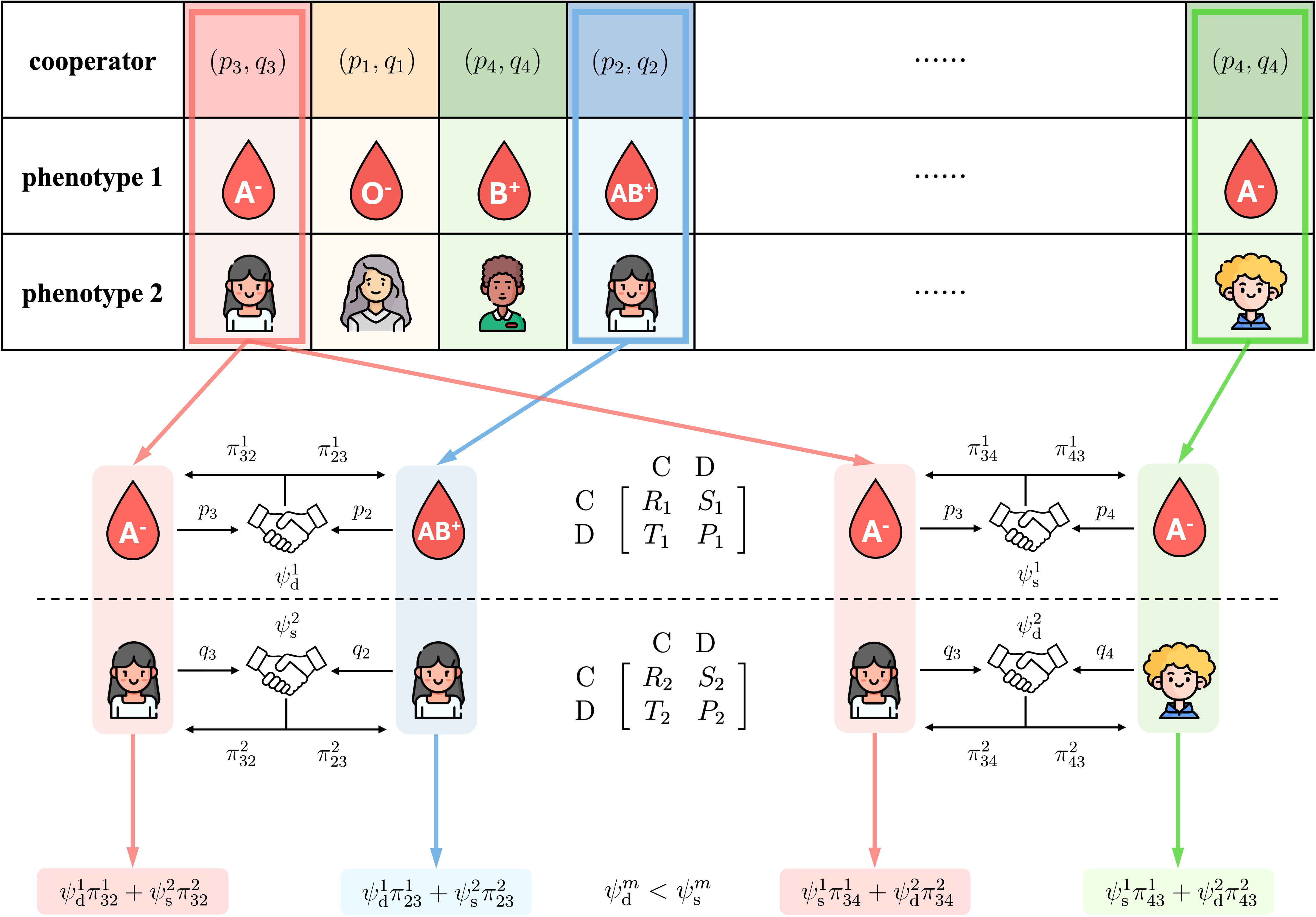}
	\captionsetup{font = small}
	\caption{\textbf{Illustration of the payoff calculation on a two-layer network.} The first layer corresponds to strategies $p_i$ with blood type as the trait, while the second layer corresponds to strategies $q_i$ with hair color as the trait. The interaction probabilities $\psi_{\mathrm{s}}^m$ between individuals sharing the same phenotype and $\psi_{\mathrm{d}}^m$ between individuals holding different phenotypes are defined in \eqref{eq:Psi}. According to \eqref{eq:payoff}, the payoff for strategy $3$ interacting with strategy $2$ on the first layer is $\pi_{32}^1 = R_1p_3p_2 + S_1p_3(1-p_2) + T_1(1-p_3)p_2 + P_1(1-p_3)(1-p_2)$. The other $\pi_{ij}^m$ values can be calculated similarly.}
\label{fig:payoff_calculation}
\end{figure}

A detailed illustration of the payoff calculation is provided in Fig.~\ref{fig:payoff_calculation}. On the first layer, if an individual with blood type $\text{A}^-$ encounters an individual with blood type $\text{AB}^+$, the interaction occurs with probability $\psi_{\mathrm{d}}^1$. Here, we adopt a more precise classification of blood types, where the symbols ``$+$'' and ``$-$'' denote Rh-positive and Rh-negative types, respectively. In contrast, two individuals both carrying blood type $\text{A}^-$ interact on the first layer with probability $\psi_{\mathrm{s}}^1$. As an example, the expected payoff of an $\text{A}^-$ individual adopting strategy $p_3$ when interacting with an $\text{AB}^+$ individual adopting strategy $p_2$ is given by
\begin{equation}
\psi_{\mathrm{d}}^1[R_1p_3p_2 + S_1p_3(1 - p_2) + T_1(1 - p_3)p_2 + P_1(1 - p_3)(1 - p_2)].
\end{equation} 
Similarly, on the second layer, individuals with different hair colors interact with probability $\psi_{\mathrm{d}}^2$, whereas individuals sharing the same hair color interact with probability $\psi_{\mathrm{s}}^2$. For instance, the expected payoff of a black-haired individual adopting strategy $q_3$ when interacting with another black-haired individual adopting strategy $q_2$ is 
\begin{equation}
\psi_{\mathrm{s}}^2[R_2q_3q_2 + S_2q_3(1 - q_2) + T_2(1 - q_3)q_2 + P_2(1 - q_3)(1 - q_2)].
\end{equation}
Notably, interactions on the first layer do not affect interactions on the second layer.

\subsubsection{Updating rule}
\label{Updating rule}

The system evolves according to the \textbf{Moran process}~\cite{moran1958random,moran1962statistical}, incorporating mutations. Individuals reproduce in proportion to their fitness~\cite{smith1982evolution,hofbauer1998evolutionary}, which is defined as $f_i = \exp(\beta\pi_i)$, where $\pi_i$ is the expected payoff of an individual with strategy $i$ and $\beta$ is the selection strength. A large $\beta$ corresponds to strong selection, in which fitness differences generated by payoffs dominate the evolutionary dynamics, whereas a small $\beta$ represents weak selection, in which these differences have a limited effect. In what follows, we focus on the weak selection regime and extend our analysis to strong selection in Section~\ref{section:strong_selection}. 

We distinguish between interaction and replacement graphs, consistent with the framework introduced by Ohtsuki et al.~\cite{ohtsuki2007breaking}. The matrices $\Psi^m$ define the interaction graphs on which individuals obtain their payoffs, whereas reproductive replacement occurs across the entire population. At each update, the offspring of a parent selected in proportion to fitness replaces an individual chosen uniformly at random from the entire population. Thus, the interaction graphs are structured by phenotypic homophily and the replacement graph is complete. Biologically, this setup represents a population in which phenotype recognition shapes social interactions, while offspring dispersal and replacement occur throughout the population. Socially, individuals interact preferentially with similar others but can learn from successful role models beyond their interaction partners. Related interpretations appear in evolutionary models based on phenotypes: Antal et al.~\cite{antal2009evolution} study cooperation conditional on phenotypic similarity in a well-mixed population, while Fu et al.~\cite{fu2012evolution} model homophilic interactions with evolutionary updating interpreted as reproduction or social learning. In our model, global updating allows strategies and phenotypes to spread across the population even when interactions are restricted by homophily.

During reproduction, an offspring inherits the strategy of the parent with probability $1- u$ or adopts a randomly chosen strategy (including that of the parent) with probability $u$. Based on the updating mechanism, the expected changes in the frequency of strategy $i$ due to selection and mutation in a single generation are
\begin{align}
	\Delta x_i^{\mathrm{sel}} = x_i \left( \dfrac{f_i}{\sum_j Nx_jf_j} +1 - \dfrac{1}{N} \right) - x_i, \qquad \Delta x_i^{\mathrm{mut}} = \dfrac{1}{N} \left( \dfrac{1}{n} - x_i \right),
\label{eq:frequency_changes}
\end{align}
respectively.

For phenotypic mutations, we consider two cases: (i) \textbf{non-concurrent mutations across layers}, in which each phenotype mutates separately, and (ii) \textbf{concurrent mutations across layers}, in which all phenotypes mutate simultaneously. In the non-concurrent case, for each layer, the offspring inherits the phenotype of the parent with probability $1 - v$, or adopts a randomly chosen phenotype (including that of the parent) with probability $v$. In the concurrent case, the offspring inherits the entire combination of phenotypes from the parent with probability $1 - v$ or adopts a combination of phenotypes drawn at random from all possible combinations (again including that of the parent) with probability $v$.

\subsection{Strategy selection condition}
\label{section:subsection_strategy_selection_condition}

By averaging the change in the frequency of strategy $i$ due to selection over all possible population states, we obtain its expected change at the stationary state. At the lowest-order approximation in $\beta$, this yields
\begin{align}
	\langle \Delta x_i^{\text{sel}} \rangle = \dfrac{\beta}{N^2} \left( \sum_m \sum_j \langle \varphi_{ij}^m\pi_{ij}^m \rangle - \sum_m \sum_j \sum_r \langle x_i \varphi_{jr}^m\pi_{jr}^m \rangle \right).
\label{eq:x_sel}
\end{align}
According to the perturbation method in evolutionary games~\cite{antal2009evolution,antal2009mutation}, under weak selection, the average $\langle \, \cdot \, \rangle$ can be evaluated over the stationary population state under neutral evolution, that is, when $\beta = 0$.

The total expected change in the mutation-selection equilibrium is
\begin{align}
	\langle \Delta x_i^{\mathrm{tot}} \rangle = (1-u) \langle \Delta x_i^{\text{sel}} \rangle + u \langle \Delta x_i^{\text{mut}} \rangle = 0.
\label{eq:x_tot}
\end{align}
Substituting \eqref{eq:frequency_changes} into the above expression, the stationary frequency (or abundance) of strategy $i$ can be written as
\begin{align}
	\langle x_i \rangle = \dfrac{1}{n} + \dfrac{\beta(1-u)}{Nu} \left( \sum_m \sum_j \langle \varphi_{ij}^m \pi_{ij}^m \rangle - \sum_m \sum_j \sum_r \langle x_i \varphi_{jr}^m\pi_{jr}^m \rangle \right).
\label{eq:x}
\end{align}
We say that natural selection favors strategy $i$ if its abundance exceeds the neutral expectation $1/n$. Consequently, the condition for strategy $i$ to be favored by selection becomes
\begin{align}
	\sum_m \sum_j \langle \varphi_{ij}^m \pi_{ij}^m \rangle - \sum_m \sum_j \sum_r \langle x_i \varphi_{jr}^m\pi_{jr}^m \rangle > 0.
\label{eq:selection_condition}
\end{align}

Inspired by prior work~\cite{fu2012evolution, antal2009mutation, tarnita2011multiple}, we further simplify the computation of the strategy selection condition. In the neutral stationary state, all individuals are statistically equivalent. As a result, exchanging indices does not affect averages such as $\langle x_i \rangle$, $\langle x_i x_j \rangle$, or $\langle x_ix_jx_r \rangle$. To reduce redundancy, we adopt the following conventions for double, triple, and quadruple summations over phenotypes:
\begin{equation}
\begin{aligned}
	\sum_{k_1=1}^{r_1}\sum_{k_2=1}^{r_2} x_i^{k_1k_2} x_j^{k_1k_2} &\;\;\equiv\;\; x_i^{k_1k_2} x_j^{k_1k_2}, \\
	\sum_{k_1=1}^{r_1}\sum_{k_2=1}^{r_2}\sum_{\substack{l_2=1 \\ l_2 \ne k_2}}^{r_2} x_i^{k_1k_2} x_j^{k_1l_2} &\;\;\equiv\;\; x_i^{k_1k_2} x_j^{k_1l_2}, \\
	\sum_{k_1=1}^{r_1}\sum_{\substack{l_1=1 \\ l_1 \ne k_1}}^{r_1}\sum_{k_2=1}^{r_2} x_i^{k_1k_2} x_j^{l_1k_2} &\;\;\equiv\;\; x_i^{k_1k_2} x_j^{l_1k_2}, \\
	\sum_{k_1=1}^{r_1}\sum_{\substack{l_1=1 \\ l_1 \ne k_1}}^{r_1}\sum_{k_2=1}^{r_2} \sum_{\substack{l_2=1 \\ l_2 \ne k_2}}^{r_2} x_i^{k_1k_2} x_j^{l_1l_2}
	&\;\;\equiv\;\; x_i^{k_1k_2} x_j^{l_1l_2}.
\end{aligned}
\label{eq:xx}
\end{equation}

We further introduce the following shorthand notations for the payoffs on layer $m$ ($m = 1, 2$):
\begin{equation}
\begin{aligned}
	\bar{\pi}^m = \dfrac{1}{n^2} \sum_j \sum_r \pi_{jr}^m, \qquad \bar{\pi}_{\ast\ast}^m = \dfrac{1}{n} \sum_j &\pi_{jj}^m, \qquad \bar{\pi}_{i\ast}^m = \dfrac{1}{n} \sum_j \pi_{ij}^m, \qquad \bar{\pi}_{\ast i}^m = \dfrac{1}{n} \sum_j \pi_{ji}^m.
\label{eq:pi_ast}
\end{aligned}
\end{equation}
Using these definitions and exploiting index symmetry, the selection condition for strategy $i$ can be written in the compact form
\begin{align}
	\sum_m \left[ \lambda_1^m \left( \pi_{ii}^m - \bar{\pi}_{\ast\ast}^m \right) + \lambda_2^m \left( \bar{\pi}_{i\ast}^m - \bar{\pi}_{\ast i}^m \right) + \lambda_3^m \left( \bar{\pi}_{i\ast}^m -\bar{\pi}^m \right) \right] > 0.
\label{eq:lambda_condition}
\end{align}
The coefficients $\lambda_1^m$, $\lambda_2^m$, and $\lambda_3^m$ for $m = 1, 2$ are given by
\begin{equation}
\begin{dcases}
	\lambda_1^1 \propto \langle x_i x_j^{k_1 k_2} x_j^{k_1 k_2} \rangle + \langle x_i x_j^{k_1 k_2} x_j^{k_1 l_2} \rangle - \langle x_i x_j^{k_1 k_2} x_r^{k_1 k_2} \rangle - \langle x_i x_j^{k_1 k_2} x_r^{k_1 l_2} \rangle, \\
	\lambda_1^2 \propto \langle x_i x_j^{k_1 k_2} x_j^{k_1 k_2} \rangle + \langle x_i x_j^{k_1 k_2} x_j^{l_1 k_2} \rangle - \langle x_i x_j^{k_1 k_2} x_r^{k_1 k_2} \rangle - \langle x_i x_j^{k_1 k_2} x_r^{l_1 k_2} \rangle, \\
	\lambda_2^1 \propto \langle x_i x_i^{k_1 k_2} x_j^{k_1 k_2} \rangle + \langle x_i x_i^{k_1 k_2} x_j^{k_1 l_2} \rangle - \langle x_i x_j^{k_1 k_2} x_r^{k_1 k_2} \rangle - \langle x_i x_j^{k_1 k_2} x_r^{k_1 l_2} \rangle, \\
	\lambda_2^2 \propto \langle x_i x_i^{k_1 k_2} x_j^{k_1 k_2} \rangle + \langle x_i x_i^{k_1 k_2} x_j^{l_1 k_2} \rangle - \langle x_i x_j^{k_1 k_2} x_r^{k_1 k_2} \rangle - \langle x_i x_j^{k_1 k_2} x_r^{l_1 k_2} \rangle, \\
	\lambda_3^1 \propto n \left( \langle x_i x_j^{k_1 k_2} x_r^{k_1 k_2} \rangle + \langle x_i x_j^{k_1 k_2} x_r^{k_1 l_2} \rangle \right), \\
	\lambda_3^2 \propto n \left( \langle x_i x_j^{k_1 k_2} x_r^{k_1 k_2} \rangle + \langle x_i x_j^{k_1 k_2} x_r^{l_1 k _2} \rangle \right).
\end{dcases}
\label{eq:lambda}
\end{equation}

\section{Coalescent theory}
\label{section:coalescent_theory}

We compute the coefficients $\lambda_i^m$ in \eqref{eq:lambda} using coalescent theory~\cite{wakeley2009coalescent} under the assumptions of a large population size ($N \gg 1$) and weak selection ($\beta \to 0$). For brevity, we only provide a concise outline of the calculation here; detailed derivations can be found in previous studies~\cite{antal2009evolution, antal2009mutation, tarnita2011multiple, fu2012evolution}. As an illustration, the term $\langle x_i x_j^{k_1k_2} x_j^{k_1k_2} \rangle$ represents the probability that, when three individuals are sampled uniformly at random from the population, two of them adopt the same strategy and simultaneously share identical phenotypes across both network layers. The remaining triplet correlations admit analogous interpretations.

\subsection{Probability density of coalescent time}

\begin{figure}[htpb!]
	\centering
	\includegraphics[width=0.9\columnwidth]{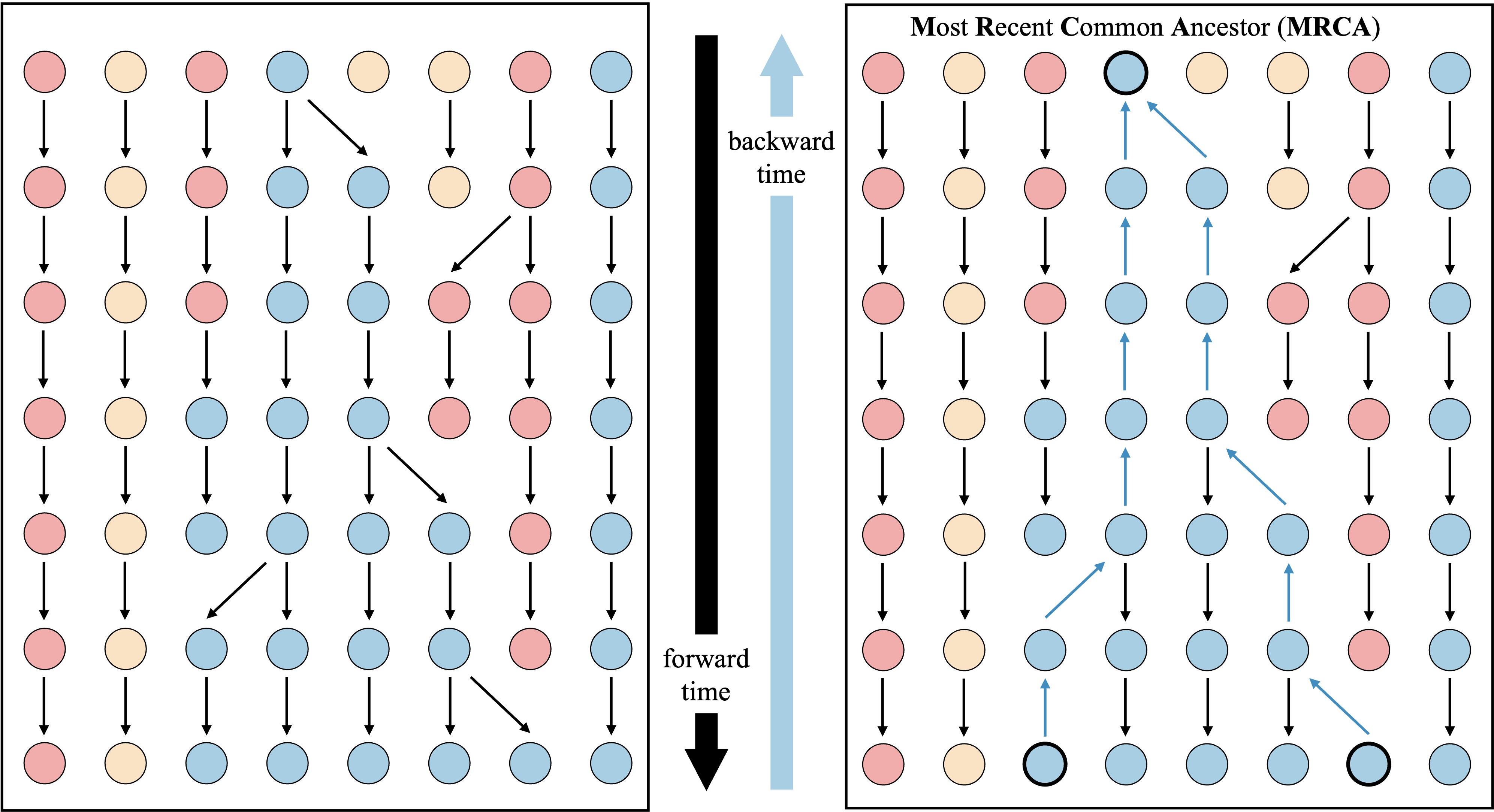}
	\captionsetup{font=small}
	\caption{\textbf{An example of MRCA tracing.} The left panel represents forward-time evolution, while the right panel illustrates the backward-time process in which two lineages coalesce to their MRCA after several generations. Individuals shown in the same color adopt the same strategy, whereas different colors indicate different strategies. This color convention is used in all figures in this section.}
\label{fig:MRCA}
\end{figure}

The central idea of coalescent theory is that, when tracing lineages backward in time, any two individuals in a finite population will eventually coalesce to a common ancestor after a finite number of generations (see Fig.~\ref{fig:MRCA}). The ancestor is known as the \textbf{most recent common ancestor (MRCA)}~\cite{wakeley2009coalescent}.

In the Moran process, the probability that two lineages coalesce in a single generation (i.e., two sampled individuals share the same parent) is
\begin{align}
	\dfrac{1}{\binom{N}{2}} \cdot \left( 1 - \dfrac{1}{N} \right) = \dfrac{2}{N^2}.
\end{align}
The probability can be understood as the product of two events: (i) a reproducing individual and its offspring are selected, and (ii) the reproducing individual is not chosen to die in this update. Consequently, the number of generations until two individuals coalesce, denoted by $T$, follows a geometric distribution with mean $N^2/2$:
\begin{align}
	P(T = t)=\left( 1 - \dfrac{2}{N^2} \right)^{t - 1}\cdot \dfrac{2}{N^2}.
\end{align}
The cumulative distribution function of $T$ is therefore 
\begin{align}
	P(T \le t) = \sum_{i = 1}^t P(T = i) = \dfrac{2}{N^2} \sum_{i = 1}^t \left( 1-\dfrac{2}{N^2} \right)^{i - 1} = 1-\left( 1 - \dfrac{2}{N^2} \right)^{t}.
\end{align}

\begin{figure}[htpb!]
	\centering
	\includegraphics[width=0.6\columnwidth]{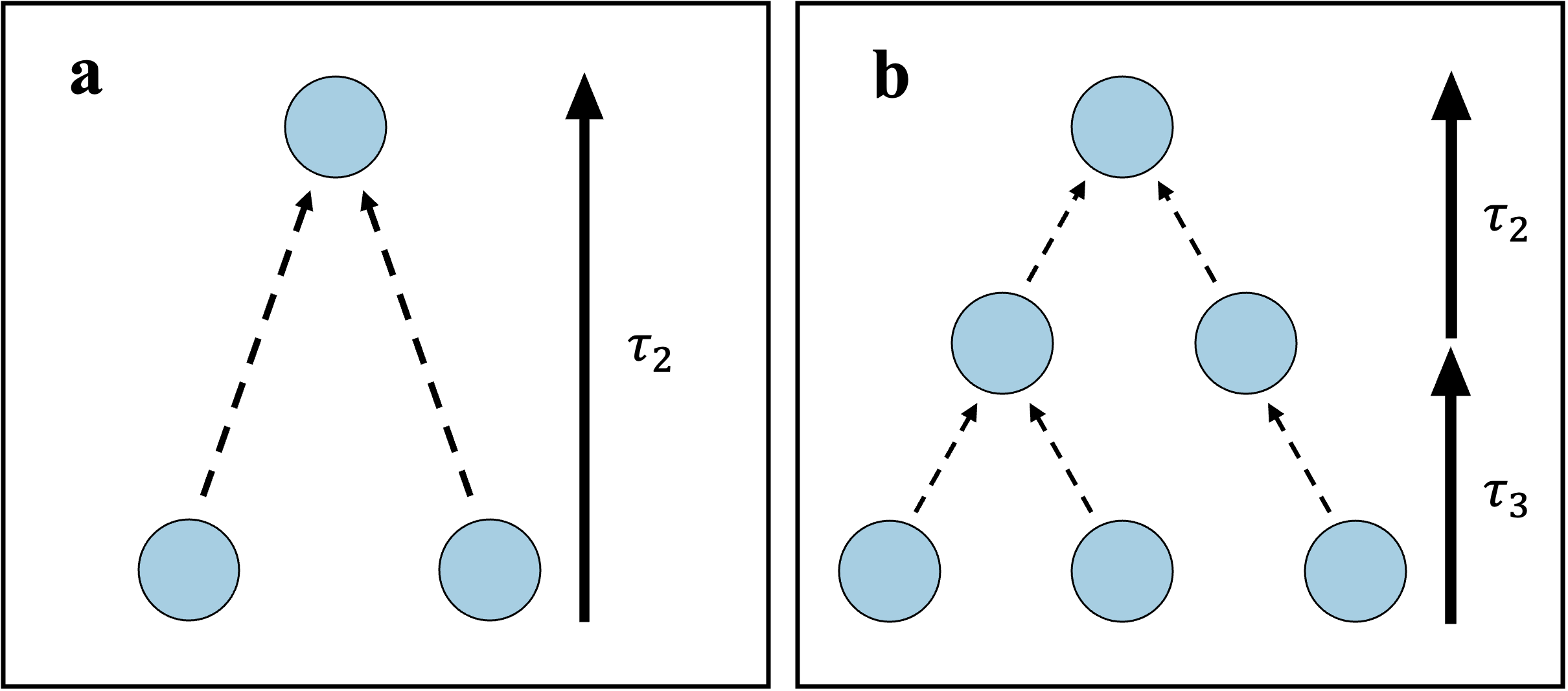}
	\captionsetup{font=small}
	\caption{\textbf{Coalescent times.} (\textbf{a}) Pairwise coalescence of two lineages. (\textbf{b}) Sequential coalescence of three lineages, consisting of an initial merger followed by a second coalescent event.}
\label{fig:2_3_coalescence}
\end{figure}

In the limit of a large population size ($N \gg 1$), this expression converges to the cumulative distribution function of an exponential distribution:
\begin{align}
	P(T \le t) = 1-\left( 1-\dfrac{2}{N^2} \right)^{t} = 1- \left( 1-\dfrac{2}{N^2} \right)^{-\frac{N^2}{2}\cdot \left( -\frac{2t}{N^2} \right)} \to 1-e^{-\frac{2t}{N^2}}.
\end{align}
We rescale the time as $\tau_2 = 2t/N^2$ (see Fig.~\ref{fig:2_3_coalescence} (\textbf{a})), which corresponds to the standard \emph{coalescent time}~\cite{wakeley2009coalescent}. The probability density function of $\tau_2$ is
\begin{align}
	T_2(\tau_2) = \dfrac{\mathrm{d}}{\mathrm{d} \tau_2}\left(1-e^{-\tau_2}\right) = e^{-\tau_2}.
\label{eq:T_2}
\end{align}
After a certain coalescent time $\tau_2$, any two individuals will have merged at their MRCA. 

For any three individuals, the coalescent process proceeds in two stages (see Fig.~\ref{fig:2_3_coalescence} (\textbf{b})). First, two of the three lineages coalesce at time $\tau_3$, with a rate three times that of a pairwise coalescence. The resulting lineage then coalesces with the remaining one at time $\tau_2$. Accordingly, the joint probability density function of the coalescent times $\tau_2$ and $\tau_3$ is given by
\begin{align}
	T_3(\tau_2, \tau_3) = e^{-\tau_2} \cdot \dfrac{\mathrm{d}}{\mathrm{d} \tau_3}\left( 1-e^{-3\tau_3} \right) = 3e^{-3\tau_3}e^{-\tau_2}.
\label{eq:T_3}
\end{align}

\subsection{Probability of identical strategy}

We now focus on the probability that two individuals share the same strategy given a coalescent time $\tau_2$. In the Moran process, the strategy along each lineage mutates at a rate
\begin{align}
	\dfrac{N^2}{2} \times \dfrac{1}{N} \times u = \dfrac{Nu}{2} = \dfrac{\mu}{2},
\label{eq:mutation_rate}	
\end{align}
where $\mu = Nu$ is the rescaled mutation rate. This rate can be interpreted as the product of two events after time rescaling: (i) an individual is selected with probability $1/N$, and (ii) the selected individual mutates with probability $u$. 

To account for mutations along multiple lineages, we adopt the classical ``light bulb problem'' as an analogy. The lifetime of a single light bulb follows an exponential distribution with density $f(t) = \lambda e^{-\lambda t}$, where $\lambda$ is the failure rate. The probability that it survives beyond time $t$ is
\begin{align}
	P(T_{\text{bulb}} > t) = 1 - P(T_{\text{bulb}} \le t) = 1- \int_0^t \lambda e^{-\lambda x} \mathrm{d}x = e^{-\lambda t}.
\end{align}
For two identical and independent bulbs, the probability that both survive beyond time $t$ is just the product of their individual survival probabilities:
\begin{align}
	P(T_{\text{two bulbs}} > t) = P(T_{\text{bulb}} > t) \times P(T_{\text{bulb}} > t) =  e^{-\lambda t} \times e^{-\lambda t} = e^{-2\lambda t}.
\end{align}

\begin{figure}[htpb!]
	\centering
	\includegraphics[width=0.6\columnwidth]{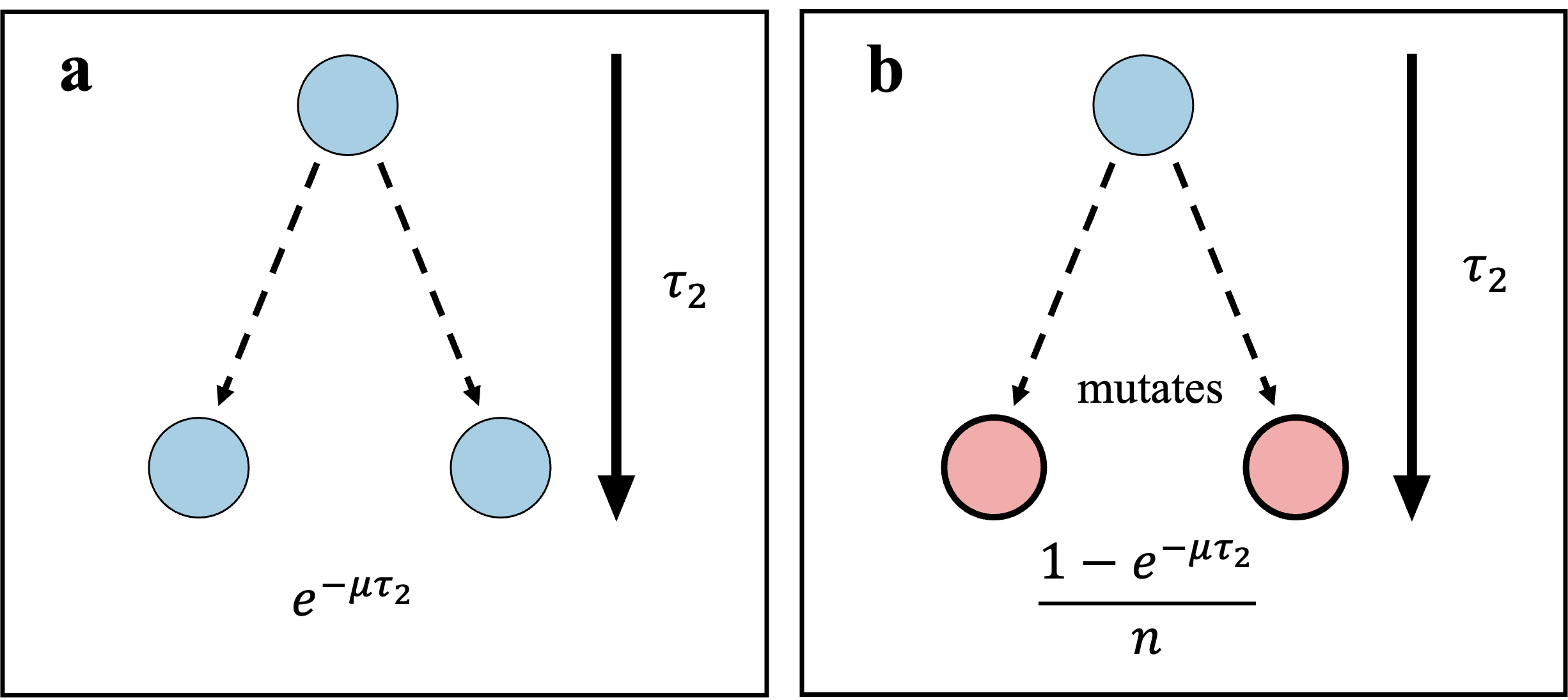}
	\captionsetup{font=small}
	\caption{\textbf{Scenarios for two individuals sharing the same strategy after coalescent time $\tau_2$.} (\textbf{a}) Neither lineage undergoes mutation during $\tau_2$. (\textbf{b}) At least one lineage mutates during $\tau_2$.}
\label{fig:s_2}
\end{figure}

We treat each lineage as a light bulb and interpret mutation events as bulb failures. In this context, the failure rate $\lambda$ corresponds to the mutation rate of a single lineage, $\mu/2$, as defined in \eqref{eq:mutation_rate}. Therefore, for two lineages, the probability that neither mutates over the coalescent time $\tau_2$ is $e^{-\mu\tau_2}$, while the probability that at least one mutation occurs is $1 - e^{-\mu\tau_2}$. These two scenarios are depicted in Fig.~\ref{fig:s_2} (\textbf{a}) and Fig.~\ref{fig:s_2} (\textbf{b}), respectively. In the latter case, regardless of whether one or both lineages mutate, the probability that these two individuals share the same strategy is $1/n$. To sum up, the probability that two randomly chosen individuals adopt the same strategy after coalescent time $\tau_2$ is
\begin{align}
	s_2(\tau_2) = e^{-\mu\tau_2} + \dfrac{1-e^{-\mu\tau_2}}{n}.
\label{eq:s_2}
\end{align}

\begin{figure}[htpb!]
	\centering
	\includegraphics[width=0.9\columnwidth]{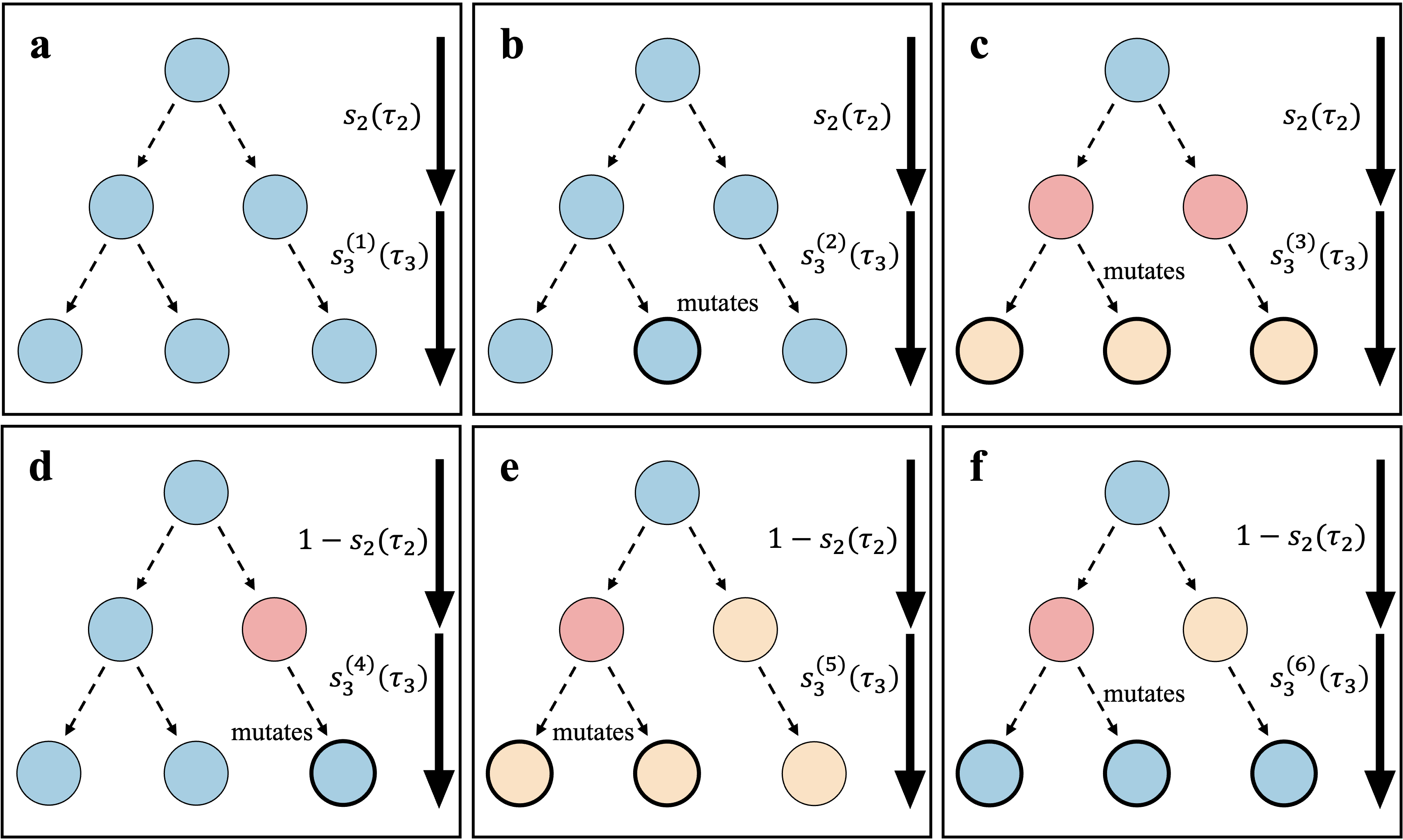}
	\captionsetup{font=small}
	\caption{\textbf{Scenarios prior to the first coalescent event at time $\tau_3$.} If the two lineages that coalesce first share the same strategy, three subcases may occur: (\textbf{a}) none of the three lineages mutates; (\textbf{b}) exactly one lineage mutates; (\textbf{c}) at least two lineages mutate. If the two lineages that coalesce first adopt different strategies, two subcases may occur: (\textbf{d}) exactly one lineage mutates; (\textbf{e}) \& (\textbf{f}) at least two lineages mutate.}
\label{fig:s_3}
\end{figure}

Similarly, we derive the probability that three randomly chosen individuals share the same strategy at time $\tau_2 + \tau_3$. Conditional on the first coalescent event occurring at time $\tau_3$, there are two possible scenarios: with probability $s_2(\tau_2)$, the two lineages that coalesce first already share the same strategy, or with probability $1-s_2(\tau_2)$, they adopt different strategies. In the former case, three subcases arise: none of the three individuals mutates, exactly one individual mutates, or at least two individuals mutate. These situations correspond to panels (\textbf{a}) -- (\textbf{c}) of Fig.~\ref{fig:s_3}. The conditional probabilities that all three individuals adopt the same strategy after $\tau_2 + \tau_3$ are
\begin{equation}
\begin{dcases}
s_3^{(1)} (\tau_3) = e^{-3\mu\tau_3/2}, \\
s_3^{(2)} (\tau_3) = 3\left(1-e^{-\mu\tau_3/2} \right)e^{-\mu\tau_3}/n, \\
s_3^{(3)} (\tau_3) = \left[1-e^{-3\mu\tau_3/2}- 3\left(1-e^{-\mu\tau_3/2} \right)e^{-\mu\tau_3} \right]/n^2.
\end{dcases}
\end{equation}
In the latter case, two subcases arise: exactly one individual mutates, or at least two individuals mutate. These are illustrated in panel (\textbf{d}) and panels (\textbf{e}) and (\textbf{f}) of Fig.~\ref{fig:s_3}, respectively. The corresponding conditional probabilities that all three individuals adopt the same strategy after $\tau_2 + \tau_3$ are
\begin{equation}
\begin{dcases}
s_3^{(4)} (\tau_3) = \left(1-e^{-\mu\tau_3/2} \right)e^{-\mu\tau_3}/n, \\
s_3^{(5)} (\tau_3) + s_3^{(6)} (\tau_3) = \left[1-e^{-3\mu\tau_3/2}- 3\left(1-e^{-\mu\tau_3/2} \right)e^{-\mu\tau_3} \right]/n^2.
\end{dcases}
\end{equation}

Putting these cases together, the probability that three randomly chosen individuals share the same strategy at time $\tau_2 + \tau_3$ can be written as
\begin{equation}
\begin{aligned}
	s_3(\tau_2, \tau_3) &= s_2(\tau_2) \left[s_3^{(1)}(\tau_3) + s_3^{(2)}(\tau_3) + s_3^{(3)}(\tau_3)\right] + \left[1 - s_2(\tau_2)\right]\left[s_3^{(4)}(\tau_3) + s_3^{(5)}(\tau_3) + s_3^{(6)}(\tau_3)\right] \\
	&= \dfrac{s_2(\tau_2)}{n^2}\left[1 + 3(n - 1)e^{-\mu\tau_3} + (n - 1)(n - 2)e^{-3/2\mu\tau_3}\right] \\
	&+ \dfrac{1 - s_2(\tau_2)}{n^2}\left[1 + (n - 3)e^{-\mu\tau_3} - (n - 2)e^{-3/2\mu\tau_3} \right].
\end{aligned}
\label{eq:s_3}
\end{equation}

\subsection{Calculation of triplet correlation}

Since the two layers, namely, the two phenotype sets, may exhibit different types of inter-layer relationships (see the next section), we first present a general computational framework. The explicit expressions will be derived later for specific types of interactions between the two layers. Let $Q_{11}(\tau)$ denote the probability that two randomly chosen individuals share the same phenotypes on both layers after a fixed time $\tau$. Similarly, let $Q_{10}(\tau)$ denote the probability that they share the same phenotype on the first layer but differ on the second layer, and let $Q_{01}(\tau)$ denote the probability that they differ on the first layer but share the same phenotype on the second layer. Based on these definitions, some of the pair correlations can be written as
\begin{equation}
\begin{aligned}
	\langle x_i^{k_1k_2} x_{\ast}^{k_1 k_2} \rangle &= \dfrac{1}{n} \int_{0}^{\infty} T_2(\tau_2)Q_{11}(\tau_2) \mathrm{d}\tau_2, \qquad
	\langle x_i^{k_1k_2} x_i^{k_1 k_2} \rangle = \dfrac{1}{n} \int_{0}^{\infty} T_2(\tau_2)s_2(\tau_2)Q_{11}(\tau_2) \mathrm{d}\tau_2, \\
	\langle x_i^{k_1k_2} x_{\ast}^{k_1 l_2} \rangle &= \dfrac{1}{n} \int_{0}^{\infty} T_2(\tau_2)Q_{10}(\tau_2) \mathrm{d}\tau_2, \qquad
	\langle x_i^{k_1k_2} x_i^{k_1 l_2} \rangle = \dfrac{1}{n} \int_{0}^{\infty} T_2(\tau_2)s_2(\tau_2)Q_{10}(\tau_2) \mathrm{d}\tau_2, \\
	\langle x_i^{k_1k_2} x_{\ast}^{l_1 k_2} \rangle &= \dfrac{1}{n} \int_{0}^{\infty} T_2(\tau_2)Q_{01}(\tau_2) \mathrm{d}\tau_2, \qquad
	\langle x_i^{k_1k_2} x_i^{l_1 k_2} \rangle = \dfrac{1}{n} \int_{0}^{\infty} T_2(\tau_2)s_2(\tau_2)Q_{01}(\tau_2) \mathrm{d}\tau_2.
\end{aligned}
\end{equation}

In these expressions, the factor $1/n$ reflects the probability of randomly sampling an individual adopting strategy $i$. In line with the convention introduced in \eqref{eq:xx} and the notation ``$\ast$'' used in \eqref{eq:pi_ast}, we refer to $\sum_{j=1}^n\sum_{k_1=1}^{r_1}\sum_{k_2=1}^{r_2} x_i^{k_1k_2} x_j^{k_1 k_2}$ compactly as $x_i^{k_1k_2} x_{\ast}^{k_1 k_2}$. The same shorthand is applied throughout the subsequent derivations. The remaining pair correlations can be obtained by symmetry arguments~\cite{fu2012evolution,antal2009mutation}:
\begin{equation}
\begin{aligned}
	\langle x_i^{k_1k_2} x_j^{k_1 k_2} \rangle &= \dfrac{1}{n-1} \left( \langle x_i^{k_1 k_2} x_{\ast}^{k_1 k_2} \rangle - \langle x_i^{k_1 k_2} x_i^{k_1 k_2} \rangle \right), \\
	\langle x_i^{k_1k_2} x_j^{k_1 l_2} \rangle &= \dfrac{1}{n-1} \left( \langle x_i^{k_1 k_2} x_{\ast}^{k_1 l_2} \rangle - \langle x_i^{k_1 k_2} x_i^{k_1 l_2} \rangle \right), \\
	\langle x_i^{k_1k_2} x_j^{l_1 k_2} \rangle &= \dfrac{1}{n-1} \left( \langle x_i^{k_1 k_2} x_{\ast}^{l_1 k_2} \rangle - \langle x_i^{k_1 k_2} x_i^{l_1 k_2} \rangle \right). 
\end{aligned}
\end{equation}

\begin{figure}[htpb!]
	\centering
	\includegraphics[width=0.9\columnwidth]{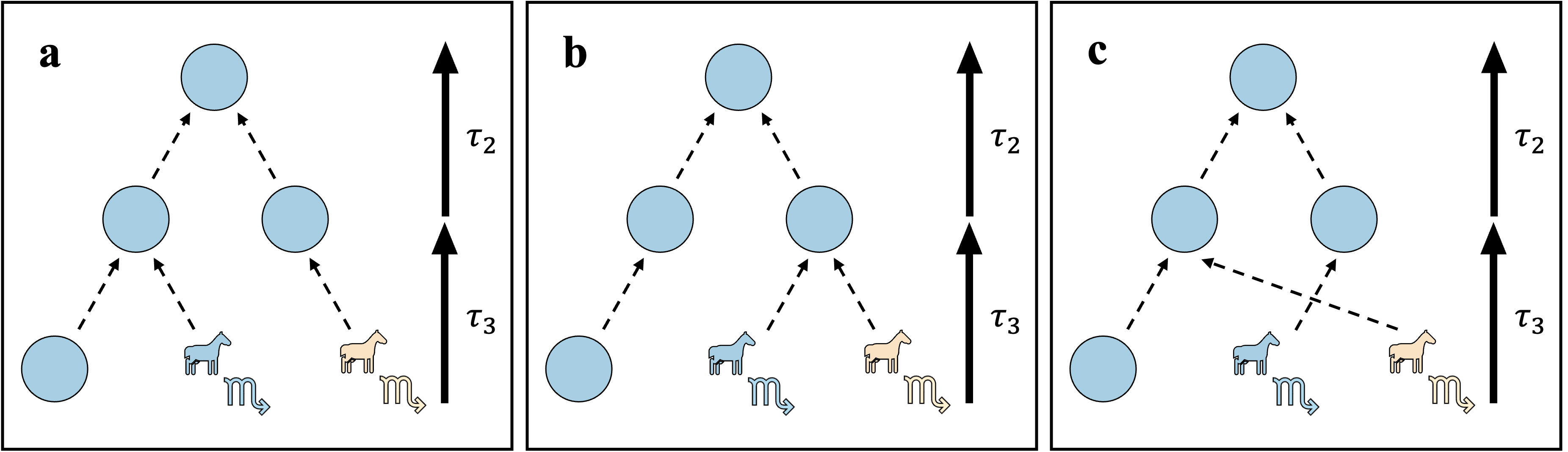}
	\captionsetup{font=small}
	\caption{\textbf{Coalescent scenarios contributing to the triplet correlation $\langle x_i x_i^{k_1k_2} x_{\ast}^{k_1k_2} \rangle$.}  (\textbf{a}) An individual adopting strategy $i$ with arbitrary phenotypes coalesces first with an individual of the same strategy $i$ and with phenotypes $(k_1, k_2)$. (\textbf{b}) An individual adopting strategy $i$ with phenotypes $(k_1, k_2)$ coalesces first with an individual of arbitrary strategy and with the same phenotypes $(k_1, k_2)$. (\textbf{c}) An individual adopting strategy $i$ with arbitrary phenotypes coalesces first with an individual of arbitrary strategy and with phenotypes $(k_1, k_2)$. The Chinese zodiac and Western astrological signs serve solely as visual labels to distinguish phenotypes on the two layers. For example, $(k_1, k_2)$ may be represented by ``Horse'' and ``Sagittarius,'' respectively. These symbols are used for graphical clarity and do not imply that zodiac or astrological signs are heritable.}
\label{fig:triplet_iiast}
\end{figure}

We are now ready to compute the triplet correlations. As illustrative examples, we focus on $\langle x_i x_i^{k_1k_2} x_{\ast}^{k_1k_2} \rangle$ and $\langle x_i x_i^{k_1k_2} x_i^{k_1k_2} \rangle$. For the correlation $\langle x_i x_i^{k_1k_2} x_{\ast}^{k_1k_2} \rangle$, three distinct coalescent scenarios may occur: (a) an individual adopting strategy $i$ with arbitrary phenotypes coalesces first with an individual of the same strategy $i$ and with phenotype $k_1$ on the first layer and $k_2$ on the second layer, (b) an individual adopting strategy $i$ with phenotypes $(k_1, k_2)$ coalesces first with an individual of arbitrary strategy and with the same phenotypes $(k_1, k_2)$, and (c) an individual adopting strategy $i$ with arbitrary phenotypes coalesces first with an individual of arbitrary strategy and with phenotypes $(k_1, k_2)$. These scenarios are depicted in panels (\textbf{a}) -- (\textbf{c}) of Fig.~\ref{fig:triplet_iiast}. The corresponding probabilities are given by $s_2(\tau_3)Q_{11}(\tau_2+\tau_3)/3$, $s_2(\tau_2+\tau_3)Q_{11}(\tau_3)/3$, and $s_2(\tau_2+\tau_3)Q_{11}(\tau_2+\tau_3)/3$, respectively. Collecting these contributions yields
\begin{equation}
\begin{aligned}
	\langle x_i x_i^{k_1 k_2} x_{\ast}^{k_1 k_2} \rangle &= \dfrac{1}{3n} \int_{0}^{\infty}\int_{0}^{\infty} T_3(\tau_2,\tau_3)(\ast)_{11}(\tau_2,\tau_3) \mathrm{d}\tau_2 \mathrm{d}\tau_3, \\
	(\ast)_{11}(\tau_2,\tau_3) &\equiv s_2(\tau_3)Q_{11}(\tau_2 + \tau_3) + s_2(\tau_2+\tau_3)Q_{11}(\tau_3) + s_2(\tau_2 + \tau_3)Q_{11}(\tau_2 + \tau_3).
\label{eq:triplet_iiast}
\end{aligned}
\end{equation}

\begin{figure}[htpb!]
	\centering
	\includegraphics[width=0.6\columnwidth]{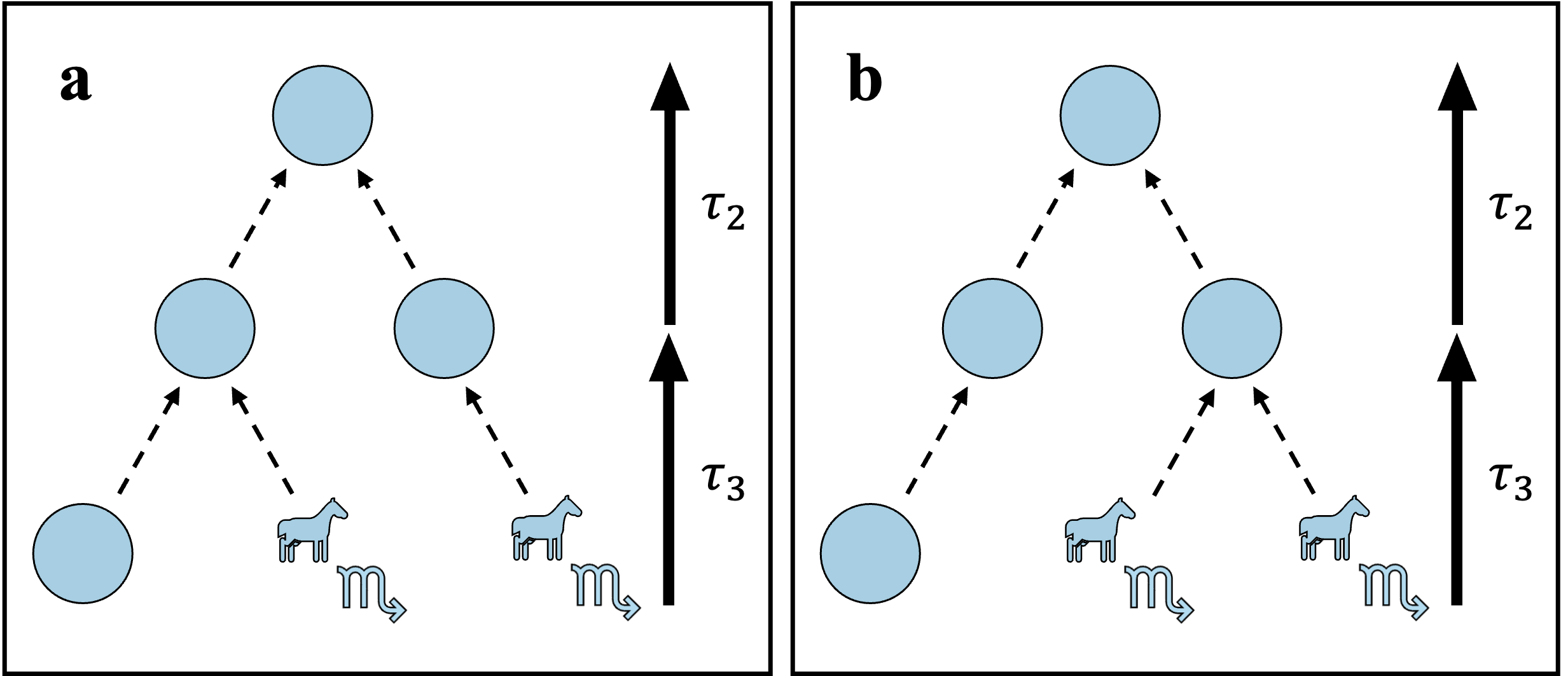}
	\captionsetup{font=small}
	\caption{\textbf{Coalescent scenarios contributing to the triplet correlation $\langle x_i x_i^{k_1k_2} x_i^{k_1k_2} \rangle$.}  (\textbf{a}) An individual adopting strategy $i$ with arbitrary phenotypes coalesces first with an individual of the same strategy $i$ and with phenotypes $(k_1, k_2)$. (\textbf{b}) Two individuals, both adopting strategy $i$ with phenotypes $(k_1, k_2)$, coalesce first.}
\label{fig:triplet_iii}
\end{figure}

For $\langle x_i x_i^{k_1k_2} x_i^{k_1k_2} \rangle$, two distinct scenarios may occur: (a) an individual adopting strategy $i$ with arbitrary phenotypes coalesces first with an individual of the same strategy $i$ and with phenotypes $(k_1, k_2)$, and (b) two individuals, both adopting strategy $i$ with phenotypes $(k_1, k_2)$, coalesce first. These scenarios are depicted in panels (\textbf{a}) and (\textbf{b}) of Fig.~\ref{fig:triplet_iii}, with the corresponding probabilities given by $2s_3(\tau_2, \tau_3)Q_{11}(\tau_2+\tau_3)/3$ and $s_3(\tau_2, \tau_3)Q_{11}(\tau_3)/3$, respectively. Accordingly, we obtain
\begin{align}
	\langle x_i x_i^{k_1 k_2} x_i^{k_1 k_2} \rangle = \dfrac{1}{3n} \int_{0}^{\infty}\int_{0}^{\infty} T_3(\tau_2,\tau_3)s_3(\tau_2,\tau_3) \left[Q_{11}(\tau_3) + 2Q_{11}(\tau_2 + \tau_3)\right] \mathrm{d}\tau_2 \mathrm{d}\tau_3.
\label{eq:triplet_iii}
\end{align}

Similarly, we obtain the following expressions:
\begin{equation}
\begin{aligned}
	\langle x_i x_i^{k_1 k_2} x_{\ast}^{k_1 l_2} \rangle &= \dfrac{1}{3n} \int_{0}^{\infty}\int_{0}^{\infty} T_3(\tau_2,\tau_3)(\ast)_{10}(\tau_2,\tau_3) \mathrm{d}\tau_2 \mathrm{d}\tau_3, \\
	\langle x_i x_i^{k_1 k_2} x_{\ast}^{l_1 k_2} \rangle &= \dfrac{1}{3n} \int_{0}^{\infty}\int_{0}^{\infty} T_3(\tau_2,\tau_3)(\ast)_{01}(\tau_2,\tau_3) \mathrm{d}\tau_2 \mathrm{d}\tau_3, \\
	\langle x_i x_i^{k_1 k_2} x_i^{k_1 l_2} \rangle &= \dfrac{1}{3n} \int_{0}^{\infty}\int_{0}^{\infty} T_3(\tau_2,\tau_3)s_3(\tau_2,\tau_3) \left[ Q_{10}(\tau_3)+2Q_{10}(\tau_2+\tau_3)\right] \mathrm{d}\tau_2 \mathrm{d}\tau_3, \\	
	 \langle x_i x_i^{k_1 k_2} x_i^{l_1 k_2} \rangle &= \dfrac{1}{3n} \int_{0}^{\infty}\int_{0}^{\infty} T_3(\tau_2,\tau_3)s_3(\tau_2,\tau_3) \left[ Q_{01}(\tau_3)+2Q_{01}(\tau_2+\tau_3)\right] \mathrm{d}\tau_2 \mathrm{d}\tau_3, \\
	(\ast)_{10}(\tau_2,\tau_3) &= s_2(\tau_3)Q_{10}(\tau_2+\tau_3)+s_2(\tau_2+\tau_3)Q_{10}(\tau_3) +s_2(\tau_2+\tau_3)Q_{10}(\tau_2+\tau_3), \\
	(\ast)_{01}(\tau_2,\tau_3) &= s_2(\tau_3)Q_{01}(\tau_2+\tau_3)+s_2(\tau_2+\tau_3)Q_{01}(\tau_3) +s_2(\tau_2+\tau_3)Q_{01}(\tau_2+\tau_3).
\end{aligned}
\label{eq:correlation}
\end{equation}
The remaining triplet correlations can be derived by symmetry arguments over strategies~\cite{fu2012evolution, antal2009mutation}. When the two focal individuals share the same phenotypes on both layers of the network, we obtain 
\begin{equation}
\begin{aligned}
\langle x_i x_i^{k_1 k_2} x_j^{k_1 k_2} \rangle &= \dfrac{1}{n-1} \left( \langle x_i x_i^{k_1 k_2} x_{\ast}^{k_1 k_2} \rangle - \langle x_i x_i^{k_1 k_2} x_i^{k_1 k_2} \rangle \right), \\
\langle x_i x_j^{k_1 k_2} x_j^{k_1 k_2} \rangle &= \dfrac{1}{n-1} \left( \langle x_i^{k_1 k_2} x_i^{k_1 k_2} \rangle - \langle x_i x_i^{k_1 k_2} x_i^{k_1 k_2} \rangle \right), \\
\langle x_i x_j^{k_1 k_2} x_r^{k_1 k_2} \rangle &= \dfrac{1}{n-2} \left( \langle x_i^{k_1 k_2} x_j^{k_1 k_2} \rangle - 2 \langle x_i x_i^{k_1 k_2} x_j^{k_1 k_2} \rangle \right).
\end{aligned}
\label{eq:correlation_more}
\end{equation}
Moreover, analogous expressions hold when the two focal individuals share the same phenotype on the first layer but differ on the second, or vice versa. In these cases, the subscripts $k_1k_2$ are replaced by $k_1l_2$ or $l_1k_2$, respectively.

\subsection{Calculation of strategy selection condition}

The strategy selection condition derived for discrete strategies can be naturally extended to continuous strategies~\cite{fu2012evolution, antal2009mutation}. To this end, we consider $n$ discrete strategies $(p_i, q_i)$ arranged on a regular grid in $[0, 1]^2$. As the grid is refined and $n \to \infty$, we obtain the continuous strategy space $(p, q) \in [0, 1]^2$. Based on \eqref{eq:payoff}, for an individual adopting strategy $(p, q)$ against another individual adopting strategy $(p', q')$, we define the payoffs on the two layers as
\begin{equation}
\begin{dcases}
	A(p, p') = R_1pp'+S_1p(1-p')+T_1(1-p)p'+P_1(1-p)(1-p'), \\
	B(q, q') = R_2qq'+S_2q(1-q')+T_2(1-q)q'+P_2(1-q)(1-q').
\end{dcases}
\end{equation}

Furthermore, the summations in \eqref{eq:pi_ast} can be replaced by integrals in the continuous limit. In analogy with the discrete strategy selection condition in \eqref{eq:lambda_condition}, we obtain the relative probability density functions of $p$ and $q$ as
\begin{equation}
\begin{aligned}
	D^1(p) &= \lambda_1^1 \left( A(p, p) - \int_0^1 A(p', p') \mathrm{d}p' \right) + \lambda_2^1 \left( \int_0^1A(p, p') \mathrm{d}p' - \int_0^1 A(p', p) \mathrm{d}p' \right) \\
	&+ \lambda_3^1 \left( \int_0^1A(p, p') \mathrm{d}p' - \int_0^1 \int_0^1 A(p, p') \mathrm{d}p\mathrm{d}p' \right), \\
	D^2(q) &= \lambda_1^2 \left( B(q, q) - \int_0^1 B(q', q') \mathrm{d}q' \right) + \lambda_2^2 \left( \int_0^1B(q, q') \mathrm{d}q' - \int_0^1 B(q', q) \mathrm{d}q' \right) \\
	&+ \lambda_3^2 \left( \int_0^1B(q, q') \mathrm{d}q' - \int_0^1 \int_0^1 B(q, q') \mathrm{d}q\mathrm{d}q' \right).
\end{aligned}
\label{eq:Dpq}
\end{equation}
The coefficients $\lambda_1^m$,  $\lambda_2^m$, and $\lambda_3^m$ ($m = 1, 2$) are identical to those derived for discrete strategies, up to a rescaling by the number of strategies $n$. In the continuous limit, the condition for strategy $(p, q)$ to be selected is therefore
\begin{equation}
D^1(p) + D^2(q) > 0.
\end{equation}

The population is expected to evolve toward cooperation on a given layer (or on both layers) of the network when the population averages satisfy $\langle \, p \, \rangle > 1/2$ and/or $\langle \, q \, \rangle > 1/2$. This condition is equivalent to 
\begin{equation}
\int_0^1 \int_0^1 p[D^1(p) + D^2(q)] \mathrm{d}p\mathrm{d}q > 0,  \qquad \int_0^1 \int_0^1 q[D^1(p) + D^2(q)] \mathrm{d}p\mathrm{d}q > 0.
\end{equation}
Notably, according to \eqref{eq:Dpq}, we have
\begin{equation}
\int_0^1 D^1(p)dp = 0, \qquad \int_0^1 D^2(q)dq = 0.
\end{equation}
As a result, natural selection favors cooperation on the first layer and/or the second layer if and only if
\begin{align}
	\int_0^1 pD^1(p) \mathrm{d}p > 0, \qquad \int_0^1 qD^2(q) \mathrm{d}q > 0.
\end{align}
The above decoupling relies on the additive contribution of the two layers to the selection gradient. These inequalities can be recast in terms of the structure coefficients $\sigma_m$ as the $\sigma$-rule, introduced by Tarnita et al.~\cite{tarnita2009strategy} and subsequently extended by McAvoy and Wakeley~\cite{mcavoy2022evaluating}:
\begin{align}
	\sigma_m R_m + S_m > T_m + \sigma_m P_m, \qquad m=1,2.
\label{eq:sigma_rule}
\end{align}

Notably, the layerwise conditions are obtained to first order in the selection strength $\beta$. Writing $\pi_i = \pi_i^1 + \pi_i^2$, where $\pi_i^m$ denotes the payoff contribution from layer $m$ for strategy $i$, we have
\begin{equation}
f_i = \exp[\beta(\pi_i^1 + \pi_i^2)] = 1 + \beta(\pi_i^1 + \pi_i^2) + \frac{\beta^2}{2}(\pi_i^1 + \pi_i^2)^2 + \mathcal{O}(\beta^3).
\end{equation}
The first-order term is additive across layers, whereas the second-order term contains the mixed contribution $\beta^2\pi_i^1\pi_i^2$. Thus, higher-order effects can couple the contributions of the two layers, and the layerwise separation need not persist beyond weak selection (see Section~\ref{section:strong_selection} for an examination of stronger selection). 

\section{Types of dependencies between phenotypes}
\label{section:types_of_dependencies}

In the absence of mutations, the two phenotypes of an individual are inherited from the parent. Thus, the relationship between phenotypes across the two network layers only affects the initialization of the population and the mutation process in simulations. The possible scenarios are summarized in Table~\ref{tab:relationship}. For precise definitions of \emph{concurrent} mutation and \emph{non-concurrent} mutation, we refer the reader to the previous section describing the updating rule. Briefly, mutations may occur separately on each layer or simultaneously on both layers.

\begin{table}[htbp]
\centering
\caption{Classification of possible dependencies between phenotypes across layers.}
\vspace{5pt}
\begin{tabular}{|c|c|c|}
\hline
\diagbox{Dependency type}{Mutation temporality} & non-concurrent & concurrent \\
\hline
independence & \textbf{\ref{subsection:independence}} &  \\
\cline{1-2}
unidirectional influence (non-reciprocal epistasis) & \textbf{\ref{subsection:unidirection}} & \textbf{\ref{subsection:concurrent}}  \\
\cline{1-2}
bidirectional influence (reciprocal epistasis) & \textbf{\ref{subsection:bidirection}} &  \\ 
\hline
\end{tabular}
\label{tab:relationship}
\end{table}

Before proceeding, we emphasize that all examples discussed in this section are used solely as descriptive analogies motivated by findings in biology and social science. They are intended to highlight different types of dependencies between heritable traits and do not imply any normative, value-based, or discriminatory interpretations. Our modeling framework is agnostic to the specific biological or social meaning of the phenotypes and focuses exclusively on their dependency structures.

\begin{figure}[htpb!]
	\centering
	\includegraphics[width=0.75\columnwidth]{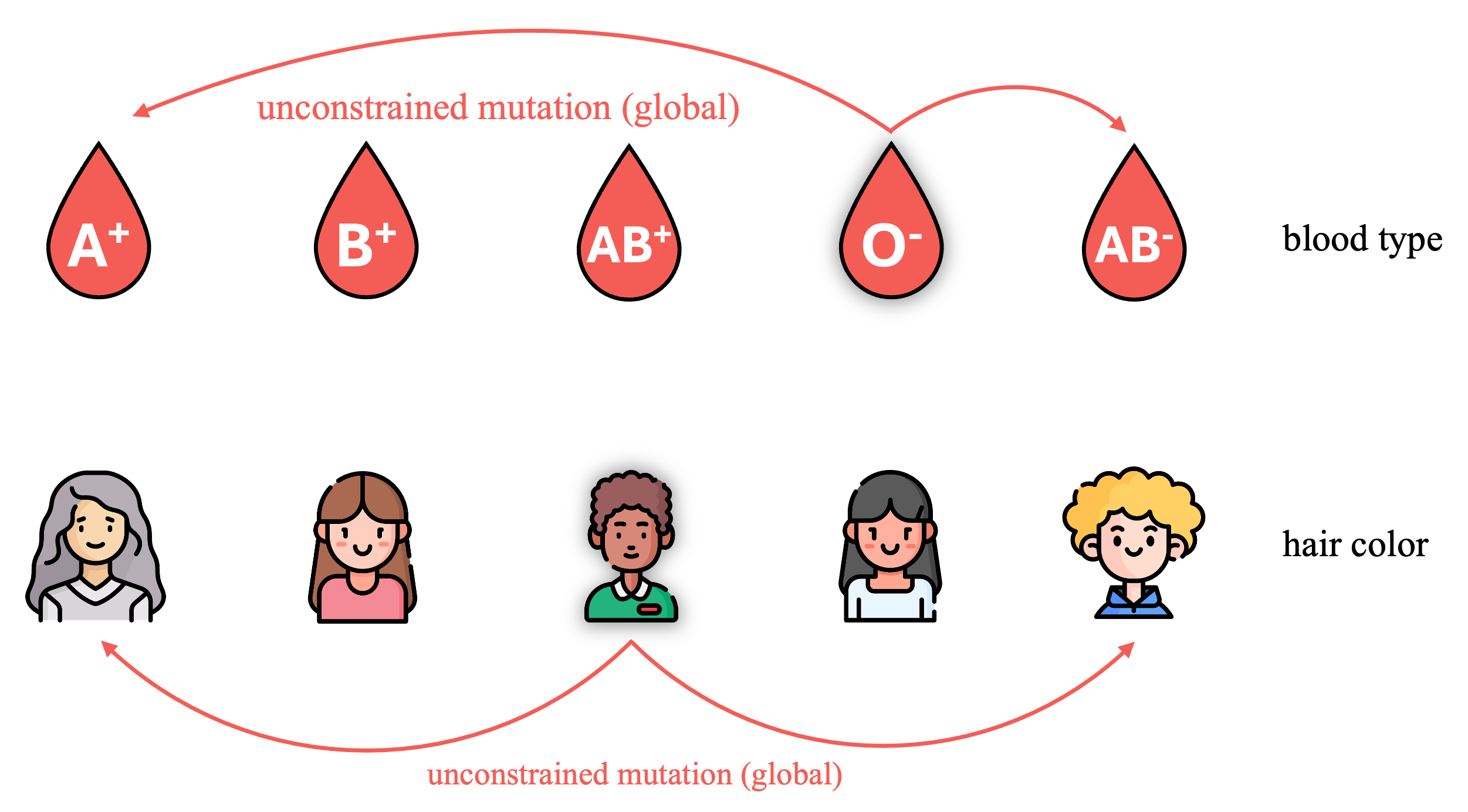}
	\captionsetup{font=small}
	\caption{\textbf{Independence between phenotypes.} An illustrative example of two heritable but unrelated traits assigned to different layers of a multiplex network. Mutations affecting one phenotype do not alter the other.}
\label{fig:blood_hair}
\end{figure}

The simplest type of dependency arises when the phenotypes on the two network layers are statistically unrelated (\textbf{independence}). In this case, we consider two heritable but independent traits: blood type and hair color, as illustrated in Fig.~\ref{fig:blood_hair}. On the one hand, blood type is determined by genetic loci such as the ABO system located on chromosome 9~\cite{amundadottir2009genome} and the Rh system located on chromosome 1~\cite{avent2000rh}. On the other hand, hair color is primarily governed by genes including \emph{MC1R}, \emph{OCA2}, and \emph{TYRP1}~\cite{branicki2011model, sulem2007genetic}, which are located on different chromosomes. As a result, mutations affecting one phenotype do not influence the other, consistent with the assumption of independence.

Nevertheless, biological traits need not vary independently, as genetic and developmental interactions can couple their variation~\cite{otto1997deleterious, armbruster2014integrated}. As a particular example, genetic modifiers of mutation show how an inherited background can influence subsequent variation. In mice, natural variation at a locus containing the DNA-repair gene \emph{Mutyh} is associated with differences in germline mutation rates and spectra~\cite{sasani2022natural}. Similarly, inherited biallelic \emph{MUTYH} variants in humans are associated with elevated somatic mutation burdens and altered mutational signatures in normal tissues~\cite{robinson2022inherited}. These findings motivate the consideration of background-dependent mutational processes. 

\begin{figure}[htpb!]
    \centering
    \includegraphics[width=0.75\columnwidth]{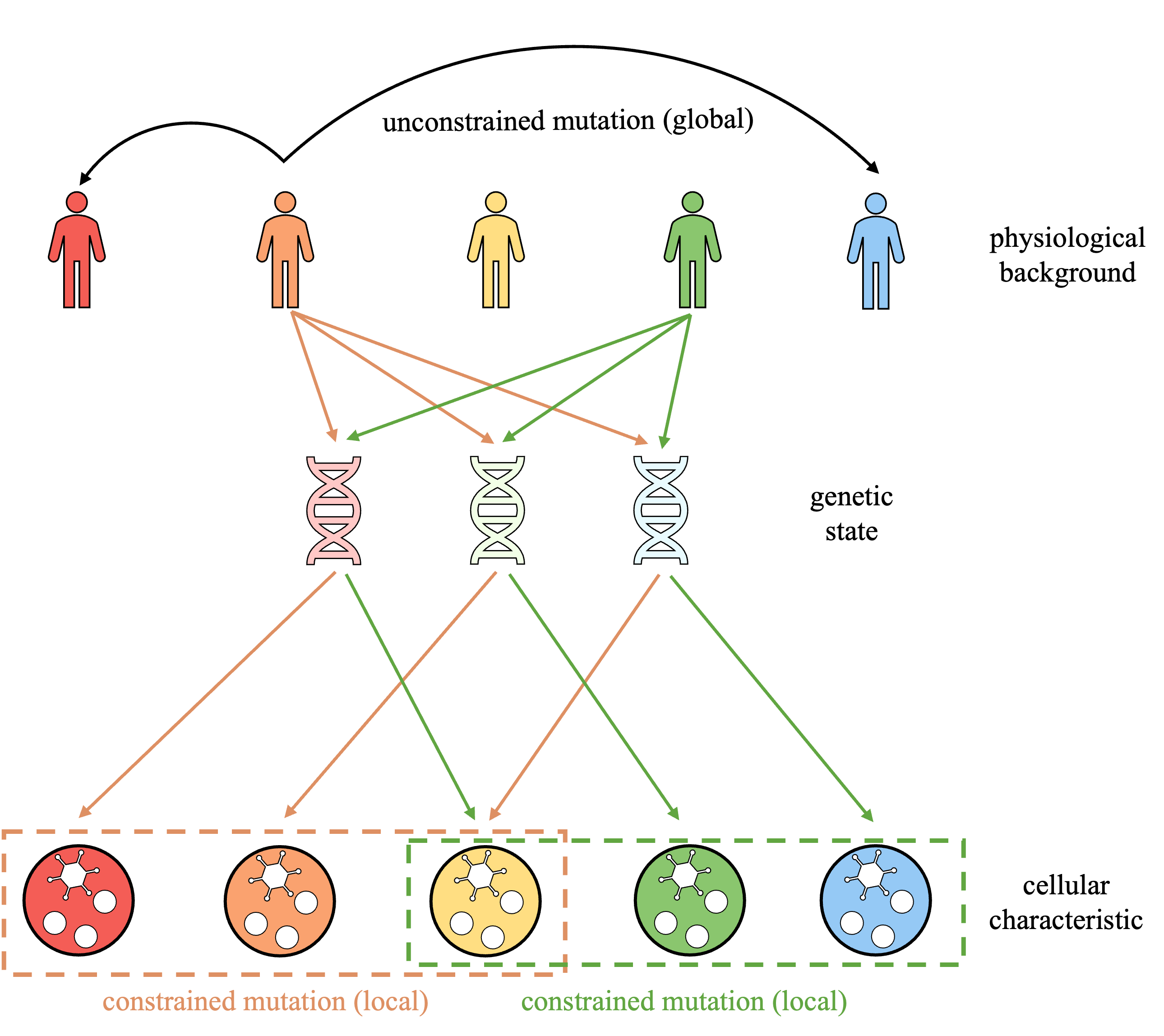}
    \captionsetup{font=small}
    \caption{\textbf{Unidirectional influence between phenotypes.} The top row represents active-layer phenotypes, interpreted as physiological backgrounds, and the bottom row represents passive-layer cellular phenotypes. Both phenotype spaces are ordered from left to right. The middle row represents three genetic states (each of them produces a cellular phenotype determined by the background), which do not constitute a third layer. Orange and green arrows show the mappings under the corresponding backgrounds, and dashed boxes delimit the accessible cellular phenotypes. Under the orange background, the three genetic states map to red, orange, and yellow cellular phenotypes; under the green background, they map to yellow, green, and blue, respectively. A mutation on the active layer from orange to green shifts the background two positions to the right. With the genetic state retained, the phenotype on the passive layer also shifts two positions to the right: red becomes yellow, orange becomes green, and yellow becomes blue. By contrast, a mutation among the genetic states at a fixed background changes the passive phenotype only within the corresponding dashed box, leaving the active phenotype unchanged.}
    \label{fig:gene_pheno}
\end{figure}

We represent such dependence through a simplified model of \textbf{unidirectional influence}, which we refer to here as \textbf{non-reciprocal epistasis}. The first layer acts as the active layer, and the second as the passive layer. The active-layer phenotype determines a set of admissible passive-layer phenotypes. During initialization, the active phenotype is chosen freely, and the passive phenotype is selected from the corresponding admissible set. During evolution, mutations on the active layer are unrestricted by the passive layer. Whenever the phenotype on the active layer changes, the admissible set on the passive layer shifts in the same direction and by the same number of positions. The current phenotype on the passive layer shifts together with the admissible set, retaining its relative position within it. By contrast, a mutation originating on the passive layer changes its phenotype only within the current admissible set and leaves the phenotype on the active layer unchanged. Figure~\ref{fig:gene_pheno} depicts these rules through a background-dependent genotype-phenotype mapping.

\begin{figure}[htpb!]
	\centering
	\includegraphics[width=0.75\columnwidth]{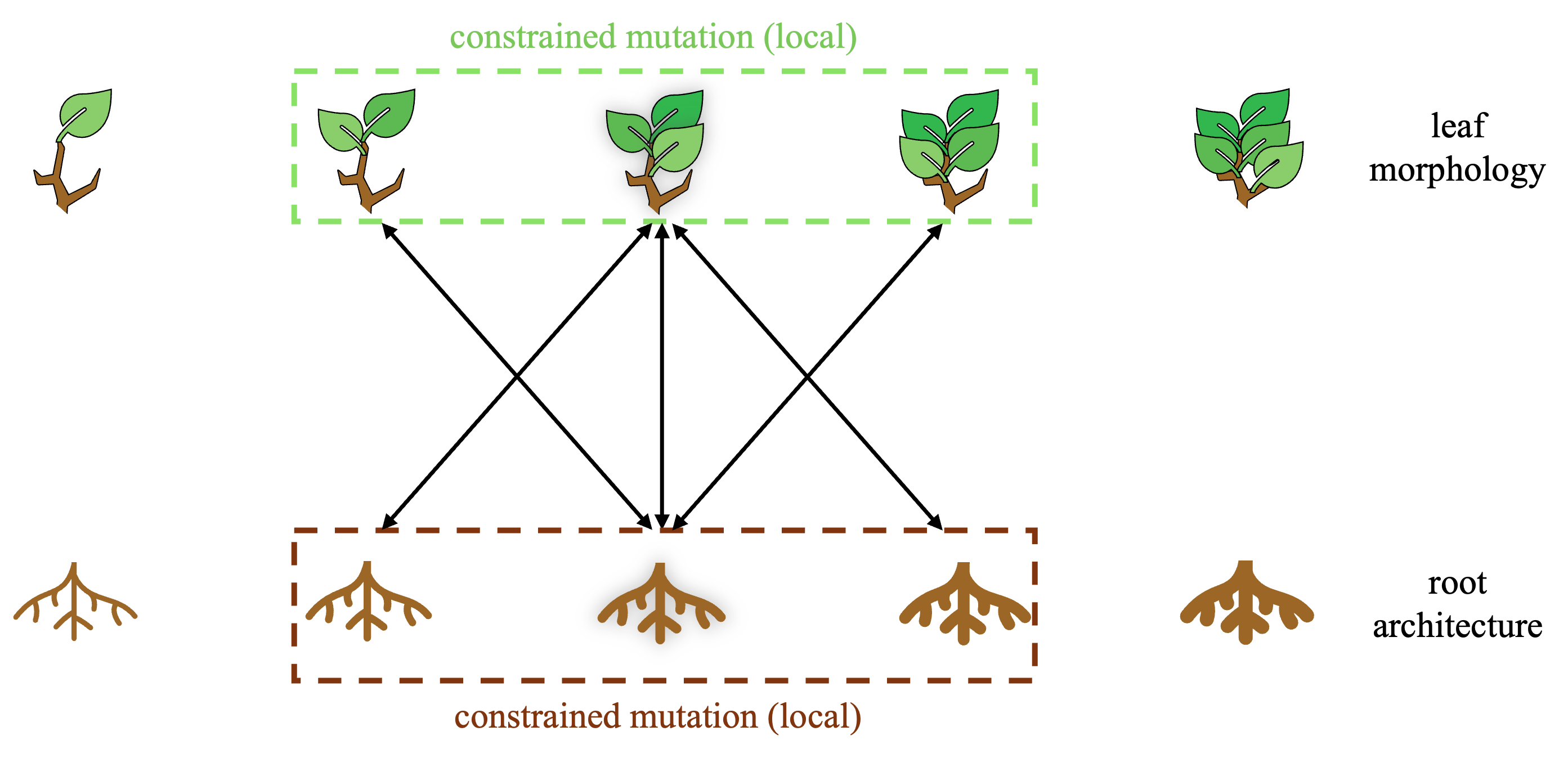}
	\captionsetup{font=small}
	\caption{\textbf{Bidirectional influence between phenotypes in plants.} The phenotype on the first layer represents leaf morphology, including type, surface area, and microscopic features like stomatal density. The phenotype on the second layer represents root architecture. Each phenotype on one layer maps to a limited set of admissible phenotypes on the other, reflecting a reciprocal dependence.}
\label{fig:Leaf_Root}
\end{figure}

We further consider bidirectional coupling between phenotypes on the two layers (\textbf{reciprocal epistasis} or \textbf{mutual regulation}). As shown in Fig.~\ref{fig:Leaf_Root}, leaf morphology and root architecture in plants exert reciprocal influence on one another: leaf morphology can shape the nutrient demand and thereby induce adjustments in root architecture, such as root length and density, while the robustness of the root system, determined by nutrient acquisition, can in turn affect leaf growth and development. Taken together, these interactions exemplify coordinated evolution and structural compensation between aboveground and belowground organs~\cite{fortunel2012leaf}. In analogy with this biological setting, we model reciprocal epistasis by assuming that the phenotypes on the two layers mutually constrain each other. Likewise, this bidirectional constraint applies both to phenotype initialization and to subsequent mutations. The admissible phenotypes on one layer are restricted by the phenotype currently expressed on the other layer, and any phenotypic mutation on either layer is therefore confined to a subset determined by its counterpart.

\begin{figure}[htpb!]
	\centering
	\includegraphics[width=0.75\columnwidth]{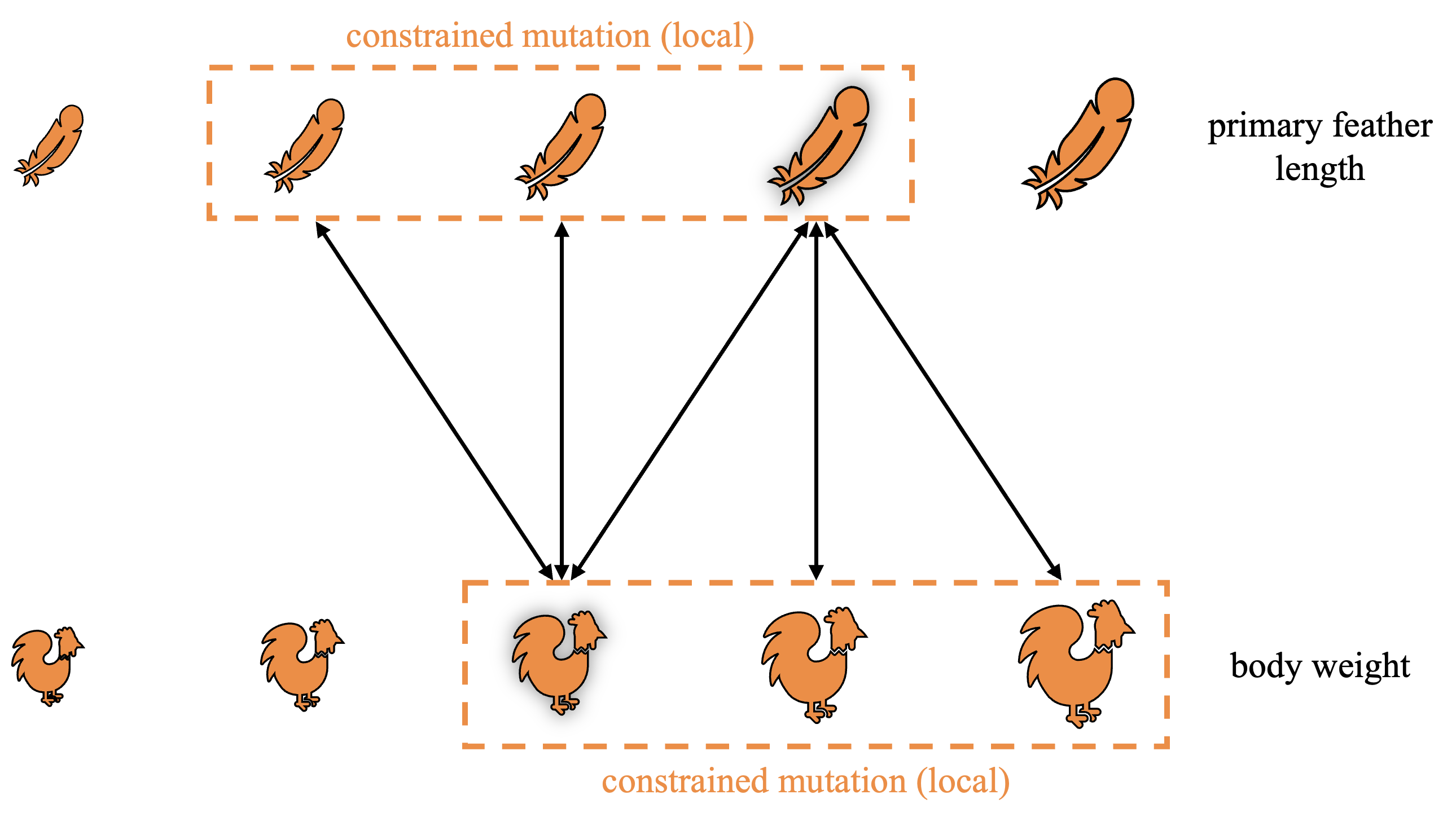}
	\captionsetup{font=small}
	\caption{\textbf{Bidirectional influence between phenotypes in animals.} The phenotype on the first layer represents the length of primary feathers, increasing from left to right. The phenotype on the second layer represents body weight, also increasing from left to right. Each phenotype on one layer maps to a limited set of admissible phenotypes on the other, reflecting a reciprocal dependence.}
\label{fig:feather_weight}
\end{figure}

Another example of reciprocal epistasis is given in Fig.~\ref{fig:feather_weight} from animal studies based on selective breeding research in domestic fowls~\cite{hu1999heritabilities}. Empirical evidence indicates a strong genetic correlation between primary feather length and body weight in fowls, suggesting that these two traits are not independently inherited. Moreover, primary feather traits are associated with mating behavior, reflecting their role in courtship and sexual selection. Similarly, comparable forms of mutual regulation have been documented in animal social systems, where individual behavioral traits and social roles coevolve through feedback mechanisms~\cite{drews1993concept}. Empirical studies across multiple species, including primates (e.g., baboon and rhesus macaques) and social carnivores (e.g., wolves), indicate that social position can influence the expression of aggressive behavior, while behavioral interactions in turn reshape social structure~\cite{cheney2007baboon, simons2022agonism, mech2003wolf}. These examples show that reciprocal epistasis is not restricted to morphological traits, but can also arise in broader biological and social contexts through bidirectional constraint and feedback.

\begin{figure}[htpb!]
	\centering
	\includegraphics[width=0.75\columnwidth]{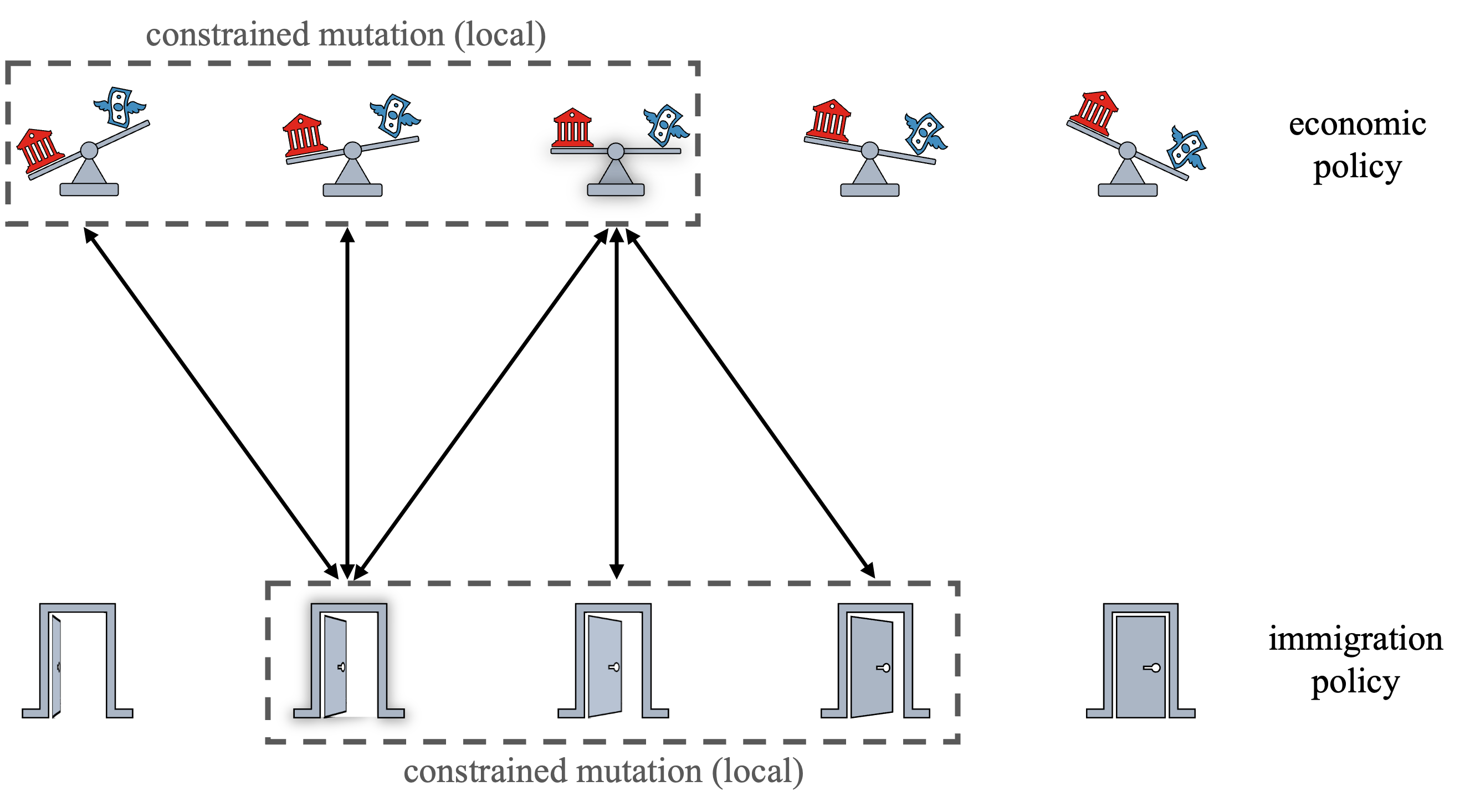}
	\captionsetup{font=small}
	\caption{\textbf{Bidirectional influence between phenotypes in a human social system.} The phenotype on the first layer is associated with economic attitude, ordered from strongly pro-regulation to strongly pro-market. The phenotype on the second layer is associated with immigration attitude, ordered from strongly pro-immigration to strongly anti-immigration. An individual's positions across these dimensions may exhibit mutual dependence, reflecting coevolution under social interaction rather than independent change.}
\label{fig:Policy}
\end{figure}

More generally, if we regard an individual's stance on a given issue, namely, the degree of orientation along a commonly used left-right ideological spectrum, as an abstract phenotype, then bidirectional dependencies between such phenotypes can naturally arise. Here, the term ``phenotype'' is used purely in a modeling sense to denote persistent, heritable, or imitable individual states, without any normative or evaluative implication. For concreteness, attitudes may be discretized using a conventional ordinal scale, such as a five-point spectrum ranging from strongly left-leaning to strongly right-leaning, with intermediate and neutral positions~\cite{bryson1998end}. Taking attitudes toward economic policy and immigration policy as two dimensions (Fig.~\ref{fig:Policy}), the joint distribution of these two phenotypes captures the coexistence of diverse ideological profiles within a population. From a modeling perspective, this abstraction provides a structured framework for studying the coevolution of attitudes under social interaction and influence, allowing one to analyze how collective states may shift, stabilize, or become polarized over time, without attributing any intrinsic value to specific positions.

In what follows, we provide detailed calculations for each scenario of phenotype dependency summarized in Table~\ref{tab:relationship}. For clarity, each case is analyzed separately, and the corresponding strategy selection conditions are derived step by step.

\subsection{Non-concurrent mutation and phenotypic independence}
\label{subsection:independence}

In this case, the treatment of phenotypes closely parallels that of strategies when computing the probabilities of identity. Specifically, we derive the probability that two randomly chosen individuals share the same phenotype on each network layer after their coalescent time $\tau$. We denote these probabilities by $f_1(\tau)$ and $f_2(\tau)$ for the first and second layers, respectively. For the phenotype on the $m$-th layer, by analogy with \eqref{eq:s_2}, we obtain
\begin{align}
	f_m(\tau) = e^{-\nu\tau} + \dfrac{1-e^{-\nu\tau}}{r_m}, \qquad m = 1,2,
\end{align}
where $\nu = Nv$ is the rescaled mutation rate of the phenotype. Consequently, the joint probabilities take the form
\begin{equation}
\begin{dcases}
	Q_{11}(\tau) = f_1(\tau)f_2(\tau), \\
	Q_{10}(\tau) = f_1(\tau)\left[ 1-f_2(\tau) \right], \\
	Q_{01}(\tau) = \left[ 1-f_1(\tau) \right] f_2(\tau).
\end{dcases}
\label{eq:independence}
\end{equation}

Under the assumption that phenotypes on the two network layers mutate separately (at different times) and independently, the structure coefficients $\sigma_m$ in \eqref{eq:sigma_rule} can be derived as
\begin{align}
	\sigma_m = \dfrac{(\mu+\nu+1) \left[ r_m(\mu+2\nu+3)+\nu(\mu+\nu+2) \right]}{(\mu+\nu+3)\left[ r_m(\mu+1)+\nu(\mu+\nu+2) \right]}, \qquad m=1,2.
\label{eq:sigma_independence}
\end{align}
In practice, however, the phenotype space need not be finite. To examine whether our conclusion depends on the assumption of a finite number of phenotypes, we consider the extreme case of $r_m \to \infty$. In this limit, \eqref{eq:sigma_independence} reduces to
\begin{align}
	\sigma_m = \sigma^\ast = \dfrac{(\mu+\nu+1) (\mu+2\nu+3)}{(\mu+\nu+3)(\mu+1)}, \qquad m=1,2.
\label{eq:sigma_infinite}
\end{align}
That being said, even when the phenotype space is unbounded, the $\sigma$-rule remains valid, and the structure coefficient $\sigma_m$ is still well defined for any $m$. The detailed substitutions and intermediate steps are algebraically involved and are therefore provided in the accompanying code (see Section~\ref{section:code_data}). The same procedure applies to subsequent cases.

Building on the seminal work of Fu et al. for single-layer networks~\cite{fu2012evolution}, where each individual is characterized by a single trait, we adapt their analytical framework to our setting by generalizing the payoff structure (with payoff matrix $A = [R, S, T, P]$) while preserving the underlying evolutionary dynamics. In this setting, interactions occur only between individuals sharing the same phenotype, and an individual adopting strategy $i$ cooperates with probability $p_i$. The resulting $\sigma$-rule for a single-layer network takes the form 
\begin{align}
\sigma R + S > T + \sigma P.
\end{align} 
The corresponding structure coefficient $\sigma$ is given by
\begin{align}
	\sigma = \dfrac{(\mu+\nu+1) \left[ M(\mu+2\nu+3)+\nu(\mu+\nu+2) \right]}{(\mu+\nu+3)\left[ M(\mu+1)+\nu(\mu+\nu+2) \right]},
\label{eq:sigma}
\end{align}
where $M$ denotes the number of phenotypes. This expression recovers the established result in~\cite{fu2012evolution} when the two-layer network collapses into a single layer, in the sense that the two layers encode identical sets of phenotypes with $M = r_1 = r_2$.

\begin{figure}[htpb!]
	\centering
	\includegraphics[width=\columnwidth]{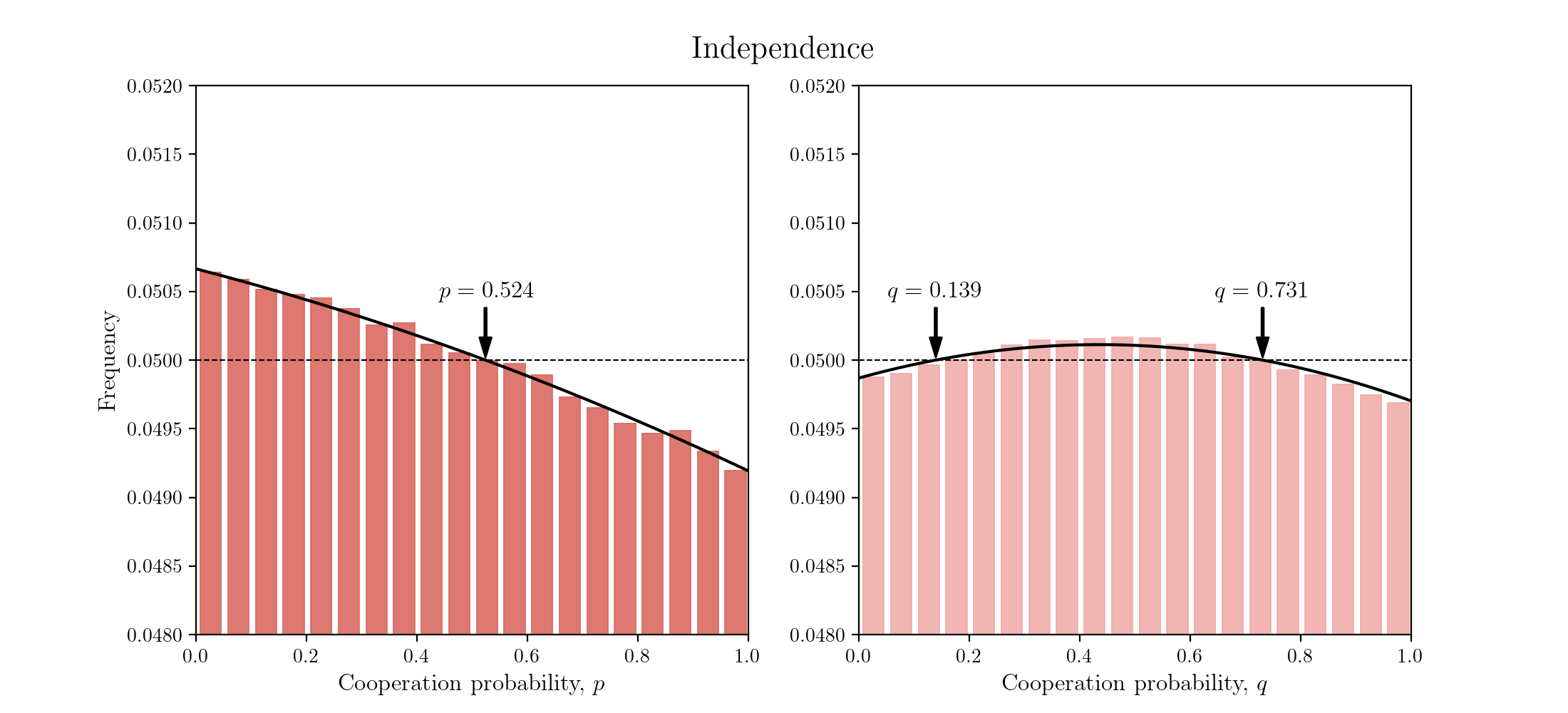}
	\captionsetup{font=small}
	\caption{\textbf{Simulation results under non-concurrent mutation and phenotypic independence.} In each panel, the black solid curve depicts the theoretical prediction for the stationary distribution, while the black dashed line indicates the neutral expectation, that is, the uniform proportion $1/n$. Their intersection defines the theoretical critical point at which the predicted abundance equals the neutral baseline. The histograms display results from agent-based simulations. Parameters: $N = 50$, $n = 20 \times 20 = 400$, $\beta = 0.001$, $r_1 = r_2 = 3$, $u = 0.04$, $v = 0.02$, $[R_1,S_1,T_1,P_1] =  [3, 0, 5, 1]$ and $[R_2, S_2, T_2, P_2] = [3, 1, 5, 0]$.}
\label{fig:simulation_independence}
\end{figure}

To validate our analytical results, we conducted agent-based Monte Carlo simulations. A comparison between simulation outcomes and theoretical predictions is presented in Fig.~\ref{fig:simulation_independence}.

\subsection{Non-concurrent mutation and unidirectional influence}
\label{subsection:unidirection}

In the presence of unidirectional influence, phenotypic mutations on the first layer of the network are assumed to remain unrestricted. To ensure computational consistency and, in particular, the validity of the symmetry condition of the coalescent framework~\cite{fu2012evolution, antal2009mutation}, we further assume that the number of phenotypes in the second layer satisfies
\begin{align}
	r_2 \geq r_1+2K.
\label{eq:r2_condition}
\end{align}
The condition is a boundary-free embedding convention rather than a biological restriction. We embed the $r_1$ modifier backgrounds as centers in the ordered target space and define each accessible target set of size $2K + 1$. The minimal finite construction has $r_2 = r_1 + 2K$.

The probability that two randomly sampled individuals share the same phenotype on the first layer after their coalescent time $\tau$ is
\begin{align}
	g_1(\tau) = e^{-\nu\tau} + \dfrac{1-e^{-\nu\tau}}{r_1}.
\end{align}
In contrast, phenotypic mutations on the second layer are subject to epistatic constraints imposed by the first layer. Specifically, a mutation on the second layer is restricted to a subset of $2K + 1$ admissible phenotypes: the phenotype of comparable level on the first layer and its $2K$ nearest neighbors (see Fig.~\ref{fig:gene_pheno}), where $K \in \mathbb{N}$. Thus, the probability that two randomly sampled individuals share the same phenotype on the second layer after their coalescent time $\tau$ is
\begin{align}
	g_2(\tau) = e^{-\nu\tau} + \dfrac{1-e^{-\nu\tau}}{2K+1}.
\end{align}
Under the assumption that mutations on the two layers occur separately in time, the joint probabilities can be written as
\begin{equation}
\begin{dcases}
	Q_{11}(\tau) = g_1(\tau)g_2(\tau), \\
	Q_{10}(\tau) = g_1(\tau)\left[ 1-g_2(\tau) \right], \\
	Q_{01}(\tau) = \left[ 1-g_1(\tau) \right] g_2(\tau).
\end{dcases}
\end{equation}

After straightforward but lengthy calculations, we obtain
\begin{equation}
\begin{dcases}
\sigma_1 = \dfrac{(\mu+\nu+1) \left[ r_1(\mu+2\nu+3)+\nu(\mu+\nu+2) \right]}{(\mu+\nu+3)\left[ r_1(\mu+1)+\nu(\mu+\nu+2) \right]}, \\
\sigma_2 =  \dfrac{(\mu+\nu+1) \left[ (2K+1)(\mu+2\nu+3)+\nu(\mu+\nu+2) \right]}{(\mu+\nu+3)\left[ (2K+1)(\mu+1)+\nu(\mu+\nu+2) \right]}.
\end{dcases}
\label{eq:sigma_unidirection}
\end{equation}
The expression for the structure coefficient $\sigma_1$ coincides with \eqref{eq:sigma_independence}: it depends solely on the total number of phenotypes $r_1$ on the first layer, and is entirely independent of the constraint parameter $K$. Meanwhile, the structure coefficient $\sigma_2$ depends explicitly on $K$, reflecting the epistatic restriction imposed by the first layer, but is independent of the total number of phenotypes $r_2$ on the second layer. 

As in \eqref{eq:sigma_infinite}, when $r_1 \to \infty$, $\sigma_1$ converges to the same limiting value $\sigma^\ast$. Moreover, even when $r_2 \to \infty$, the expression for $\sigma_2$ remains unchanged, since only the locally admissible mutation range of size $2K + 1$ enters the coalescent identity probabilities. In this continuous limit, where both phenotype spaces become unbounded, the technical condition in \eqref{eq:r2_condition} is no longer required.

\begin{figure}[htpb!]
	\centering
	\includegraphics[width=\columnwidth]{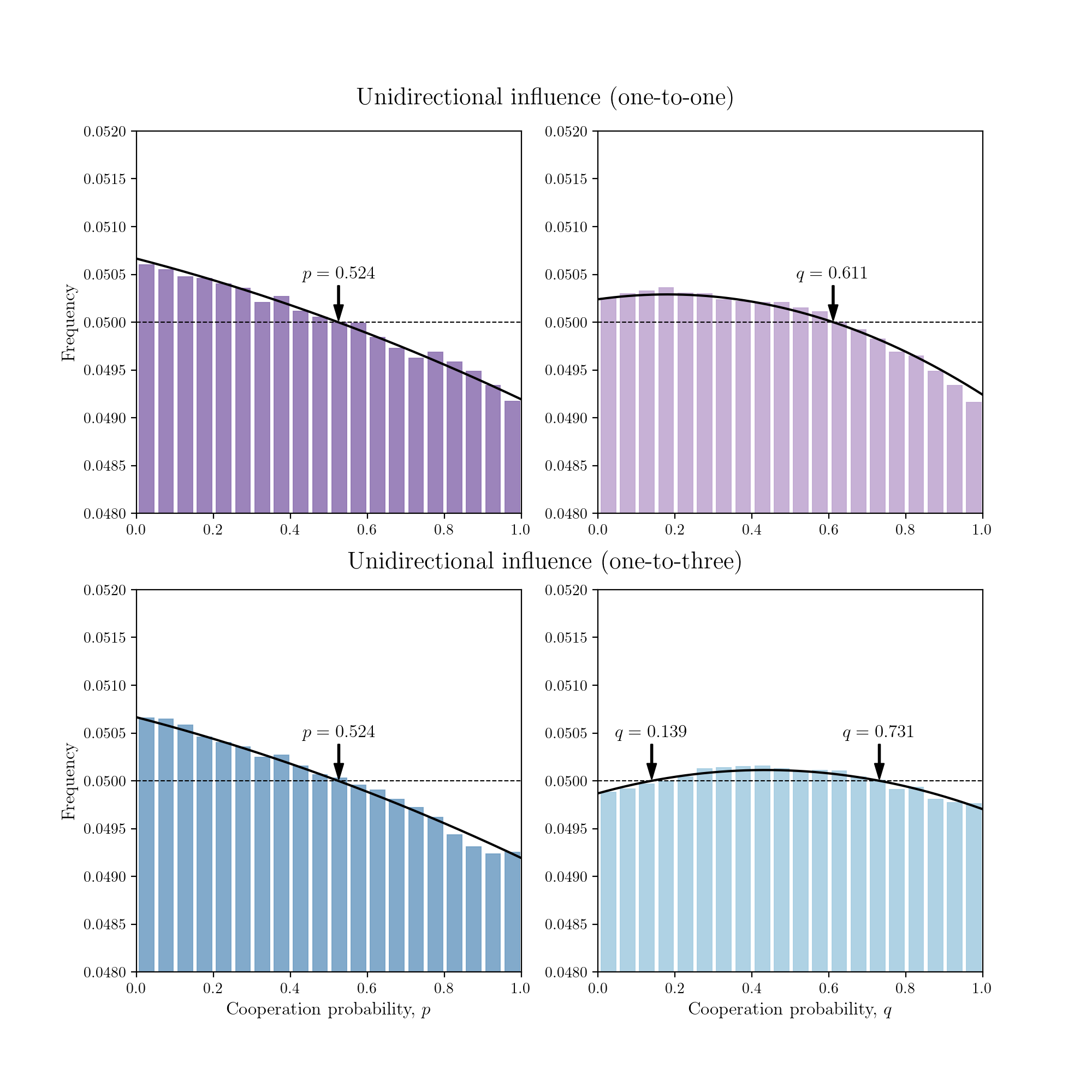}
	\captionsetup{font=small}
	\caption{\textbf{Simulation results under non-concurrent mutation and unidirectional influence.} 
In each panel, the black solid curve depicts the theoretical prediction for the stationary distribution, while the black dashed line indicates the neutral expectation, that is, the uniform proportion $1/20$. Their intersection defines the theoretical critical point at which the predicted abundance equals the neutral baseline. The histograms display results from agent-based simulations. The top two panels correspond to $K = 0$, and the bottom two panels correspond to $K = 1$. Parameters: $N = 50$, $n = 20 \times 20 = 400$, $\beta = 0.001$, $r_1 = 3$, $r_2 = 3$ (one-to-one) or $r_2 = 5$ (one-to-three), $u = 0.04$, $v = 0.02$, $[R_1,S_1,T_1,P_1] =  [3, 0, 5, 1]$ and $[R_2, S_2, T_2, P_2] = [3, 1, 5, 0]$.}
\label{fig:simulation_unidirection}
\end{figure}

A comparison between simulation outcomes and theoretical predictions is shown in Fig.~\ref{fig:simulation_unidirection}, which further confirms that the abundances of strategies on the second layer vary with the constraint parameter $K$.

\subsection{Non-concurrent mutation and bidirectional influence}
\label{subsection:bidirection}

Under bidirectional constraints between the two layers, each phenotype on the first layer corresponds to $2K + 1$ admissible phenotypes on the second layer, and conversely, each phenotype on the second layer is associated with $2K + 1$ admissible phenotypes on the first layer (see Figs.~\ref{fig:Leaf_Root},~\ref{fig:feather_weight}, and~\ref{fig:Policy} for illustrative examples). 

As shown in the previous two subsections, our framework naturally accommodates an infinite number of phenotypes. Accordingly, to maintain symmetry between layers under bidirectional regulation, we assume that both layers contain infinitely many phenotypes. This assumption guarantees that, for any phenotype on one layer, there always exist $2K+1$ corresponding phenotypes on the other layer.

Since the relationship between the two layers is bidirectional, neither layer plays a privileged hierarchical role. Nevertheless, in the calculation, one must impose an evaluation order to determine whether a phenotypic mutation occurs during the coalescent time $\tau$. We begin by computing $Q_{11}(\tau)$. Because the mutation space on each layer is restricted to $2K + 1$ admissible phenotypes, the probability is invariant with respect to the order in which the two layers are evaluated. Accordingly, defining
\begin{align}
	h(\tau) = e^{-\nu\tau} + \dfrac{1-e^{-\nu\tau}}{2K + 1},
\end{align}
we obtain
\begin{align}
	Q_{11}(\tau) = h^2(\tau).
\end{align}

We then compute $Q_{10}(\tau)$ and $Q_{01}(\tau)$. Still, the evaluation requires specifying an order, although the two layers are symmetric. We first consider the case in which the mutation status is determined on the first layer during $\tau$. Because of the bidirectional constraint, the admissible mutation space on the first layer is restricted to $2K + 1$ phenotypes. The probability that the first-layer phenotypes coincide while the second-layer phenotypes differ hence equals $h(\tau)[1 - h(\tau)]/2$. It is noteworthy that since either phenotype can mutate with equal probability, this contribution is weighted by $1/2$. We then consider the reverse case in which the mutation status is determined on the second layer during $\tau$. The same reasoning applies, leading to an identical contribution of $h(\tau)[1 - h(\tau)]/2$. Summing the two mutually exclusive cases, we obtain
\begin{align}
	Q_{10}(\tau) = h(\tau)[1 - h(\tau)].
\end{align}
By symmetry between the two layers, it follows immediately that $Q_{01}(\tau) = Q_{10}(\tau)$. 

\begin{figure}[htpb!]
	\centering
	\includegraphics[width=0.9\columnwidth]{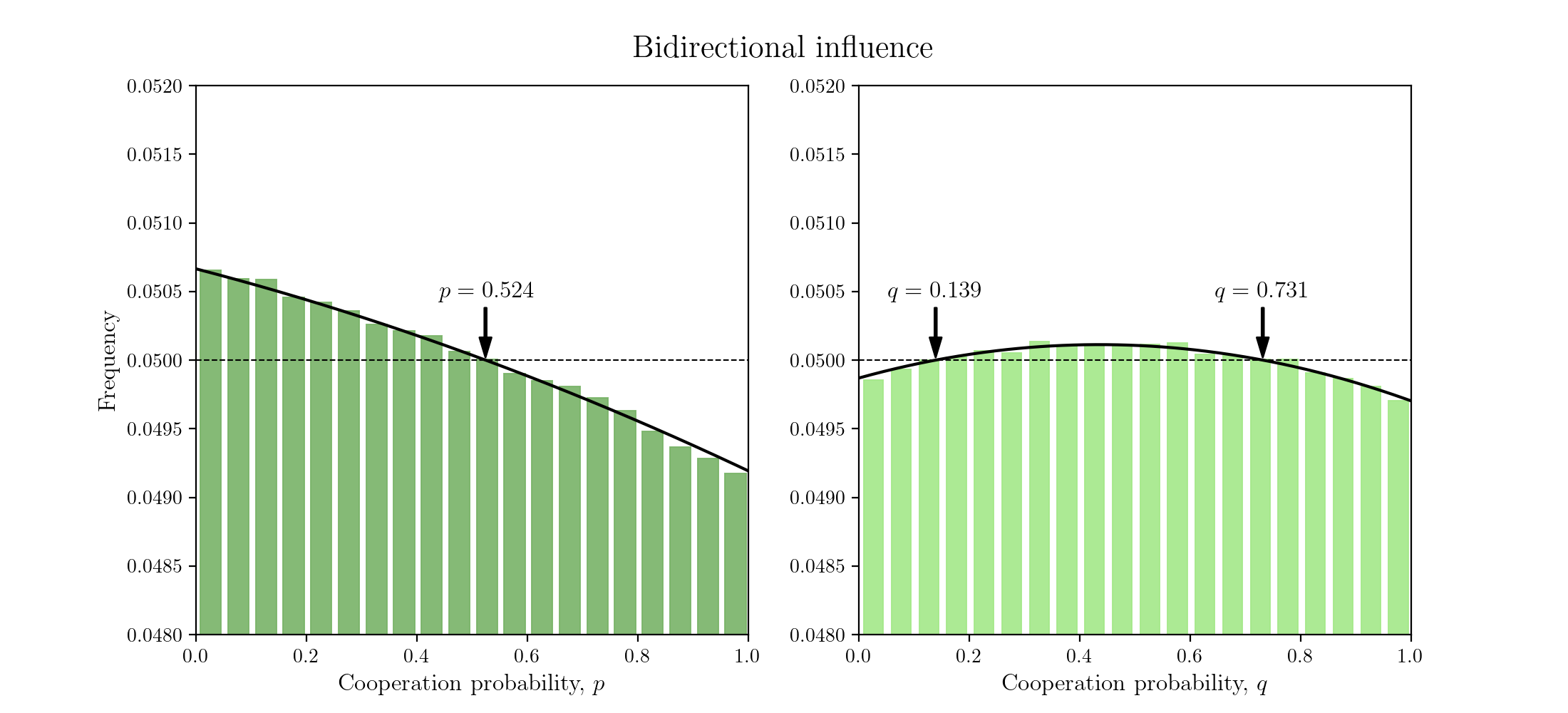}
	\captionsetup{font=small}
	\caption{\textbf{Simulation results under non-concurrent mutation and bidirectional influence.} In each panel, the black solid curve depicts the theoretical prediction for the stationary distribution, while the black dashed line indicates the neutral expectation, that is, the uniform proportion $1/20$. Their intersection defines the theoretical critical point at which the predicted abundance equals the neutral baseline. The histograms display results from agent-based simulations. Parameters: $N = 50$, $n = 20 \times 20 = 400$, $K = 1$, $\beta = 0.001$, $u = 0.04$, $v = 0.02$, $[R_1,S_1,T_1,P_1] =  [3, 0, 5, 1]$ and $[R_2, S_2, T_2, P_2] = [3, 1, 5, 0]$.}
\label{fig:simulation_bidirection}
\end{figure}

We finally obtain
\begin{align}
	\sigma_m = \dfrac{(\mu+\nu+1) \left[ (2K+1)(\mu+2\nu+3)+\nu(\mu+\nu+2) \right]}{(\mu+\nu+3)\left[ (2K+1)(\mu+1)+\nu(\mu+\nu+2) \right]}, \qquad m=1,2.
\label{eq:sigma_bidirection}
\end{align}
where both $\sigma_1$ and $\sigma_2$ are independent of the total number of phenotypes. A comparison between simulation outcomes and theoretical results is shown in Fig.~\ref{fig:simulation_bidirection}. As a special example, we consider $K = 0$, for which $\sigma_m = 1$ for $m = 1, 2$, indicating that the population update process is equivalent to the case without phenotypic mutations. Indeed, substituting $\nu = 0$ into \eqref{eq:sigma_independence} also yields $\sigma_m = 1$, thereby confirming the consistency between the two results.

\subsection{Concurrent mutation}
\label{subsection:concurrent}

Let $\mathcal{H}_1$ denote the set of all possible phenotypes on the first layer, and $\mathcal{H}_2$ the set of all possible phenotypes on the second layer. Since updates and mutations of phenotypes on both layers occur concurrently, we treat each ordered pair of phenotypes across the two layers as a single composite phenotype, forming the set $\mathcal{H}$. For instance, under the independence assumption,
\begin{align}
	\mathcal{H} = \left\{ (k_1, k_2) \, \left | \, k_1 \in \mathcal{H}_1, k_2 \in \mathcal{H}_2 \right. \right\}.
\end{align}
We then consider all possible pairs of composite phenotypes arising from two randomly sampled individuals, which defines the set $\mathcal{H}^2$:
\begin{align}
	\mathcal{H}^2 = \left\{ (k_1, k_2) \times (l_1, l_2) \, \left | \, (k_1, k_2),  (l_1, l_2) \in \mathcal{H} \right. \right\}.
\end{align}

We denote the size of $\mathcal{H}$ by $H$, i.e., $H = \left | \mathcal{H} \right |$ and thus $ \left | \mathcal{H}^2 \right |= H^2$. For instance, in the independent case, $H = r_1r_2$. We also classify subsets of pairs in $\mathcal{H}^2$. Let $H_{10}$ denote the number of pairs that differ only on the second layer, i.e., of the form $(k_1, k_2) \times (k_1, l_2)$ with $l_2 \ne k_2$. Similarly,  let $H_{01}$ denote the number of pairs that differ only on the first layer, i.e., of the form $(k_1, k_2) \times (l_1, k_2)$ with $l_1 \ne k_1$. Based on these quantities, we then define
\begin{align}
	\delta_{1} = \dfrac{H_{10}}{H}, \qquad \delta_{2} = \dfrac{H_{01}}{H}.
\end{align}
It follows that
\begin{equation}
\begin{dcases}
	Q_{11}(\tau) = e^{-\nu\tau} + \dfrac{1-e^{-\nu\tau}}{H}, \\
	Q_{10}(\tau) = \dfrac{\delta_{1}}{H} \left( 1-e^{-\nu\tau} \right), \\
	Q_{01}(\tau) = \dfrac{\delta_{2}}{H} \left( 1-e^{-\nu\tau} \right).
\end{dcases}
\end{equation}

Finally, we obtain
\begin{align}
	\sigma_m = \dfrac{(\mu+\nu+1) \left[ H(\mu+2\nu+3)/(1+\delta_m)+\nu(\mu+\nu+2) \right]}{(\mu+\nu+3)\left[ H(\mu+1)/(1+\delta_m)+\nu(\mu+\nu+2) \right]}.
\label{eq:sigma_concurrent}
\end{align}
In the limiting case $\delta_m = 0$, where composite phenotypes do not differ across layers, this expression reduces to the single-layer result given in \eqref{eq:sigma}. Moreover, analogous to the non-concurrent case, the three distinct scenarios can be summarized as follows:
\begin{itemize}
 	\item Independence: $H = r_1r_2$, $H_{10} = r_1(r_2-1)r_2$, and $H_{01} = (r_1-1)r_1r_2$. After simplification, we obtain $H/(1+\delta_m) = r_m$. This result coincides with that obtained in Subsection~\ref{subsection:independence}, indicating that under independence, concurrent mutation does not alter the condition under which cooperation is favored.
	\item Unidirectional influence: $H = (2K+1)r_1$ and $H_{10} = 2K(2K+1)r_1$, which yields $H/(1+\delta_1) = r_1$. The value of $H_{01}$ depends on the relation between $r_1$ and $2K+1$. When $r_1 \ge 2K+1$, we obtain
	\begin{equation}
	\begin{aligned}
		H_{01} &= 2\sum_{i=1}^{2K} (i-1)i + 2K(2K+1)(r_1-2K) = 2K(2K + 1)\left[r_1 - (2K + 1) + \frac{2(2K + 1)-1}{3}\right].
	\end{aligned}
	\end{equation}
	When $r_1 \le 2K$, we obtain
	\begin{equation}
	\begin{aligned}
		H_{01} &= 2\sum_{i=1}^{r_1-1} (i-1)i + (r_1-1)r_1(2K+2-r_1) = (r_1 -1)r_1\left[(2K + 1) - r_1 + \frac{2r_1 - 1}{3}\right].
	\end{aligned}
	\end{equation}
This result differs from that in Subsection~\ref{subsection:unidirection}. The condition for the emergence of cooperation in the constrained layer is jointly decided by both $r_1$ and $K$.
	\item Bidirectional influence: in this case, both layers are required to have infinitely many phenotypes. In the limit $r_m \to \infty$ for $m = 1, 2$, the total number of phenotype combinations satisfies $H \to \infty$, whereas $\delta_m \to 2K$ for both layers. By substituting these limits into \eqref{eq:sigma_concurrent}, we recover the same result as in \eqref{eq:sigma_infinite},
	\begin{align}
		\sigma_m = \dfrac{(\mu+\nu+1) (\mu+2\nu+3)}{(\mu+\nu+3)(\mu+1)}, \qquad m=1,2.
	\end{align}
\end{itemize}

\begin{figure}[htpb!]
	\centering
	\includegraphics[width=0.9\columnwidth]{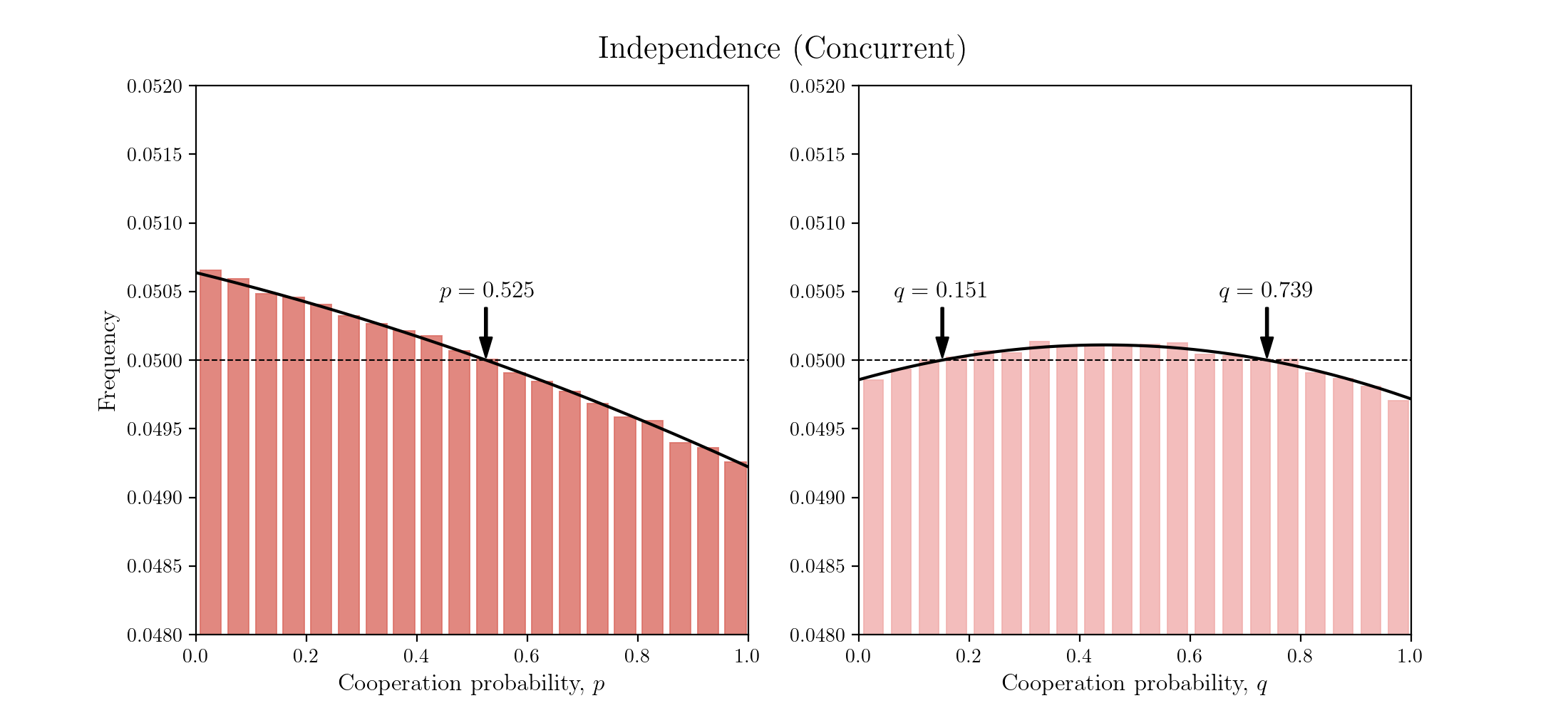}
	\captionsetup{font=small}
	\caption{\textbf{Simulation under concurrent mutation and independence.} In each panel, the black solid curve depicts the theoretical distribution, while the black dashed line marks the expected proportion of each strategy under neutral drift ($1/20$). Their intersection identifies the theoretical critical point where the distribution meets the uniform baseline. The histogram displays the simulation results. Parameters: $N = 50$, $n = 20 \times 20 = 400$, $\beta = 0.001$, $r_1 = r_2 = 3$, $u = 0.04$, $v = 0.02$, $[R_1,S_1,T_1,P_1] =  [3, 0, 5, 1]$ and $[R_2, S_2, T_2, P_2] = [3, 1, 5, 0]$.}
\label{fig:simulation_concurrent}
\end{figure}

A comparison between simulation outcomes and theoretical results for the independence case is shown in Fig.~\ref{fig:simulation_concurrent}.

\section{Greater diversity shepherds cooperation}
\label{section:Extended_Analysis}

As a starting point, we consider the basic $\sigma$-rule, $\sigma R + S > T + \sigma P$. When mutual cooperation is better than mutual defection, namely, $R > P$ , this condition can be equivalently rewritten as
\begin{align}
	\sigma > \dfrac{T - S}{R - P}.
\end{align}
For a given payoff structure, this inequality shows that larger values of $\sigma$ relax the condition under which cooperation is favored, by lowering the required relative benefit of cooperation over defection. We hence focus on identifying the mechanisms that increase the structure coefficient $\sigma_m$ in the scenarios analyzed above. In what follows, we examine how it depends on the parameters governing phenotypic diversity and coupling, as given in \eqref{eq:sigma_independence}, \eqref{eq:sigma_unidirection}, \eqref{eq:sigma_bidirection}, and \eqref{eq:sigma_concurrent}, all of which take the form of rational functions of $r_m$ and/or $2K + 1$.

We consider a linear rational function
\begin{align}
	\sigma (x) = \dfrac{ax+b}{cx+d},
\end{align}
where $a, b, c, d > 0$, and $x > 0$. Differentiating with respect to $x$ gives
\begin{align}
	\sigma' (x) = \dfrac{ad-bc}{(cx+d)^2}.
\end{align}
Therefore, $\sigma (x)$ is an increasing function of $x$ whenever $ad - bc>0$. In Eqs.~\eqref{eq:sigma_independence}, \eqref{eq:sigma_unidirection}, and \eqref{eq:sigma_bidirection}, the structure coefficient $\sigma_m$ can be written in this form by identifying $x$ with either $r_m$ or $2K + 1$ and with
\begin{equation}
\begin{dcases}
	a = (\mu+\nu+1)(\mu+2\nu+3), \\
	b = \nu(\mu+\nu+1)(\mu+\nu+2), \\
	c = (\mu+\nu+3)(\mu+1), \\
	d = \nu(\mu+\nu+3)(\mu+\nu+2).
\end{dcases}
\end{equation}
A direct calculation yields
\begin{align}
	ad - bc = 2\nu(\nu+1)(\mu+\nu+1)(\mu+\nu+2)(\mu+\nu+3) > 0,
\end{align}
which establishes that $\sigma_m$ increases monotonically with $r_m$ or $2K+1$ in these cases.

We next consider the constrained layer in the cases of concurrent mutation under unidirectional influence, where $\sigma_2$ depends on both $r_1$ and $K$. When $r_1 \ge 2K+1$, explicit calculation (see the supplementary code) shows that $\partial \sigma_2 / \partial K > 0$ and
\begin{align}
	\dfrac{\partial \sigma_2}{\partial r_1} = C_1 \cdot \left[ -2(2K + 1)^2 + 3r_1(2K + 1) + 2 \right], 
\end{align}
where $C_1$ denotes a positive factor. It follows that $\partial \sigma_2 / \partial r_1 > 0$. When $r_1 \le 2K$, a similar calculation yields $\partial \sigma_2 / \partial r_1 > 0$ and 
\begin{align}
	\dfrac{\partial \sigma_2}{\partial K} = C_2 \cdot \left[ -2r_1^2 + 3(2K+1)r_1 + 2 \right],
\end{align}
where $C_2$ again denotes a positive factor. Consequently, $\partial \sigma_2 / \partial K > 0$.

We note that the total number of possible phenotypes (or their combinations) depends on the parameters $r_m$ and $K$ under different scenarios. These parameters, therefore, collectively characterize the degree of phenotypic diversity in the population. The analytical results indicate that increasing phenotypic diversity enlarges the space of possible evolutionary configurations, thereby facilitating the emergence and maintenance of cooperation in social dilemmas satisfying $R > P$. This finding is consistent with previous studies that highlight the positive role of diversity in promoting cooperation~\cite{wu2017coevolutionary, santos2012role}.

\section{Impact of mutation probabilities on cooperation}
\label{section:mutation_cooperation}

Having examined the conditions for the emergence of cooperation, we now investigate how mutation probabilities influence the average cooperation level in the stationary state. We consider the case of non-concurrent mutation with independent phenotypes as an example and focus on the average cooperation probability $\langle p \rangle$ on the first layer for illustration. The analysis for the second layer is analogous. Since $\langle p \rangle$ is a multivariate function, we evaluate its partial derivative with respect to the mutation rates to characterize their effects. As a side note, the average cooperation probability $\langle p \rangle$ can be written as
\begin{equation}
\begin{aligned}
\langle p \rangle &= \frac{1}{2} + \frac{\beta(1 - u)}{Nu} \int_0^1 pD^1(p) \mathrm{d}p  =  \frac{1}{2} + C \cdot (1 - u) \cdot \\
&
\begin{rcases}
\begin{dcases}-v(u + v)^2(T_1 - R_1 + P_1 - S_1)N^3 \\
 + v(u + v)[2(R_1 - P_1)r_1 + 3(R_1 - P_1) - 5(T_1 - S_1)]N^2 \\
 - u(u + v)(T_1 - R_1 + P_1 - S_1)r_1N^2 \\
 + \mathcal{O}(N)
 \end{dcases}
 \end{rcases}
 .
 \end{aligned}
\end{equation}

\subsection{Strategy mutation probability}
\label{section:strategy_mutation_probability}

\begin{figure}[htpb!]
	\centering
	\includegraphics[width=\columnwidth]{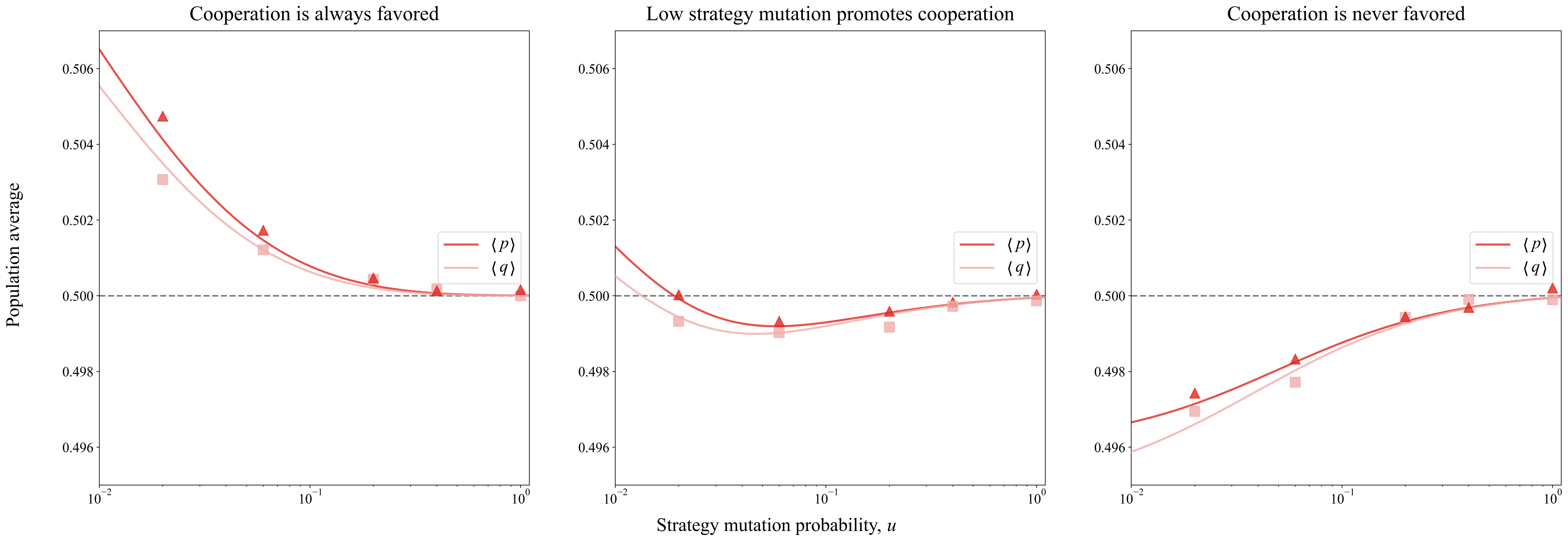}
	\captionsetup{font=small}
	\caption{\textbf{Cooperation levels (population averages) as functions of the strategy mutation probability in finite populations for the independence and non-concurrent case.} Solid curves denote theoretical predictions, where the darker curve represents $\langle p \rangle$ and the lighter curve represents $\langle q \rangle$. Markers indicate simulation results averaged over $10^9$ generations, with triangles for $\langle p \rangle$ and squares for $\langle q \rangle$. Panel-specific parameters: (left) $N = 100$, $r_1 = 15$ and $r_2 = 10$, $[R_1, S_1, T_1, P_1] = [R_2, S_2, T_2, P_2] = [15, 0, 16, 1]$, $v = 0.3$; (middle) $N = 100$, $r_1 = 15$ and $r_2 = 10$, $[R_1, S_1, T_1, P_1] = [R_2, S_2, T_2, P_2] = [3, 0, 4, 1]$, $v= 0.025$; (right) $N = 50$, $r_1 = 4$ and $r_2 = 3$, $[R_1,S_1,T_1,P_1] = [R_2, S_2, T_2, P_2] = [3, 0, 5, 1]$, $v= 0.02$. Common parameters: $n = 20 \times 20 = 400$, $\beta = 0.001$, and for simulations $u \in \{0.02, 0.06, 0.2, 0.4, 1\}$.}
\label{fig:langle_p_q_u_different_parameter}
\end{figure}

In contrast to the large population limit ($N \gg 1$), where the average level of cooperation exhibits a unified monotonic dependency on the mutation probability $u$, finite populations display a much wider variety of evolutionary regimes in the prisoner's dilemma. As shown in Fig.~\ref{fig:langle_p_q_u_different_parameter}, the monotonicity of the average cooperation level depends sensitively on the corresponding parameter configurations. (i) Monotonically decreasing (left panel): selection always favors cooperation, but its average level declines as higher mutation probabilities drive the population to the neutral limit. (ii) U-shaped (middle panel): cooperation is favored only at low mutation probabilities; the average level initially decreases with $u$ but subsequently rebounds, revealing a non-monotonic dependency. (iii) Monotonically increasing (right panel): selection always opposes cooperation, yet the average level increases as higher mutation probabilities drive the population to the neutral limit.

\begin{figure}[htpb!]
	\centering
	\includegraphics[width=\columnwidth]{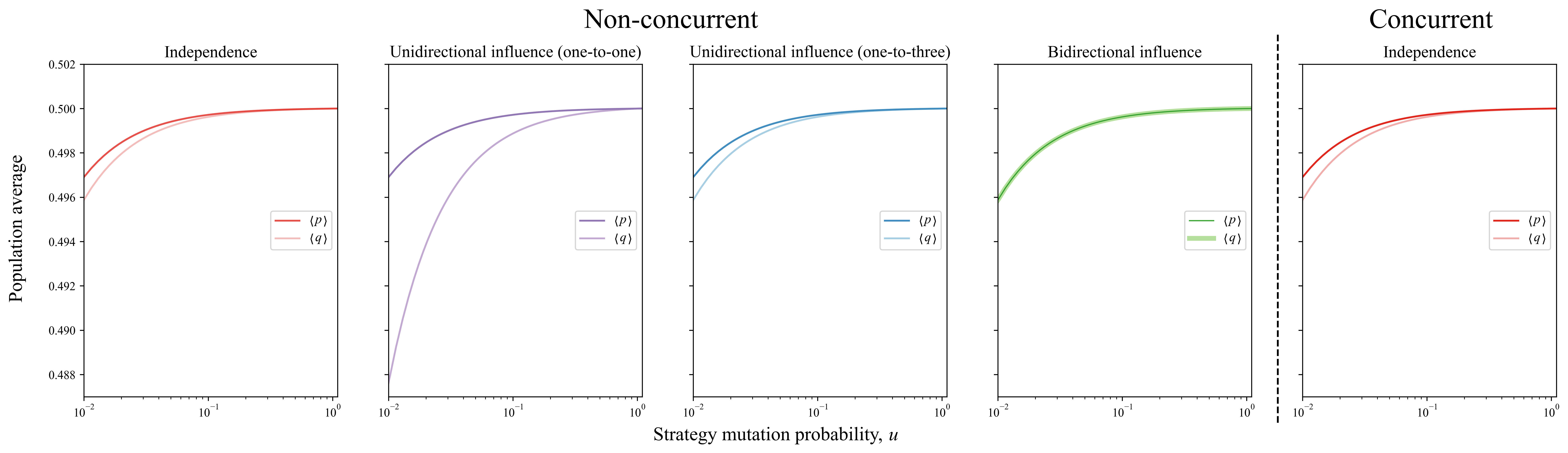}
	\captionsetup{font=small}
	\caption{\textbf{Cooperation levels (population averages) as functions of the strategy mutation probability in large populations.} Each column corresponds to a specific type of interlayer phenotypic relationship, as indicated at the top. Solid curves denote theoretical predictions, where the darker curve represents $\langle p \rangle$ and the lighter curve represents $\langle q \rangle$. Parameters: $N = 5 \times 10^6$, $n = 20 \times 20 = 400$, $\beta = 0.001$, $r_1 = 3$ and $r_2 = 3, 3, 5, 3$ from left to right (excluding the bidirectional influence case), $[R_1, S_1, T_1, P_1] = [R_2, S_2, T_2, P_2] = [3, 0, 5, 1]$, $v= 0.02$.}
\label{fig:langle_p_q_u}
\end{figure}

Turning to the large population limit, we find that the dependencies observed in finite populations subside. The partial derivative of $\langle p \rangle$ with respect to the strategy mutation probability $u$ is given by
\begin{align}
	\dfrac{\partial \langle p \rangle}{\partial u} = C_3 \cdot \left[ v(u+v)^4(T_1 - R_1 + P_1 - S_1)N^6 + \mathcal{O}(N^5) \right], 
\end{align}
where $C_3$ denotes a positive factor. Given that the population size is assumed to be large ($N \gg 1$), the leading-order term in $N$ dominates the expression. The sign of the partial derivative is thus governed by the coefficient of $N^6$, namely the payoff combination $T_1 - R_1 + P_1 - S_1$. In particular, for the prisoner's dilemma with payoffs satisfying $T_1 > R_1 > P_1 > S_1$, the average cooperation probability $\langle p \rangle$ increases with $u$. Under the conventional payoff matrix $[3, 0, 5, 1]$, the dependence of $\langle p \rangle$ and $\langle q \rangle$ on $u$ is illustrated in Fig.~\ref{fig:langle_p_q_u}.

\subsection{Phenotypic mutation probability}
\label{section:phenotypic_mutation_probability}

The partial derivative of $\langle p \rangle$ with respect to the phenotypic mutation probability $v$ is given by
\begin{align}
	\dfrac{\partial \langle p \rangle}{\partial v} = C_4 \cdot p(v), 
\end{align}
where $C_4$ denotes a positive factor and $p(v)$ is a quartic polynomial in $v$. The sign of the partial derivative is hence determined by that of $p(v)$.

We compute the second derivative of $p(v)$, obtaining
\begin{align}
	p''(v) = -2N^2 \left[ 12N^2(R_1 - P_1)v^2 + \mathcal{O}(v) \right].
\end{align}
For a large population size $N$ and the prisoner's dilemma with payoffs satisfying $T_1 > R_1 > P_1 > S_1$, $p''(v)$ is either negative for all $v \geq 0$, or positive for small $v$ and negative thereafter. Consequently, the first derivative $p'(v)$ is either monotonically decreasing in $v$, or first increasing and then decreasing as $v$ grows. When $v = 0$, we have
\begin{align}
	p'(0) = (Nu+1)(Nu+3) \left[ 2N(T_1 - R_1 + P_1 - S_1)u + 5T_1- 3R_1 + 3P_1 - 5S_1 \right].
\end{align}
For large $N$ and for the prisoner's dilemma, we obtain $p'(0) > 0$. Therefore, $p'(v)$ is positive for small $v$ but becomes negative for sufficiently large $v$. 

It follows that $p(v)$ initially increases and subsequently decreases for $v \geq 0$. Since $v$ represents the phenotypic mutation probability, it is restricted to the interval $[0,1]$. To determine the sign of $p(v)$ over $v \in [0, 1]$, it therefore suffices to examine the boundary values at $v = 0$ and $v = 1$. At $v = 0$, we obtain
\begin{align}
	p(0) = (Nu+1)(Nu+2)(Nu+3) \left[ N(T_1 - R_1 + P_1 - S_1)u + 3T_1 - R_1 + P_1 - 3S_1\right],
\end{align}
which is positive under the condition stated above. At $v = 1$, however, the sign of $p(1)$ can be either positive or negative. For example, when $P_1 \to R_1$, we have
\begin{align}
	p(1) = (T_1 - S_1)(Nu+N+3)^2(Nu+2N+2)(Nu+1) > 0,
\end{align}
whereas when $S_1 \to P_1$ and $R_1 \to T_1$, we get
\begin{align}
	p(1) = -2(T_1 - P_1)\left[(u+1)^2N^4 + \mathcal{O}(N^3) \right] < 0.
\end{align}
Accordingly, the average cooperation level $\langle p \rangle$ as a function of $v$ over $v \in [0,1]$ may either increase monotonically or increase initially and then decrease as $v$ grows.

\begin{figure}[htpb!]
	\centering
	\includegraphics[width=\columnwidth]{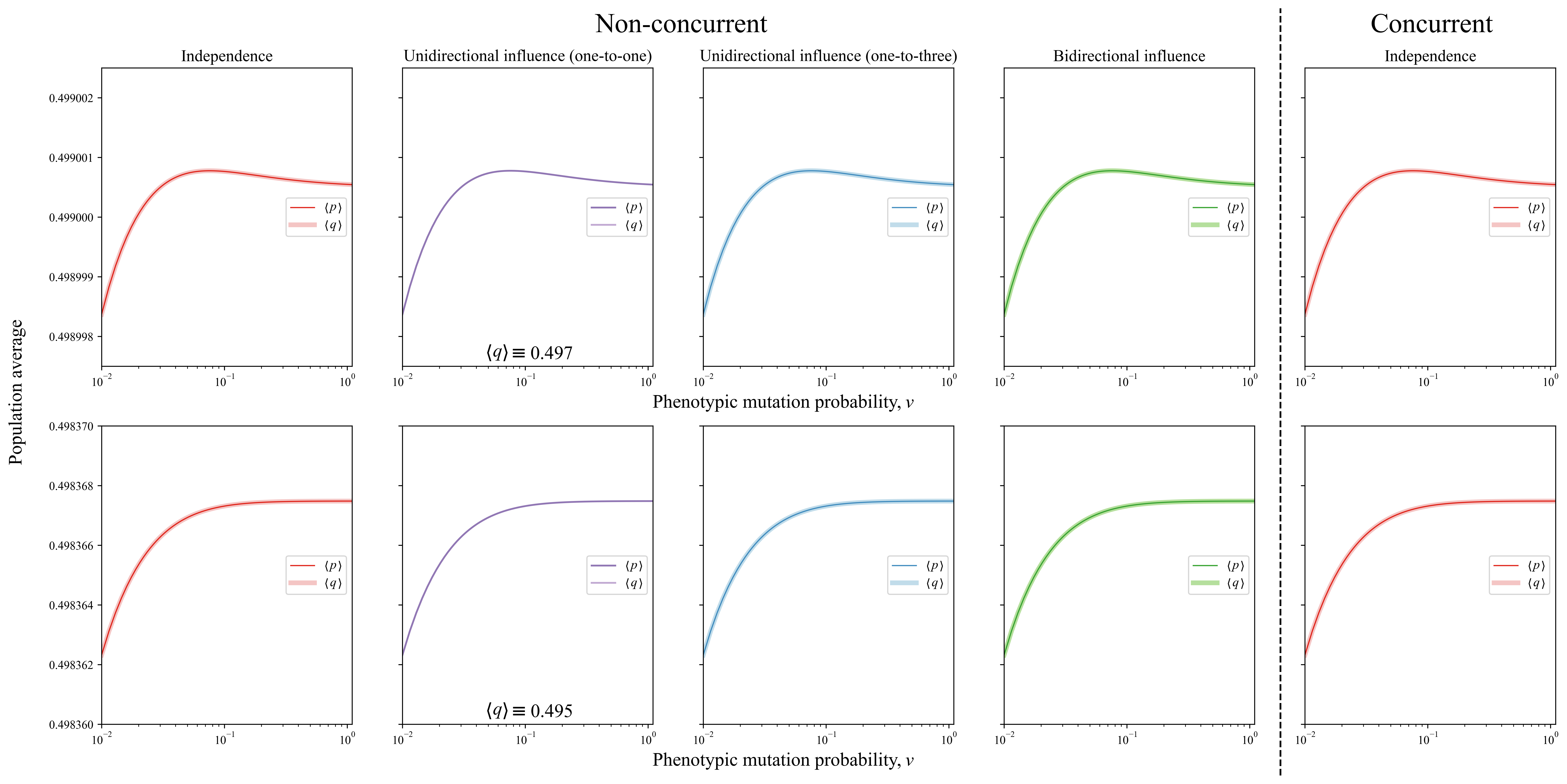}
	\captionsetup{font=small}
	\caption{\textbf{Cooperation levels (population averages) as functions of the phenotypic mutation probability in large populations.} Each column corresponds to a specific type of interlayer phenotypic relationship, as indicated at the top. Solid curves denote theoretical predictions, where the darker curve represents $\langle p \rangle$ and the lighter curve represents $\langle q \rangle$. Parameters: $N = 5 \times 10^6$, $n = 20 \times 20 = 400$, $\beta = 0.001$, $r_1 = 3$ and $r_2 = 3, 3, 5, 3$ from left to right (excluding the bidirectional influence case), $[R_1, S_1, T_1, P_1] = [R_2, S_2, T_2, P_2] = [3, 0, 5, 1]$ for the first row, $[R_1, S_1, T_1, P_1] = [R_2, S_2, T_2, P_2] = [1.1, 0, 5, 1]$ for the second row, $u= 0.04$.}
\label{fig:langle_p_q_v}
\end{figure}

We show the dependence of $\langle p \rangle$ and $\langle q \rangle$ on $v$ for two different parameter settings in Fig.~\ref{fig:langle_p_q_v}. In the top row, corresponding to a more favorable environment with $R_m = 3$, both $\langle p \rangle$ and $\langle q \rangle$ increase initially and then decrease as $v$ grows. In the bottom row, corresponding to a harsher environment with $R_m = 1.1$, both quantities increase monotonically with $v$.

\section{Comparison between synchronous and asynchronous updating processes}
\label{section:Comparison}

\subsection{Wright-Fisher process}

In this section, we analyze the evolutionary dynamics under the Wright-Fisher process~\cite{fisher1930genetical,wright1931evolution}, in which generations are non-overlapping and the entire population updates synchronously. The population size remains constant, and individuals reproduce in proportion to their fitness. Since the population is well-mixed, the computation of individual fitness and average strategy abundance follows a procedure analogous to that used for the Moran process. However, due to the synchronous updating scheme, the pair and triplet correlations, denoted by $\langle \, \cdot \, \rangle$, differ from those in the asynchronous setting and hence need to be rederived. 

We consider neutral drift in a population of size $N$ under the Wright-Fisher process. In each generation, the population produces exactly $N$ offspring in total, and each offspring independently chooses its parent from the previous generation with equal probability $1/N$. Consequently, for any given offspring, the probability that it descends from a particular individual is $1/N$, and for two sampled offspring, the probability that they coalesce in a single generation, that is, both lineages are inherited from the same parent, is $1/N$. Let $T$ denote the number of generations until the two lineages coalesce. Then $T$ follows a geometric distribution,
\begin{align}
	P(T = t) = \left( 1-\dfrac{1}{N} \right)^{t-1}\cdot \dfrac{1}{N}.
\end{align}
For large population size ($N \gg 1$), the cumulative distribution function of $T$ is
\begin{align}
	P(T \le t) = \sum_{i=1}^t P(T = i) = \dfrac{1}{N} \sum_{i=1}^t \left( 1-\dfrac{1}{N} \right)^{i-1} = 1-\left( 1-\dfrac{1}{N} \right)^{t} \to 1-e^{-\frac{t}{N}}.
\end{align}

By rescaling time as $\tau = t/N$, we obtain the \emph{coalescent time} for the Wright-Fisher process. Analogous to the Moran process, in the limit of $N \gg 1$, the probability density functions in continuous time $\tau$ are given by
\begin{equation}
\begin{aligned}
	T_2(\tau_2) &= \dfrac{\mathrm{d}}{\mathrm{d}\tau_2}\left(1-e^{-\tau_2}\right) = e^{-\tau_2}, \\
	T_3(\tau_2, \tau_3) &= e^{-\tau_2} \cdot  \dfrac{\mathrm{d}}{\mathrm{d}\tau_3}\left(1-e^{-3\tau_3}\right) = 3e^{-(\tau_2 + 3\tau_3)}.
\end{aligned}
\end{equation}
We thus observe that, after rescaling time, the coalescent time distributions under the Wright-Fisher process are formally identical to those obtained for the Moran process.

Along the ancestral lineages of any two sampled individuals, strategy mutations occur at a rate $\mu = 2Nu$, while phenotypic mutations occur at a rate $\nu = 2Nv$. As a concrete example, we consider the case of non-concurrent mutation and independent traits (see Subsection~\ref{subsection:independence} in Section~\ref{section:types_of_dependencies} and Fig.~\ref{fig:blood_hair}). Following a derivation similar to that in previous sections, we obtain
\begin{equation}
\begin{dcases}
	Q_{11}(\tau) = w_1(\tau)w_2(\tau), \\
	Q_{10}(\tau) = w_1(\tau)\left[ 1 - w_2(\tau) \right], \\
	Q_{01}(\tau) = \left[ 1 - w_1(\tau) \right] w_2(\tau),
\end{dcases}
\end{equation}
where
\begin{align}
	w_m(\tau) = e^{-\nu\tau}+\dfrac{1-e^{-\nu\tau}}{r_m}, \qquad m = 1, 2.
\end{align}
Accordingly, the triplet correlation $\langle x_i x_i^{k_1 k_2} x_i^{k_1 k_2} \rangle$ can be expressed as
\begin{align}
	\langle x_i x_i^{k_1 k_2} x_i^{k_1 k_2} \rangle = \dfrac{1}{3n} \int_{0}^{\infty}\int_{0}^{\infty} T_3(\tau_2,\tau_3)s_3(\tau_2,\tau_3) \left[Q_{11}(\tau_3) + 2Q_{11}(\tau_2+\tau_3)\right] \mathrm{d}\tau_2 \mathrm{d}\tau_3.
\label{X_i_X_i_k1k2_X_i_k1k2_tau'}
\end{align}
The remaining correlation terms can be derived analogously, with appropriate modifications to \eqref{eq:correlation} and \eqref{eq:correlation_more}.

Finally, we obtain the $\sigma$-rule under the Wright-Fisher process,
\begin{align}
	\sigma_mR_m + S_m > T_m + \sigma_mP_m, \qquad m = 1,2
\label{sigma'_m-rule}
\end{align}
where
\begin{align}
	\sigma_m = \dfrac{(\mu + \nu + 1) \left[ r_m(\mu + 2\nu + 3)+\nu(\mu + \nu + 2) \right]}{(\mu + \nu + 3)\left[ r_m(\mu + 1)+\nu(\mu + \nu + 2) \right]}.
\label{W_F_sigma_rule}
\end{align}
The expression has the same form as \eqref{eq:sigma_independence} obtained under the Moran process.

In the Moran process, each generation consists of two sequential steps: one individual is selected to reproduce, and another is chosen to die, leading to an \textbf{asynchronous} updating process. In contrast, the Wright-Fisher process features random reproduction across the entire population, where the next generation is formed simultaneously from the offspring pool. This corresponds to a \textbf{synchronous} updating process. In the following subsections, we will provide a comparison between these two updating mechanisms.

\subsection{Comparison of \texorpdfstring{$\sigma$}{sigma}-rules}
\label{section:Comparison_sigma}

As noted above, \eqref{W_F_sigma_rule} is formally identical to \eqref{eq:sigma_independence}. If we further specialize to the donation game, characterized by the payoff matrix $[R, S, T, P] = [b - c, -c, b, 0]$, \eqref{W_F_sigma_rule} yields the critical benefit-to-cost ratio
\begin{align}
	\left( \dfrac{b}{c} \right)^{\ast} = \dfrac{\sigma_m + 1}{\sigma_m - 1} =\dfrac{1}{r_m - 1}(\mu + \nu + 2)+\dfrac{r_m}{r_m - 1} \dfrac{\mu^2 + 2\mu(\nu + 2) +\nu^2 + 3\nu + 3}{\nu(\mu + \nu + 2)}.
\end{align}
We emphasize that this expression exhibits a striking formal similarity to the results obtained by Antal et al.~\cite{nowak2010evolutionary, tarnita2009evolutionary} for structured populations within evolutionary game theory. This resemblance suggests that, when viewed through the lens of coalescent theory, asynchronous and synchronous updating processes can lead to consistent conditions for the emergence of cooperation.

\subsection{Comparison of evolutionary dynamics}
\label{section:Comparison_phenotype}

Under our model assumptions, phenotypic mutations occur with probability $v$ in a population of $N$ individuals. Accordingly, in the Moran process, the effective phenotypic mutation rate along a lineage is $\nu = Nv$. In the limit of infinitely many possible phenotypes, the ancestral phenotype label can be approximated as undergoing a random walk in a phenotype space driven by mutation. On this coalescent timescale, the characteristic diffusion rate of phenotype labels is of order $Nv$. Therefore, individuals tend to remain grouped within clusters whose typical size scales as $\mathcal{O}(\sqrt{Nv})$, and these clusters collectively diffuse through the phenotype space. Tracing the ancestral line of a randomly chosen individual backward in rescaled time $\tau$, the phenotype label $k$ changes by one unit at rate $\nu/2$ in either direction, corresponding to a symmetric random walk approximation.

A closely related description was obtained by Antal et al.~\cite{antal2009evolution2}, who considered phenotype evolution under the Wright-Fisher process in the limit of infinitely many phenotypes. In their framework, each phenotype diffuses through the phenotype space at rate $2Nv$, and individuals remain grouped in clusters whose typical size scales as $\mathcal{O}(\sqrt{2Nv})$. This is consistent with our rescaling here, where the effective phenotypic mutation rate is $\nu = 2Nv$. Hence, with appropriate rescaling of time, synchronous and asynchronous updating schemes yield formally consistent descriptions of phenotypic diffusion.

A similar correspondence also arises in the evolution of strategies. However, unlike phenotypes, strategy dynamics are influenced by selection in addition to mutation. While the selective component depends explicitly on the updating rule and therefore differs between the Moran and Wright-Fisher processes, the mutational component alone exhibits a direct correspondence: the effective strategy mutation rate is $\mu = Nu$ under the Moran process and $\mu = 2Nu$ under the Wright-Fisher process. This mirrors the relationship observed for phenotypic diffusion and further highlights the consistency between the two updating schemes at the level of evolutionary dynamics driven by mutations.

\section{Collapsing distinct interaction layers}
\label{section:flattening}

Our multiplex interaction model allows multiple games to be represented simultaneously on distinct layers. Collapsing these layers into a single interaction layer simplifies the multi-layer model but removes information about layer-specific interaction structure. We examine how this reduction affects the conditions for cooperation. Following Section~\ref{section:types_of_dependencies}, particularly its Subsection~\ref{subsection:concurrent}, we enumerate all possible phenotype combinations and denote their total number by $H$. For independent traits, for example, $H = r_1r_2$. In the resulting single-layer model, we assume that an individual adopting strategy $i$ cooperates with probability $p_i$. We next consider two rules for selecting interaction partners.

\subsection{Strict matching criterion}

Under the stricter rule, individuals interact only when their phenotype combinations are identical, that is, both phenotypes must match:
\begin{align}
	\psi_{k_1k_2 \circ l_1l_2} =
	\begin{cases}
		\psi_{\mathrm{s}}(=1), & k_1 = l_1 \text{ and } k_2 = l_2, \\
		\psi_{\mathrm{d}}(=0), & \text{otherwise} .
	\end{cases}
\label{eq:Psi_stricter}
\end{align}
This rule treats each distinct phenotype pair as a composite phenotype, yielding $H$ possible ones. The resulting model is formally equivalent to that introduced by Fu et al.~\cite{fu2012evolution}. For a game with payoff matrix $A = [R,S,T,P]$, the $\sigma$-rule is
\begin{align}
	\sigma R + S > T + \sigma P.
\label{eqsigma_rule_stricter}
\end{align} 
The corresponding structure coefficient $\sigma$ is given by
\begin{align}
	\sigma = \dfrac{(\mu+\nu+1) \left[ H(\mu+2\nu+3)+\nu(\mu+\nu+2) \right]}{(\mu+\nu+3)\left[ H(\mu+1)+\nu(\mu+\nu+2) \right]}.
\label{eq:sigma_stricter}
\end{align}

As shown in Section~\ref{section:Extended_Analysis}, the structure coefficient $\sigma_m$ increases with the number of phenotypes, making cooperation more readily favored by selection. The structure coefficient in \eqref{eq:sigma_stricter} has the same dependence on $H$, with $H \geq r_m$ for either layer. Thus, under the collapsed representation, strict phenotypic matching is at least as conducive to the emergence and maintenance of cooperation as the corresponding layer-specific interaction structure.

\subsection{Permissive matching criterion}

Under the more permissive rule, individuals interact whenever they share at least one phenotype. Equivalently, an interaction is excluded only when both phenotypes differ:
\begin{align}
	\psi_{k_1k_2 \circ l_1l_2} =
	\begin{cases}
		\psi_{\mathrm{d}}(=1), & k_1 = l_1 \text{ or } k_2 = l_2, \\
		\psi_{\mathrm{s}}(=0), & \text{otherwise} .
	\end{cases}
\label{eq:Psi_permissive}
\end{align}
The distribution of the cooperation probability $p$ is
\begin{equation}
\begin{aligned}
	D(p) &= \lambda_1 \left( A(p, p) - \int_0^1 A(p', p') \mathrm{d}p' \right) + \lambda_2 \left( \int_0^1A(p, p') \mathrm{d}p' - \int_0^1 A(p', p) \mathrm{d}p' \right) \\
	&+ \lambda_3 \left( \int_0^1A(p, p') \mathrm{d}p' - \int_0^1 \int_0^1 A(p, p') \mathrm{d}p\mathrm{d}p' \right), \\
\end{aligned}
\end{equation}
where the three coefficients $\lambda_i$ are given by
\begin{equation}
\begin{dcases}
	\lambda_1 \propto & \langle x_i x_j^{k_1 k_2} x_j^{k_1 k_2} \rangle + \langle x_i x_j^{k_1 k_2} x_j^{k_1 l_2} \rangle + \langle x_i x_j^{k_1 k_2} x_j^{l_1 k_2} \rangle \\
	& -  \langle x_i x_j^{k_1 k_2} x_r^{k_1 k_2} \rangle - \langle x_i x_j^{k_1 k_2} x_r^{k_1 l_2} \rangle - \langle x_i x_j^{k_1 k_2} x_r^{l_1 k_2} \rangle, \\
	\lambda_2 \propto & \langle x_i x_i^{k_1 k_2} x_j^{k_1 k_2} \rangle + \langle x_i x_i^{k_1 k_2} x_j^{k_1 l_2} \rangle + \langle x_i x_i^{k_1 k_2} x_j^{l_1 k_2} \rangle \\
	& - \langle x_i x_j^{k_1 k_2} x_r^{k_1 k_2} \rangle - \langle x_i x_j^{k_1 k_2} x_r^{k_1 l_2} \rangle - \langle x_i x_j^{k_1 k_2} x_r^{l_1 k_2} \rangle, \\
	\lambda_3 \propto & n \left( \langle x_i x_j^{k_1 k_2} x_r^{k_1 k_2} \rangle + \langle x_i x_j^{k_1 k_2} x_r^{k_1 l_2} \rangle + \langle x_i x_j^{k_1 k_2} x_r^{l_1 k _2} \rangle \right).
\end{dcases}
\label{eq:lambda_permissive}
\end{equation}

Natural selection therefore favors cooperation if and only if
\begin{align}
	\int_0^1 pD(p) \mathrm{d}p > 0.
\end{align}
This condition yields a $\sigma$-rule of the same form as \eqref{eqsigma_rule_stricter}, with
\begin{align}
	\sigma = \dfrac{(\mu + \nu + 1)(\mu + 2\nu + 1)C_{R,P}}{(\mu + \nu + 3)(\mu + 2\nu + 3)C_{S,T}}.
\end{align}
The auxiliary coefficients are
\begin{equation}
\begin{aligned}
	C_{R,P} = & - 2\mu^2\nu^2 - 6\mu\nu^3 - 10\mu\nu^2 - 4\nu^4 - 12\nu^3 - 10\nu^2 \\
	& + (r_1+r_2)(2\mu^2\nu^2 + 6\mu\nu^3 + 10\mu\nu^2 + 4\nu^4 + 12\nu^3 + 10\nu^2) \\
	& + r_1r_2(3\mu^2\nu + \mu^2 + 11\mu\nu^2 + 21\mu\nu + 6\mu + 12\nu^3 + 37\nu^2 + 36\nu + 9), \\
	C_{S,T} = & -2\mu^2\nu^2 - 6\mu\nu^3 - 6\mu\nu^2 - 4\nu^4 - 12\nu^3 - 6\nu^2 \\
	& + (r_1+r_2)(2\mu^2\nu^2 + 6\mu\nu^3 + 6\mu\nu^2 + 4\nu^4 + 12\nu^3 + 6\nu^2) \\
	& + r_1r_2(3\mu^2\nu + \mu^2 + 7\mu\nu^2 + 9\mu\nu + 2\mu + 7\nu^2 + 6\nu + 1).
\end{aligned}
\end{equation}

We compare the structure coefficient under permissive matching with that of the layer having the larger phenotype space. For $\nu > 0$, the layer-specific coefficient $\sigma_m$ in~\eqref{eq:sigma_independence} increases strictly with $r_m$. Without loss of generality, we label the layers such that $r_1 \geq r_2 \geq 1$, hence $\sigma_1 = \max\{\sigma_1,\sigma_2\}$. We then evaluate the difference:
\begin{align}
	\sigma-\sigma_1 = \dfrac{-2\nu(\nu+1)(r_1-1)(\mu+\nu+1)C_{\mathrm{nume}}}{(\mu+\nu+3)(\mu+2\nu+3)(\mu\nu+\mu r_1+\nu^2+2\nu+r_1)C_{\mathrm{deno}}}.
\end{align}
Here, $C_{\text{nume}}$ and $C_{\text{deno}}$ are
\begin{equation}
\begin{aligned}
	C_{\text{nume}} = \; & r_1(2\mu^3\nu+14\mu^2\nu+8\mu^2\nu^2+10\mu\nu^3+38\mu\nu^2+30\mu\nu+4\nu^4+24\nu^3+44\nu^2+22\nu) \\
	&+ r_1r_2(\mu^3+3\mu^2\nu+6\mu^2+2\mu\nu^2+12\mu\nu+11\mu+2\nu^2+9\nu+6) \\
	&+(r_1-r_2+1)(2\mu^2\nu^2+6\mu\nu^3+8\mu\nu^2+4\nu^4+12\nu^3+8\nu^2), \\
	C_{\text{deno}} = \;& r_1r_2(3\mu^2\nu+\mu^2+7\mu\nu^2+9\mu\nu+2\mu+7\nu^2+6\nu+1) \\
	&+(r_1+r_2-1)(2\mu^2\nu^2+6\mu\nu^3+6\mu\nu^2+4\nu^4+12\nu^3+6\nu^2).
\end{aligned}
\end{equation}
For $\mu \geq 0$ and $\nu > 0$, the ordering $r_1 \geq r_2 \geq 1$ ensures that $C_{\mathrm{nume}}$, $C_{\mathrm{deno}}$, and all remaining denominator factors are positive. Hence, whenever $r_1 > 1$, we obtain $\sigma < \sigma_1$. By symmetry, if the layer labels satisfy $r_2 > r_1$, the same argument gives $\sigma < \sigma_2$. Thus, irrespective of the labeling of the layers, $\sigma < \max\{\sigma_1,\sigma_2\}$, given $\nu>0$ and $\max\{r_1, r_2\} > 1$. Equality is recovered when $\nu = 0$ or $r_1 = r_2 = 1$, in which case all three structure coefficients equal one. Thus, under the collapsed representation, permissive phenotypic matching makes selection less favorable to the emergence and maintenance of cooperation than does the corresponding layer-specific interaction structure.

Together, these results show that the consequences of collapsing interaction layers depend on the matching rule. For a fixed payoff matrix with $R > P$, strict matching is at least as conducive to cooperation as either layer separately, whereas permissive matching is no more conducive than the layer with the larger phenotype space. In terms of admissible interaction partners, each multiplex layer lies between strict and permissive matching, with a correspondingly intermediate mean degree. This degree ordering provides a parallel to the role of interaction neighborhoods in shaping cooperation through network reciprocity~\cite{ohtsuki2006simple}.

\section{Alternative payoff formulations}
\label{section:normalized}

\subsection{Payoffs averaged over interactions}
\label{section:interaction_normalized_payoff}

The baseline payoff in~\eqref{eq:payoff_i} averages accumulated payoffs over individuals sharing a strategy, while retaining differences in their interaction opportunities. As an alternative, we consider payoffs averaged over interactions. At the strategy level, the total number of interactions involving individuals adopting strategy $i$ on layer $m$ is
\begin{equation}
	\varphi_i^m = \sum_{j=1}^n \varphi_{ij}^m .
\label{eq:total_interaction_weight}
\end{equation}
Thus, the expected number of interactions per individual taking strategy $i$ on layer $m$ is $\varphi_i^m/(Nx_i)$. The payoff averaged over interactions is defined as
\begin{align}
	\pi_i = \dfrac{1}{Nx_i} \sum_{m=1}^2\sum_{j=1}^n\dfrac{\varphi_{ij}^m \pi_{ij}^m}{\varphi_i^m/(Nx_i)} = \sum_{m=1}^2\sum_{j=1}^n\dfrac{\varphi_{ij}^m \pi_{ij}^m}{\varphi_i^m}.
\label{eq:interaction_normalized_payoff}
\end{align}
For each layer, this expression pools payoffs and interaction counts over individuals sharing strategy $i$ before taking their ratio. A layer with $\varphi_i^m = 0$ is assigned a zero contribution.

\begin{figure}[htpb!]
	\centering
	\includegraphics[width=\columnwidth]{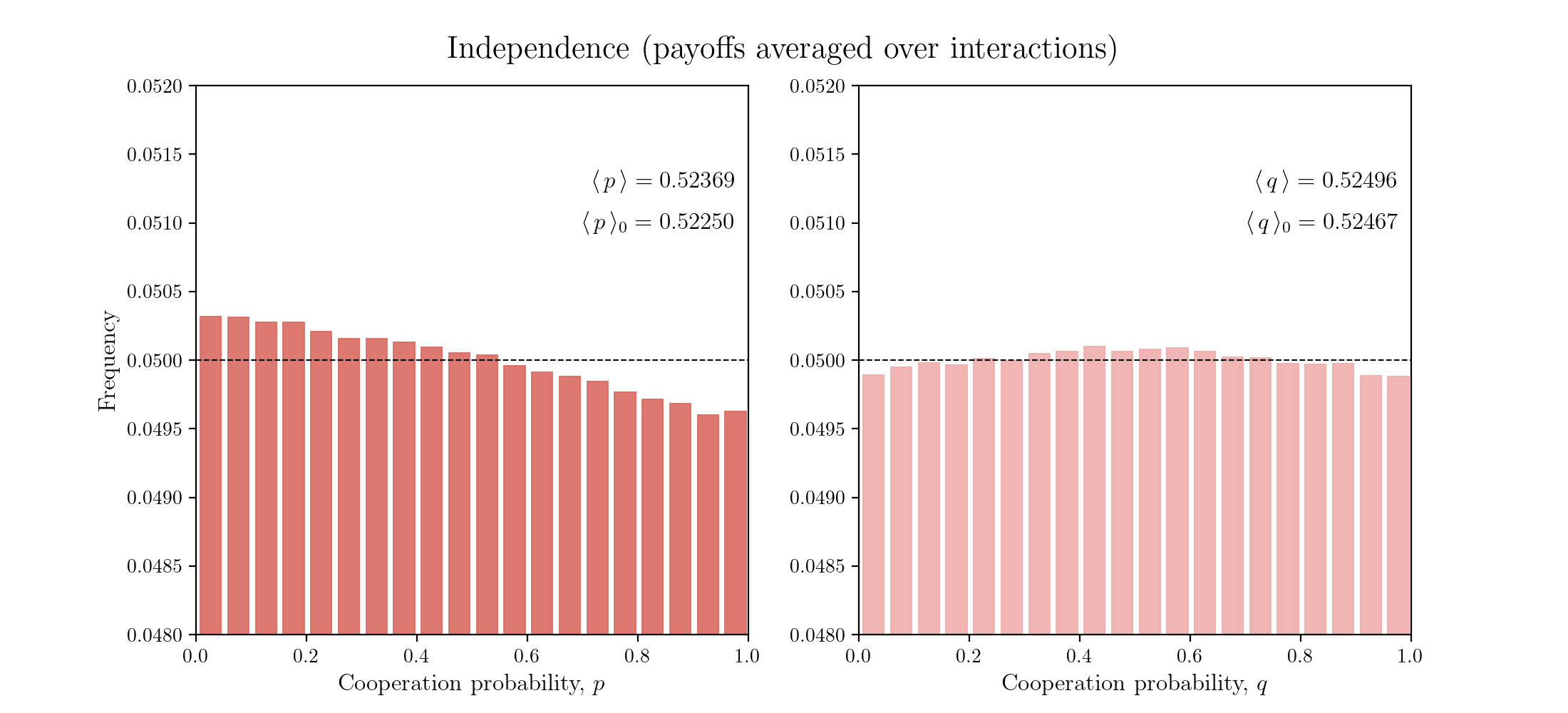}
	\captionsetup{font=small}
	\caption{\textbf{Simulation results with payoffs averaged over each individual's interactions under non-current phenotypic mutations and independence.} In each panel, the black dashed line indicates the neutral expectation, that is, the uniform proportion $1/20$. The legends $\langle p\rangle_{0}$ and $\langle q\rangle_{0}$ denote the population-averaged cooperation probabilities for the baseline setting in Fig.~\ref{fig:simulation_independence} of Section~\ref{section:types_of_dependencies}, whereas $\langle p\rangle$ and $\langle q\rangle$ denote the corresponding probabilities in the present setting. The histograms show results from agent-based simulations. Parameters: $N = 50$, $n = 20 \times 20 = 400$, $\beta = 0.001$, $r_1 = r_2 = 3$, $u = 0.04$, $v = 0.02$, $[R_1,S_1,T_1,P_1] = [3, 0, 5, 1]$, and $[R_2, S_2, T_2, P_2] = [3, 1, 5, 0]$. Simulation results were averaged over $1 \times 10^9$ generations.}
\label{fig:simulation_independence_interaction_normalized}
\end{figure}

In the simulations, we instead normalize each individual's accumulated payoff by its own interaction count. For any individual $a$, let $i_a$ denote its strategy and $k_a^m$ its phenotype on layer $m$. Under perfect homophily, its number of interaction partners and accumulated payoff on that layer are
\begin{equation}
    d_a^m = \sum_{b \neq a}\psi_{k_a^m k_b^m}^m, \qquad \pi_a^m = \sum_{b \neq a}\psi_{k_a^m k_b^m}^m\pi_{i_a i_b}^m.
\end{equation}
The payoff entering its fitness is
\begin{equation}
    \pi_a = \sum_{m = 1}^{2}\frac{\pi_a^m}{d_a^m},
    \qquad
    f_a = \exp\left(\beta\pi_a\right).
\label{eq:individual_interaction_normalized_payoff}
\end{equation}
Here, self-interactions are excluded, and a layer contributes zero when the individual has no eligible interaction partners. This individual calculation in general differs from normalization after aggregation by strategy and allows individuals sharing the same strategy to have different fitness values. Compared with the baseline results in Fig.~\ref{fig:simulation_independence}, the distributions in Fig.~\ref{fig:simulation_independence_interaction_normalized} retain their qualitative shapes but are flatter, with smaller differences in strategy abundance. In fact, averaging payoffs over interactions can reduce payoff differences between individuals, thereby weakening fitness differences even when the selection strength $\beta$ is unchanged.

\subsection{Payoff weighted by strategy frequency}

The baseline payoff $\pi_i$ in~\eqref{eq:payoff_i} averages accumulated payoffs over individuals adopting strategy $i$. To examine the effect of explicitly weighting this payoff by strategy frequency, we consider
\begin{equation}
    \tilde{\pi}_i = \dfrac{1}{N}\sum_{m=1}^2\sum_{j=1}^n\varphi_{ij}^m\pi_{ij}^m = x_i\pi_i,
    \qquad
    \tilde{f}_i = \exp\left(\beta\tilde{\pi}_i\right).
\label{eq:frequency_weighted_payoff}
\end{equation}
The baseline payoff already depends on population composition through interactions with other strategies. The additional factor $x_i$ introduces a direct dependence on the frequency of the focal strategy. Holding a positive baseline payoff fixed, a more frequent strategy receives a larger effective payoff. Thus, this weighting introduces an additional advantage for common strategies. A biological motivation for such frequency-dependent benefits comes from warning coloration: common warning signals can provide greater protection against predators because they are more readily recognized and avoided, giving individuals carrying these signals a survival advantage~\cite{mathieu2016warning}. We use this example to motivate an explicit frequency effect, represented here by the factor $x_i$.

\begin{figure}[htpb!]
	\centering
	\includegraphics[width=\columnwidth]{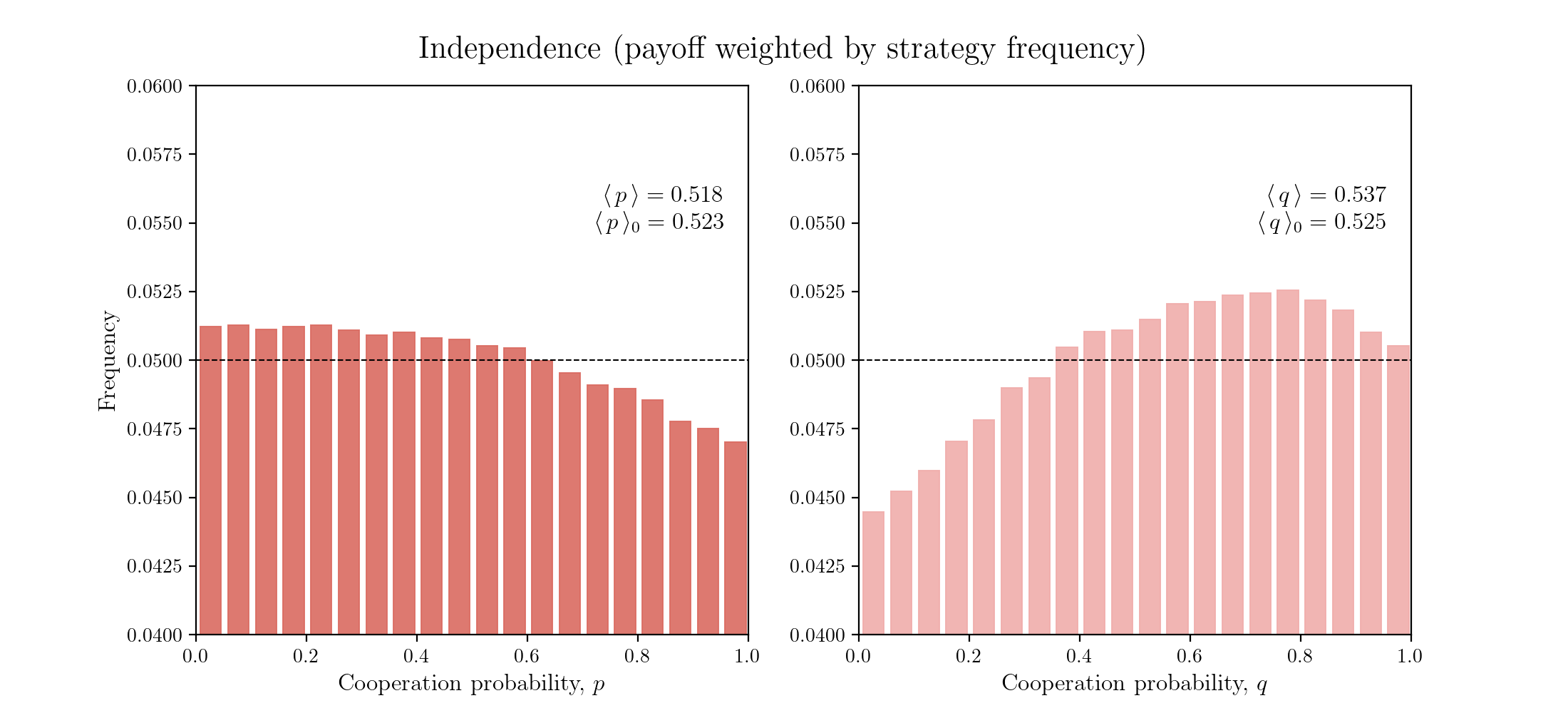}
	\captionsetup{font=small}
	\caption{\textbf{Simulation results with payoffs weighted by strategy frequency under non-concurrent phenotypic mutations and independence.} In each panel, the black dashed line indicates the neutral expectation, that is, the uniform proportion $1/20$. The legends $\langle p\rangle_{0}$ and $\langle q\rangle_{0}$ denote the population-averaged cooperation probabilities for the baseline setting considered in Fig.~\ref{fig:simulation_independence} of Section~\ref{section:types_of_dependencies}, whereas $\langle p\rangle$ and $\langle q\rangle$ denote the corresponding probabilities in the present setting. The histograms show results from agent-based simulations. Parameters: $N = 50$, $n = 20 \times 20 = 400$, $\beta = 0.001$, $r_1 = r_2 = 3$, $u = 0.04$, $v = 0.02$, $[R_1,S_1,T_1,P_1] = [3, 0, 5, 1]$, and $[R_2, S_2, T_2, P_2] = [3, 1, 5, 0]$. Simulation results were averaged over $1 \times 10^{11}$ generations.}
\label{fig:simulation_independence_unnormalized}
\end{figure}

We performed simulations using the frequency-weighted payoff formulation as shown in Fig.~\ref{fig:simulation_independence_unnormalized}. Under the parameters considered, mean cooperation decreased in the prisoner's dilemma layer and increased in the snowdrift layer, widening the difference between the two layers relative to the baseline results in Fig.~\ref{fig:simulation_independence}.

\section{Imperfect homophily}
\label{section:imperfect_homophily}

The preceding sections assumed perfect homophily: two individuals interact on a given layer if and only if they share the same phenotype on that layer. We can relax this assumption and consider imperfect homophily. Instead of the binary interaction rule in~\eqref{eq:Psi}, we use a distance-dependent interaction model. As before, phenotypes on layer $m$ are indexed by $\{1, 2, \ldots, r_m\}$ along an ordered scale, such as the number of leaves in Fig.~\ref{fig:Leaf_Root}. We define the distance between phenotypes $k$ and $l$ as $d_{kl} = |k-l| \in \{0, 1, \ldots, r_m - 1\}$. The relative probability of interaction between individuals with these phenotypes is then
\begin{equation}
	\psi_{kl}^m = \begin{cases}
		1, & r_m = 1, \\
		1 - \dfrac{d_{kl}}{r_m-1}, & r_m > 1 .
	\end{cases}
\label{eq:Psi_imperfect_homophily}
\end{equation}

\begin{figure}[htpb!]
	\centering
	\includegraphics[width=\columnwidth]{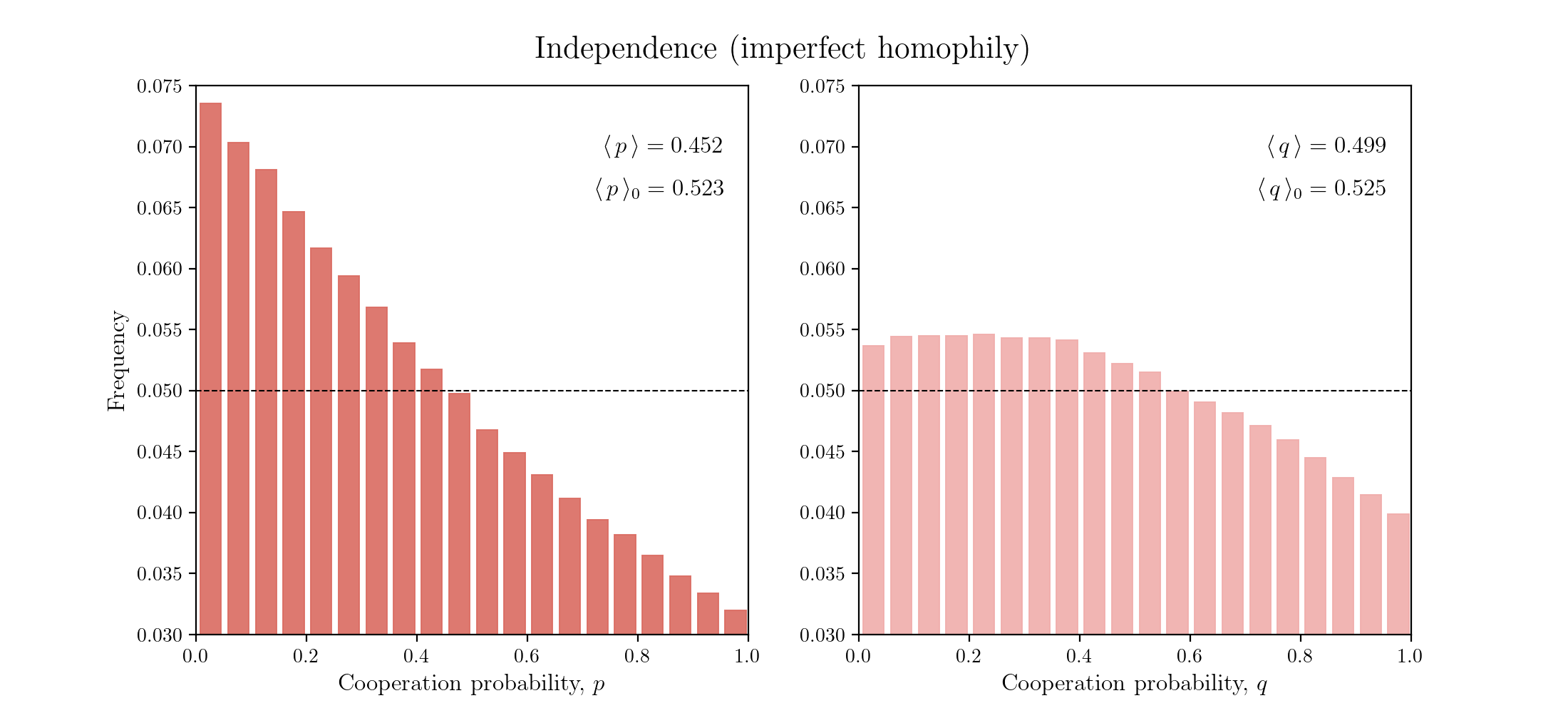}
	\captionsetup{font=small}
	\caption{\textbf{Simulation results under a distance-dependent interaction model with non-concurrent phenotypic mutations and independence.} In each panel, the black dashed line indicates the neutral expectation, that is, the uniform proportion $1/20$. The legends $\langle p\rangle_{0}$ and $\langle q\rangle_{0}$ denote the population-averaged cooperation probabilities for the baseline setting considered in Fig.~\ref{fig:simulation_independence} of Section~\ref{section:types_of_dependencies}, whereas $\langle p\rangle$ and $\langle q\rangle$ denote the corresponding probabilities in the present setting. The histograms show results from agent-based simulations. Parameters: $N = 50$, $n = 20 \times 20 = 400$, $\beta = 0.001$, $r_1 = r_2 = 3$, $u = 0.04$, $v = 0.02$, $[R_1,S_1,T_1,P_1] = [3, 0, 5, 1]$, and $[R_2, S_2, T_2, P_2] = [3, 1, 5, 0]$. Simulation results were averaged over $1 \times 10^9$ generations.}
\label{fig:simulation_independence_change_psi}
\end{figure}

We perform simulations under this imperfect-homophily interaction rule (Fig.~\ref{fig:simulation_independence_change_psi}) and compare the results with the perfect-homophily baseline (Fig.~\ref{fig:simulation_independence}). In both the prisoner's dilemma and the snowdrift game, the simulated strategy distributions shift toward lower cooperation probabilities, resulting in lower population-averaged cooperation. The distance-dependent rule retains interactions between individuals sharing the same phenotype while allowing additional interactions across phenotype groups. These results suggest that broadening the range of interaction partners can weaken the support for cooperation provided by phenotypic homophily. This pattern also parallels the classic result of network reciprocity that a larger number of neighbors raises the benefit-to-cost threshold required for cooperation to be favored~\cite{ohtsuki2006simple}.

\section{Beyond weak selection: multiplex versus isolated-layer dynamics}
\label{section:strong_selection}

Throughout the main text, we derive the layer-specific $\sigma$-rules in the limit of weak selection, where game payoffs have only a small effect on individual fitness~\cite{traulsen2007pairwise}. To determine whether isolated-layer dynamics remain predictive when selection is stronger, we compare the multiplex model with matched single-layer benchmarks in the simplest case of non-concurrent mutation and independent phenotypes.

\begin{figure}[htpb!]
	\centering
	\includegraphics[width=\columnwidth]{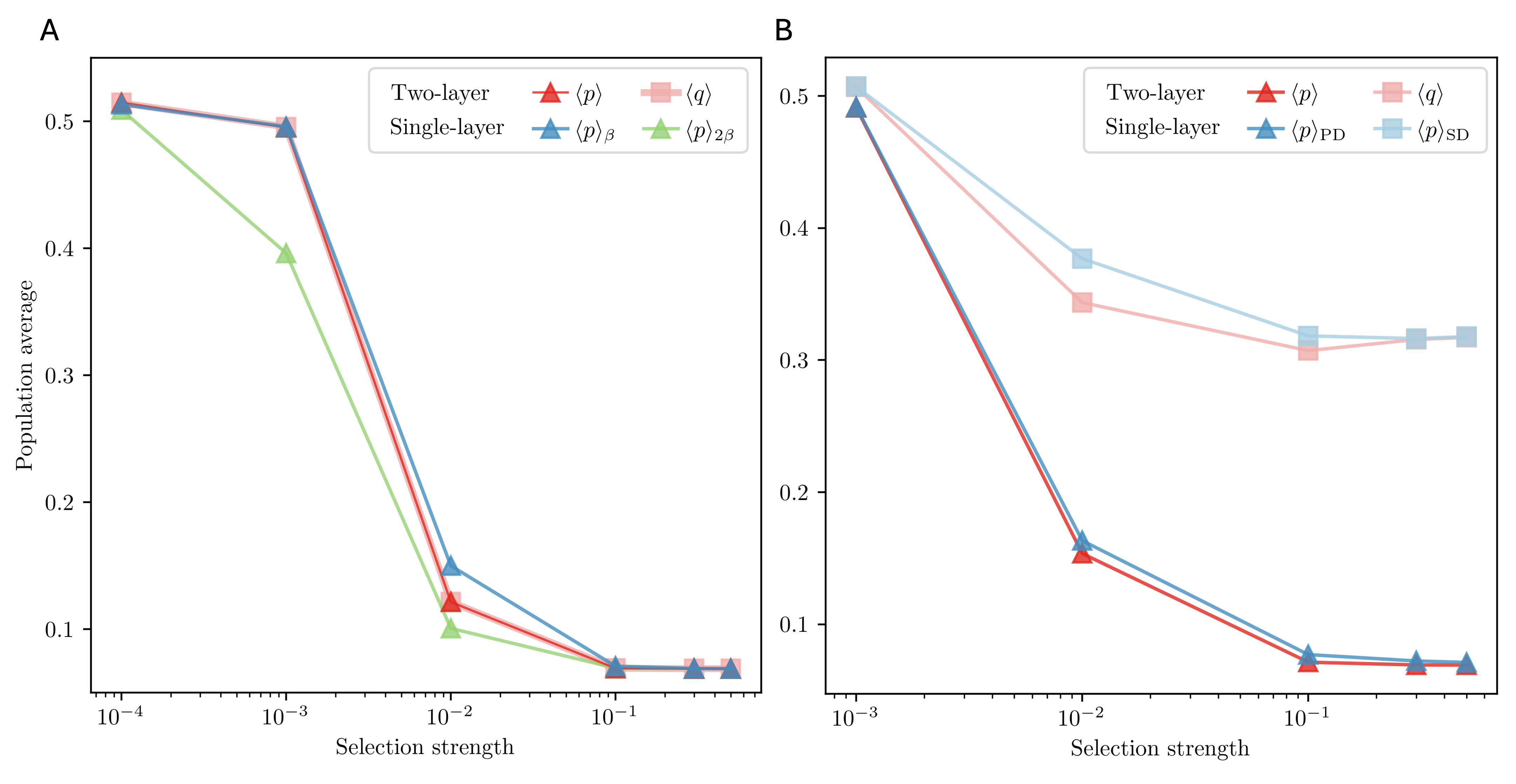}
	\captionsetup{font=small}
	\caption{\textbf{Cooperation levels (population average) in two-layer and single-layer systems across selection strengths under non-concurrent phenotypic mutations and independence.} (A) Symmetric-game benchmark. Both layers of the multiplex system use the same payoff matrix and contain the same number of phenotypes, which are assigned independently across layers. The corresponding homophilic interaction neighborhoods need not coincide. The multiplex fitness is defined as $f_i = \exp[\beta(\pi_i^1 + \pi_i^2)]$. Red triangles and light red squares denote $\langle p \rangle$ and $\langle q \rangle$ in the first and second layers of the two-layer system, respectively. Blue triangles denote the isolated single-layer result $\langle p \rangle_{\beta}$, obtained using $f_i = \exp(\beta\pi_i)$. Green triangles denote $\langle p \rangle_{2\beta}$, obtained using $f_i = \exp(2\beta\pi_i)$. Parameters: $N = 50$, $n = 20 \times 20 = 400$, $u = 0.04$, $v = 0.02$, and $\beta \in \{0.0001, 0.001, 0.01, 0.1, 0.3, 0.5\}$. For the two-layer system, $r_1 = r_2 = 3$ and $[R_1, S_1, T_1, P_1] = [R_2, S_2, T_2, P_2] = [3, 0, 5, 1]$. For the single-layer system, the number of phenotypes was $M = 3$ and the payoff structure was $[R, S, T, P] = [3, 0, 5, 1]$. (B) Heterogeneous-game benchmark. Red triangles and light red squares denote $\langle p \rangle$ and $\langle q \rangle$ in the first and second layers of the two-layer system, respectively. The first layer uses the prisoner’s dilemma (PD), whereas the second uses the snowdrift game (SD). Blue triangles and light blue squares denote $\langle p \rangle_{\mathrm{PD}}$ and $\langle p \rangle_{\mathrm{SD}}$ in the corresponding isolated single-layer PD and SD systems, respectively. Each isolated system retains the phenotype number, payoff matrix, mutation parameters, and selection strength of its corresponding multiplex layer. Parameters: $N = 50$, $n = 20 \times 20 = 400$, $u = 0.04$, $v = 0.02$, and $\beta \in \{0.001, 0.01, 0.1, 0.3, 0.5\}$. For the two-layer system, $r_1 = 3$, $r_2 = 4$, $[R_1, S_1, T_1, P_1] = [3, 0, 5, 1]$, and $[R_2, S_2, T_2, P_2] = [3, 1, 5, 0]$. All simulation results were averaged over $1\times10^{10}$ generations.}
\label{fig:simulation_selection_strength}
\end{figure}

We first consider a symmetric benchmark in which both layers use the same payoff matrix of a prisoner's dilemma and hold the same number of phenotypes, as shown in Fig.~\ref{fig:simulation_selection_strength}A. It can be expected from symmetry that the two multiplex cooperation levels, $\langle p \rangle$ and $\langle q \rangle$, are nearly identical. We compare them with an isolated single-layer system evaluated at selection strengths $\beta$ and $2\beta$. The latter is a scaled benchmark corresponding to two coincident copies of the same payoff signal; it is not assumed to be exactly equivalent to two independent interaction layers. At intermediate selection strengths, the multiplex cooperation level lay between the two single-layer benchmarks:
\begin{equation}
\langle p \rangle_{2\beta} < \langle p \rangle \simeq \langle q \rangle < \langle p \rangle_{\beta}
\end{equation}
Thus, adding a second independent domain promotes selection against cooperation relative to an isolated layer, but less than would an exact duplication of the same single-layer payoff signal. The convergence of all systems as $\beta \to 0$ is consistent with the layerwise separation derived before.

We next consider heterogeneous games, with a prisoner's dilemma on the first layer and a snowdrift game on the second layer, as shown in Fig.~\ref{fig:simulation_selection_strength}B. Under weak selection, each multiplex layer tracks its corresponding isolated single-layer benchmark. As selection increases, however, the multiplex cooperation levels become lower than the single-layer levels on both layers. The differences become most apparent at intermediate selection strengths and diminish as the dynamics approach their regimes of strong selection.

These results highlight that outside the limit of weak selection, the cooperation level in one interaction domain need not be predicted by studying that domain in isolation. Because reproduction acts on the complete strategy pair and depends on the total payoff accumulated across layers, selection generated in one layer can alter the evolutionary outcome in another. The elegant $\sigma$-rules actually characterize a nontrivial modularity of the multiplex system, whereas finite selection can generate cross-layer effects mediated by fitness.

\section{Availability of code and data}
\label{section:code_data}
All source code required to reproduce the results, together with data used for generating the figures, is available at the following GitHub repository, along with detailed usage instructions: \url{https://github.com/xingrucz/Cooperation-Multiplex}.

\end{document}